\documentclass[a4paper, amsfonts, amssymb, amsmath, reprint, showkeys, nofootinbib, twoside, superscriptaddress, accepted=2024-06-15]{quantumarticle}
\pdfoutput=1
\usepackage[english]{babel}
\usepackage[utf8]{inputenc}
\usepackage{amsthm}
\usepackage{mathtools}
\usepackage{physics}
\usepackage{xcolor}
\usepackage{graphicx}
\usepackage[left=23mm,right=13mm,top=35mm,columnsep=15pt]{geometry} 
\usepackage{adjustbox}
\usepackage{placeins}
\usepackage[T1]{fontenc}
\usepackage{lipsum}
\usepackage{csquotes}
\usepackage{tikz}
\usepackage{float}
\usetikzlibrary{quantikz}
\usepackage{overpic}
\usepackage{subfigure}
\usepackage{enumerate}
\usepackage{rotating}
\usepackage{upgreek} 
\usepackage[unicode=true, colorlinks=true]{hyperref} 
\def\quantinuumCambridge{Quantinuum, Terrington House, 13-15 Hills Road, Cambridge CB2 1NL, UK}
\def\quantinuumTokyo{Quantinuum K.K., Otemachi Financial City Grand Cube 3F, 1-9-2 Otemachi, Chiyoda-ku, Tokyo, Japan}

\begin{document}
\title{Quantum Computed Green's Functions using a Cumulant Expansion of the Lanczos Method}
\author{Gabriel Greene-Diniz}
\email{gabriel.greene-diniz@quantinuum.com}
\author{David Zsolt Manrique}
\affiliation{\quantinuumCambridge}
\author{Kentaro Yamamoto}
\affiliation{\quantinuumTokyo}
\author{Evgeny Plekhanov}
\author{Nathan Fitzpatrick}
\author{Michal Krompiec}
\affiliation{\quantinuumCambridge} 
\author{Rei Sakuma}
\affiliation{Materials Informatics Initiative, RD Technology \& Digital Transformation Center, JSR Corporation, 3-103-9, Tonomachi, Kawasaki-ku, Kawasaki, 210-0821, Kanagawa, Japan.}
\author{David Mu{\~n}oz Ramo}
\affiliation{\quantinuumCambridge}

\date{} 

\begin{abstract}
In this paper, we present a quantum computational method to calculate the many-body Green's function matrix in a spin orbital basis. We apply our approach to finite-sized fermionic Hubbard models and related impurity models within Dynamical Mean Field Theory, and demonstrate the calculation of Green's functions on Quantinuum’s H1-1 trapped-ion quantum computer. Our approach involves a cumulant expansion of the Lanczos method, using Hamiltonian moments as measurable expectation values. This bypasses the need for a large overhead in the number of measurements due to repeated applications of the variational quantum eigensolver (VQE), and instead measures the expectation value of the moments with one set of measurement circuits. From the measured moments, the tridiagonalised Hamiltonian matrix can be computed, which in turn yields the Green's function via continued fractions. While we use a variational algorithm to prepare the ground state in this work, we note that the modularity of our implementation allows for other (non-variational) approaches to be used for the ground state. 
\end{abstract}

\maketitle

\section{Introduction} \label{sec:introduction}

The Green's function (GF) is a quantity that allows access to in principle all single-particle properties, including electronic responses, and is therefore  useful for calculating spectroscopic properties, such as photoemission, photoabsorption, and conductivity \cite{fetter03, golze19}. It can also be used to capture electronic correlations in condensed matter and quantum chemical simulations, in addition to being the central quantity in many quantum embedding methods such as cluster perturbation theory and Dynamical Mean Field Theory (DMFT) \cite{senechal00, senechal02, georges96, caffarel94, avella13}. Despite its importance in these fields, the GF remains a difficult quantity to calculate accurately from first principles on classical computers. The exact GF requires knowledge of all eigenvalues and eigenvectors of the Hamiltonian, so its calculation complexity is the same as for full Hamiltonian diagonalization. There is a long history of approximate methods for the many body GF in the classical domain of quantum chemistry, with the state-of-the art often exhibiting high-degree polynomial scaling \cite{coveney23, marie23, teke19}. Quantum computation offers a novel and interesting route to obtain many-body GFs for strongly correlated systems.

In a previous study \cite{jamet_kvqa21}, an approach based on the Lanczos method was proposed to obtain the GF on quantum computers. This approach relies on a parameterised VQE method to construct the Krylov space, hence the orthonormal Lanczos basis is calculated iteratively in which a VQE-like algorithm is applied at each Lanczos iteration. This involves an overhead in the number of measurements and renders the approach more susceptible (relative to a typical ground state calculation) to the potential difficulties associated with variational optimization (such as barren plateaus) \cite{tilly22}. This is also the case for recent proposal based on variational compilation \cite{kanasugi23}. A method based on quantum subspace expansion (QSE) has also been proposed \cite{jamet_qse22}, in which the Hamiltonian and overlap matrix elements are obtained in a basis for the ground state and in another basis for the Krylov states involved in the GF. The basis sets are prepared using time evolution, and the matrix elements are subsequently measured on a quantum computer. The Krylov basis and Lanczos coefficients are then calculated iteratively on a classical computer. While an interesting approach, we note QSE requires more resources than VQE, and the reliance on trotterised time evolution may lead to a vulnerability to trotter errors as a function of the time step. A number of other recent proposals also utilise time evolution to measure the real-time GF \cite{libbi22, delre22}, as well as variational methods to capture the imaginary-time GF \cite{sakurai22, dhawan23}. We also note the approach of Kosugi and Matsushita \cite{kosugi20} in which (multi-)controlled operations between ancillas and state qubits are used to represent the ladder operators involved in the (off-)diagonal elements of the GF matrix, with a subsequent application of quantum phase estimation (QPE). Other recently proposed methods obtain the GF using an algorithm to sample a Fourier series expansion \cite{wang23}, or using a 3-qubit iTofolli gate \cite{sun23}. Such approaches are prohibitive for the noisy intermediate-scale quantum (NISQ) era in terms of circuit depth scaling. Other algorithms aimed at greater suitability for near-term quantum computers have been proposed to obtain the GF in both the time and frequency domains \cite{Endo20}, in addition to near-term algorithms for linear response properties of molecules \cite{huang22} (which require ancillary qubits in some cases).

As mentioned above, the GF is a central quantity in quantum embedding methods such as DMFT \cite{georges96}. In recent years, a number of quantum computational approaches to DMFT have been published \cite{rungger19, keen20, jaderberg20}. These works involve simplifications which increase tractability, yet decrease generalisability, such as a restriction to two sites via the single Anderson impurity model (SIAM). In this picture the quasiparticle weight has an analytical form, to which the impurity-bath interaction is related in a simple way, via the limit of infinite dimensions \cite{georges96}. Previous works have also measured the GF in the time-domain to solve the impurity model in DMFT, which can lead to a large overhead in the number of measurements per time step \cite{kreula16}, or a reliance on fault tolerant schemes such as QPE \cite{bauer16}, or the use of a simplified sinusoidal form of the GF, or other techniques \cite{keen20, jaderberg20, PhysRevA.104.032405, gomes23}. A recent approach considers low energy expansions of the GF for an alternative embedding scheme based on the rotationally invariant slave boson method, which does not require calculating the full GF \cite{besserve22}. We also note recent quantum computational approaches for the DMFT solution of the Hubbard-Holstein model \cite{backes23}, as well as novel spectral resolution techniques to measure the Anderson impurity GF \cite{cruz23}. Finally, a very interesting recent approach adopted fast-forwarding circuit protocols to time-evolve the impurity state and converge the DMFT loop on noisy hardware \cite{steckmann23}. 

We add to these previous works by providing a method to quantum compute the GF in the frequency domain that can be utilised in a hybrid quantum-classical algorithm for DMFT appropriate for NISQ. When considering which features of the GF are most important to capture in practice, the particular application of the GF calculation will clearly have an impact. Since the GF is a dynamical quantity, related to a wide range of real-world applications (such as the spectroscopic properties mentioned above), this motivates the calculation of frequency-dependent properties. GFs are also widely used in embedding applications. Therefore, we consider the following features of the GF in this work: accessing the frequency dependent spectral function of the GF (requiring an accurate representation of the imaginary part of the GF along the real frequency axis), and calculating the impurity GF in the DMFT embedding method (requiring an accurate calculation of the GF matrix elements along the imaginary axis to obtain convergence of the impurity problem).

In this work, a cumulant expansion of the Lanczos coefficients is utilised to provide measurable quantum circuits for each moment of the Hamiltonian \cite{vallury20, vallury23}. These measurable expectation values can be used to calculate the elements of the tridiagonalised Hamiltonian matrix in the Krylov basis, from which the well-known continued fraction expression of the GF can be evaluated \cite{georges96, avella13}. Thus, we extend the work of Vallury \textit{et al.} \cite{vallury20} by utilising quantum computed moments to develop a quantum algorithm for calculating the full GF matrix. As a first example, we apply our method to the calculation of the GF of the fermionic Hubbard model. We note recent studies of the ground state properties of the Hubbard model using quantum algorithms \cite{cade20, stanisic22}, in addition to ground state preparation algorithms which rely on fault tolerant quantum computation \cite{poulin09, ge19, lin20}. Rather than investigating efficient ways to find the ground state, we instead focus on a quantum approach to calculate GFs for finite chains of Hubbard sites. In addition, we show that this approach can be applied to DMFT in a straight-forward manner to iteratively calculate the quantum computed impurity GF to self-consistency. Our methodology is implemented in the InQuanto quantum computational chemistry package \cite{inquanto_web, inquanto_medium}.

Our results show that application of this approach to the H1-1 trapped ion quantum computer \cite{molmer99, h11e} results in spectral features (derived from the imaginary part of the GF) that compare well with ideal classical simulations, particularly in terms of peak positions, even without error mitigation techniques. Hence, expectation values of Hamiltonian moments are relatively robust to measurement noise for low lying Lanczos roots, which is consistent with previous applications of the method to infimum estimates of the ground state energy \cite{hollenberg96, vallury20, vallury23}. The stability of this method also extends to iterative calculations of the impurity GF in the DMFT algorithm, even when (emulated) circuit based measurements are used to evaluate the Hamiltonian moments at every DMFT iteration.

This paper is organised as follows. In section \ref{sec:methods} we describe the methods and show how the quantum computed moments can be used to obtain the GF. This is followed by a brief overview of DMFT applied to the Bethe lattice with multiple bath sites \cite{caffarel94}. In section \ref{sec:results}, we present our results. Starting with the Hubbard model, we show that our method can be applied to multiple sites and benchmark our results against the classical Lanczos method. We then demonstrate this method on quantum hardware. Following this, we utilise the quantum computed GF in a DMFT algorithm. Section \ref{sec:conclusion} provides the concluding remarks. 
\section{Methods} \label{sec:methods}

\subsection{Quantum Computed Moments for Green's Functions} \label{subsec:qgf}

In this section, we give a brief overview of the cumulant expansion of the Lanczos method, and its application to computing the GF in a quantum computational setting. The Lanczos recursion method \cite{lanczos50} can be viewed as an approximate diagonalisation scheme, resulting in a tridiagonal form of the Hamiltonian matrix which can be efficiently solved. In a typical classical Lanczos routine, the elements of the tridiagonalised matrix (referred to as Lanczos coefficients below) $\alpha_l = \langle v_l|\hat{H}| v_l \rangle$ and $\beta_l = \langle v_{l+1}|\hat{H}| v_l \rangle = \langle v_{l}|\hat{H}| v_{l+1} \rangle$ are used to numerically orthonormalise the vectors $|v_{l+1}\rangle = \frac{(\hat{H} - \alpha_{l})|v_{l}\rangle - \beta_{l}|v_{l-1} \rangle}{\beta_{l+1}}$, where $l = 1, 2, 3$, $\beta_{0} = 0$, and $\hat{H}$ is the fermionic Hamiltonian operator in second quantization. Vallury \textit{et al.} \cite{vallury20} showed that this scheme can be re-expressed for a quantum computational context. The Lanczos coefficients can be expressed using the moments of the Hamiltonian operator $\hat{H}$ with respect to some initial Lanczos vector, $\langle \hat{H}^{n} \rangle = \langle v_1|\hat{H}^{n}|v_1 \rangle$. Therefore, a quantum algorithm that can compute the moments can be used to generate the Lanczos coefficients. For example, for the upper $2 \times 2$ block of the tridiagonal matrix, i.e. $\alpha_1$,  $\alpha_2$, $\beta_1$, the quantities obtained from a quantum algorithm are the expectations values $\langle \hat{H} \rangle, \langle \hat{H}^2 \rangle,$ and $\langle \hat{H}^3 \rangle$, and the coefficients are calculated as:
\begin{equation} \label{eqn:alpha_beta}
 \begin{split}
     \alpha_1 &= \langle \hat{H} \rangle\\
     \beta_1 &= \sqrt{ \langle \hat{H}^2 \rangle - \langle \hat{H} \rangle^2 } \\
     \alpha_2 &= \frac{\langle \hat{H}^3 \rangle - 2 \langle \hat{H} \rangle \langle\hat{H}^2 \rangle + \langle \hat{H} \rangle^3}{ \langle \hat{H}^2 \rangle - \langle \hat{H} \rangle^2 }.
 \end{split}
 \end{equation}
In general, following the work of Hollenberg \textit{et al.} \cite{hollenberg93, hollenberg94}, it can be shown that the Lanczos coefficients $\alpha_l$ and $\beta_l$ can be expressed as functions of $\{\langle \hat{H}^{n} \rangle\}_{n=1..2l-1}$ and explicit expressions can be obtained via the cumulant expansion of the Lanczos coefficients \cite{vallury20, hollenberg93}.

Once the Lanczos coefficients are obtained, they can be used to calculate the zero temperature GF matrix $\mathbf{G}(\omega)$ in the continued fraction representation \cite{georges96, andrews23}. In this representation, the diagonal elements of the single particle GF take the following form when decomposed into particle and hole parts \cite{pavarini11}
\begin{equation} \label{eqn:g_diag}
\begin{multlined}
G^{\text{L}}_{i,i}(\omega) = g_{i,i}^{\text{(p)}}(\omega) + g_{i,i}^{\text{(h)}}(\omega) \\ =  \frac{n_{i,i}^{\text{(p)}}}{\omega + E_{\text{GS}} - \alpha_{1}^{\text{(p)}} - \frac{\beta_1^{\text{(p)}}}{\omega + E_{\text{GS}} - \alpha_{2}^{\text{(p)}}} - \ldots} \\ + \frac{n_{i,i}^{\text{(h)}}}{\omega - E_{\text{GS}} + \alpha_{1}^{\text{(h)}} - \frac{\beta_1^{\text{(h)}}}{\omega - E_{\text{GS}} + \alpha_{2}^{\text{(h)}}} - \ldots} ,
\end{multlined}
\end{equation}
\noindent where $E_{\text{GS}}$ is the ground state energy, and the numerators $n_{i,i}$ (defined below) are related to normalised states obtained by operating on the ground state with the fermionic creation ($\hat{f}^{\dagger}_{i}$) and annihilation ($\hat{f}_{i}$) operators, applied to spin orbital (or qubit) $i$ (Jordan-Wigner (JW) \cite{jordan28} encoding is assumed throughout this paper). These normalised states correspond precisely to the initial Lanczos vector for each diagonal GF element. Using the particle component $g_{i,i}^{\text{(p)}}(\omega)$ as an example, we can write 
\begin{equation} \label{eqn:lvec_p_ii}
    |v_1\rangle \equiv |\Psi^{\text{(p)}}_{i,i}\rangle = \frac{\hat{f}^{\dagger}_{i}|\Psi_{\text{GS}}\rangle}{\sqrt{n_{i,i}^{\text{(p)}}}},
\end{equation}
\noindent where $n_{i,i}^{\text{(p)}}=\langle\Psi_{\text{GS}}|\hat{f}_{i}\hat{f}^{\dagger}_{i}|\Psi_{\text{GS}}\rangle$, and $|\Psi_{\text{GS}}\rangle$ is the ground state represented by a quantum circuit. In Eq. \ref{eqn:g_diag} the Lanczos coefficients $\alpha_{l}^{\text{(p/h)}}, \beta_{l}^{\text{(p/h)}}$ are labelled by their particle/hole contributions, according to whether the initial Lanczos vector is obtained by operating on $|\Psi_{\text{GS}}\rangle$ with $\hat{f}^{\dagger}_{i}$ (p) or with $\hat{f}_{i}$ (h), respectively.

The off-diagonal elements are obtained in an analogous way using linear combinations of fermionic operators indexed by the matrix element. For example, this results in an initial Lanczos vector for $g_{i \ne j}^{\text{(p)}}(\omega)$
\begin{equation} \label{eqn:lvec_p_ij}
    |\Psi^{\text{(p)}}_{i \ne j}\rangle = \frac{(\hat{f}^{\dagger}_{i} + \hat{f}^{\dagger}_{j})|\Psi_{\text{GS}}\rangle}{\sqrt{n_{i \ne j}^{\text{(p)}}}},
\end{equation}
\noindent where $n_{i \ne j}^{\text{(p)}} = \langle\Psi_{\text{GS}}|(\hat{f}_{i} + \hat{f}_{j}) (\hat{f}^{\dagger}_{i} + \hat{f}^{\dagger}_{j})|\Psi_{\text{GS}}\rangle$ . Following through as above, the resulting off-diagonal element $G^{\text{L}}_{i \ne j}(\omega)$ is not the true off-diagonal element of the GF, but requires a modification which has a simple arithmetic form \cite{avella13}, shown in Eq. \ref{eqn:g_offdiag}. Utilising the symmetry of the GF matrix, we finally obtain the GF off-diagonal element as 
\begin{equation} \label{eqn:g_offdiag}
G_{i \ne j}(\omega) = \frac{1}{2}(G^{\text{L}}_{i \ne j}(\omega) - G_{i,i}(\omega) - G_{j,j}(\omega)),
\end{equation}
\noindent with $G_{i,i}(\omega) = G^{\text{L}}_{i,i}(\omega)$. Thus, a method to quantum compute the matrix $\mathbf{G}(\omega)$ is obtained, without the explicit reliance on multicontrolled ancilla qubits, time evolution, or phase estimation. On the other hand, relative to methods involving phase estimation, this approach in general requires a larger number of measurements due to the increase in the number of Hamiltonian moments with Lanczos root index $l$, hence we note trade-off between circuit depth and number of measurements when choosing between these approaches. We also note that this method does not require calling a VQE routine for each Lanczos root, nor a time-evolved subspace expansion to obtain the Lanczos basis vectors. 

In practical terms, the cumulant expansion of the Lanczos coefficients combined with the continued fraction representation of $\mathbf{G}(\omega)$ implies at least two strategies for implementation in a quantum algorithm: \textit{i)} prepare the initial Lanczos vector explicitly with a state preparation circuit and measure the moments as expectations with respect to this circuit, or \textit{ii)} measure each moment as the expectation value of an operator built by sandwiching $\hat{H}^n$ with ladder operators (or sums of ladder operators) indexed according to the initial Lanczos vector. 

As an example, consider the $n^{\text{th}}$ moment contributing to a Lanczos coefficient required for $g_{i,i}^{\text{(p)}}(\omega)$: Using \textit{i)} we would prepare $|\Psi^{\text{(p)}}_{i,i}\rangle$ and measure 

\begin{equation} \label{eqn:sand_h_exp_1}
\langle \hat{H}^{n, (\text{p})}_{i,i} \rangle = \langle \Psi^{\text{(p)}}_{i,i}| \hat{H}^n |\Psi^{\text{(p)}}_{i,i}\rangle. 
\end{equation}

\noindent Using \textit{ii)} we would prepare $|\Psi_{\text{GS}}\rangle$ and measure 

\begin{equation} \label{eqn:sand_h_exp_2}
\langle \hat{H}^{n, (\text{p})}_{i,i} \rangle = (1/n_{i,i}^{\text{(p)}})\langle \Psi_{\text{GS}}| \hat{f}_{i} \hat{H}^n \hat{f}^{\dagger}_{i} |\Psi_{\text{GS}}\rangle.
\end{equation}

Thus our methodology requires either $|\Psi^{\text{(p)}}_{i,i}\rangle$ or $|\Psi_{\text{GS}}\rangle$ to be prepared on a quantum circuit, in addition to the measurement of Hamiltonian moments. While mathematically equivalent, \textit{i)} and \textit{ii)} can lead to considerable differences in the quantum circuit resources (depending on the state preparation methods), and catering for both approaches allows for a useful flexibility in the application of the quantum Lanczos approach to GFs. As discussed in Ref. \cite{vallury20}, a naive counting of Hamiltonian terms results in an exponential growth in the number of Pauli strings with Hamiltonian power $n$. However, this can be significantly improved by the existence of large commuting sets in the Hamiltonian moment expansions \cite{tazi23}. To mitigate the rapid growth we measured the commuting Pauli terms with a single measurement circuit. However, we note there are other efficient techniques that could be also applied based on the tensor product basis (TPB) and qubit-wise commutativity \cite{vallury20, verteletskyi20}. 

The following paragraphs provide further details on strategies \textit{i)} and \textit{ii)} for executing the quantum Lanczos approach to calculate a GF, followed by an outline of how the ground state can be prepared. 

\subsubsection*{i) Initial Lanczos Vector Preparation} \label{subsubsec:lanc_vec_prep}
In the occupation number (ON) vector representation, the ground state is expressed as
\begin{equation} \label{eqn:psi_sum_cx}
    |\Psi_{\text{GS}}\rangle = \sum_x c_x |x\rangle ,
\end{equation}
\noindent where $x = \{x_i\}$ represents a set of occupations of spin orbitals $i$ for a given number of particles, and each basis configuration can be written as the ON vector of $N_i$ spin orbitals
\begin{equation} \label{eqn:x_xi}
    |x\rangle = |x_0, x_1, \ldots, x_i, \ldots, x_{N_i - 1} \rangle .
\end{equation}
Given a circuit representing the ground state, the coefficients of the ON vector (Eq. \ref{eqn:psi_sum_cx}) can then be extracted by calculating the overlap (where $\hat{U}_{\text{GS}}(\boldsymbol{\theta}_{\text{opt}})$ is a unitary applied to a reference state $|\Psi_{\text{ref}} \rangle$, as is typical of VQE)

\begin{equation} \label{cx_overlap}
\begin{split}
    c_x &= \langle x |\Psi_{\text{GS}} \rangle \\
    &=  \langle x |\hat{U}_{\text{GS}}(\boldsymbol{\theta}_{\text{opt}})|\Psi_{\text{ref}} \rangle
\end{split}
\end{equation}

\noindent for all $x$, resulting in a set of $c_x$ values, and the corresponding $\{|x \rangle\}$, hence the ON representation (Eq. \ref{eqn:psi_sum_cx}) is obtained. The ON representation of $|v_1\rangle \equiv |\Psi^{\text{(p)}}_{i,i}\rangle$ or $|v_1\rangle \equiv |\Psi^{\text{(h)}}_{i,i}\rangle$ is then obtained by applying $\hat{f}^{\dagger}_{i}$ or $\hat{f}_{i}$, respectively, to Eq. \ref{eqn:psi_sum_cx} and normalising (as in Eq. \ref{eqn:lvec_p_ii}). The resulting expansion of $|v_1\rangle$ as a linear combination of basis configurations with real coefficients can then be prepared on a quantum circuit using controlled Givens rotations as particle-conserving excitations \cite{arrazola22}.
While the procedure defined by Eqs. \ref{eqn:psi_sum_cx} - \ref{cx_overlap} in general exhibits an exponentially scaling overhead (due to the inclusion of all particle-number preserving configurations) for the representation of $|v_1\rangle$ to be exact, it is used here to demonstrate the flexibility of our procedure to allow for arbitrary preparation of the initial Lanczos vector. Hence future preparation schemes with better scaling can be easily accommodated. 

\subsubsection*{ii) Sandwiched Moment Expectation} \label{subsubsec:sand_exp}
As mentioned above, one can obtain each required term of the cumulant expansion from the measurement of an expectation, with respect to the $\hat{H}$ ground state, of a sandwiched Hamiltonian moment operator (see Eq. \ref{eqn:sand_h_exp_2}). This requires the preparation of a quantum circuit representing the ground state, which can be achieved by a VQE-optimized variational ansatz. Following the JW transformation of $\hat{f}_{i} \hat{H}^n \hat{f}^{\dagger}_{i}$ (and of $\hat{f}^{\dagger}_{i} \hat{H}^n \hat{f}_{i}$ for the hole part), the corresponding Pauli strings can then be measured with respect to the ground state circuit. In section \ref{results_sand_moments}, we report Pauli operators representing the measurable sandwiched Hamiltonian moments for GF elements of the Hubbard dimer for Lanczos coefficients corresponding to $l = 2$. 

\subsubsection*{Ground State Preparation} \label{subsubsec:gs_prep}

To prepare the ground state, a number of approaches have been proposed in recent years based on hybrid quantum-classical algorithms such as imaginary time evolution \cite{mcardle19, motta20}, or VQE \cite{tilly22}. These approaches generally involve the application of some $\hat{U}_{\text{GS}}$ to a reference state, where $\hat{U}_{\text{GS}}(\boldsymbol{\theta})$ is expressed as an ansatz which depends on parameters $\boldsymbol{\theta}$ that are variationally optimized to $\boldsymbol{\theta}_{\text{opt}}$. For the latter, a wide range of chemically intuitive (e.g. unitary coupled cluster (UCC) \cite{anand22}) and hardware efficient \cite{kandala17} ans\"{a}tze have been proposed. In addition, adaptive methods also exist in which the ansatz is constructed iteratively for a given problem \cite{grimsley19, yordanov21, gomes21}. We also note the continuing development of ground state preparation algorithms which rely on fault tolerant quantum computation \cite{poulin09, ge19, lin20}. In this work we focus on a quantum approach to the GF and assume an accurate ground state has been provided from a separate calculation. In practice, we prepared the ground state circuit using a VQE algorithm with a parameterised ansatz, but our approach also supports the classical calculation of the ground state (which would exemplify a procedure that combines classical computing for ground state properties, with quantum computing for excited state properties). This can be done as long as the classical calculation yields an expansion of the state vector in a basis of occupation configurations, since the latter in principle can always be prepared using controlled excitation gates \cite{arrazola22}, as outlined for strategy \textit{i)}. 

\subsection{Fermionic Hubbard model} \label{subsec:hubbard}

To demonstrate our approach, we quantum compute the GF of the fermionic Hubbard model, the Hamiltonian of which can be written as
\begin{equation} \label{eqn:h_hub}
\begin{split}
\hat{H}_{\text{Hub}} = &-\mu\sum_{s} \sum_{\sigma} \hat{f}^{\dagger}_{s, \sigma} \hat{f}^{ }_{s, \sigma} \\&-t\sum_{<r, s>} \sum_{\sigma} \Big(\hat{f}^{\dagger}_{r, \sigma} \hat{f}^{ }_{s, \sigma} + \hat{f}^{\dagger}_{{s, \sigma}} \hat{f}^{ }_{r, \sigma} \Big) \\&+ U\sum_{s} \hat{f}^{\dagger}_{s,\uparrow}\hat{f}^{}_{s,\uparrow} \hat{f}^{\dagger}_{s,\downarrow}\hat{f}^{}_{s,\downarrow} ,
\end{split}
\end{equation}
\noindent where $\mu$, $t$, and $U$ are the chemical potential, hopping amplitude, and on-site Coulomb interaction, respectively ($t=1$ in these calculations, which sets the energy scale throughout this paper). $s$ labels a site, $<$$r, s$$>$ denotes nearest neighbor pairs of sites, and spins are labelled by $\sigma = \ \uparrow, \downarrow$. To establish consistency with the spin orbital index $i$ of the GF matrix (see section \ref{subsec:qgf}), we use a linear mapping between the site-spin index and the spin orbital index $s,\uparrow (s, \downarrow) \mapsto i \ (i+1)$ where $s = 0, 1, 2, \dots, N_s - 1$ and $i = 2s$, hence $G_{s\sigma, \ s'\sigma'} \equiv G_{i,j}$. There are $N_s \times 1 \times 1$ Hubbard sites arranged in a linear geometry.

Since we use JW encoding throughout, the spin orbital index $i$ also corresponds to a qubit index on a quantum circuit. Fermionic operators are mapped according to JW encoding as

\begin{equation}
\begin{split}
    \hat{f}^{\dagger}_{i} &\mapsto \frac{X_i - \text{i}Y_i}{2}\prod_{d=0}^{i-1}Z_d \\
    \hat{f}^{}_{i} &\mapsto \frac{X_i + \text{i}Y_i}{2}\prod_{d=0}^{i-1}Z_d
\end{split}
\end{equation}

\noindent where $\text{i}^2 = -1$ (referring to the non-italic symbol $\text{i}$), and $P_i \in \{I_i, X_i, Y_i, Z_i\}$ refers to a Pauli rotation gate applied to qubit $i$ (for $i = 0$ the product of $Z_d$ over $d$ is omitted). This results in a qubit representation of the Hamiltonian as a sum of tensor products of Pauli operators.

Given the qubit representation of $\hat{H}_{\text{Hub}}$, the corresponding GF matrix is obtained using the methodology outlined in section \ref{subsec:qgf}. We assume the half-filled regime ($\mu = U/2$), set the initial reference state to the half-filled singlet configuration, and obtain the ground state using VQE with a parameterized ansatz. For the latter, we utilise qubit excitations \cite{yordanov21} in the Hard-core Boson representation \cite{khan22} to obtain the ground state using significantly fewer 2-qubit gates than UCC. Consider the Hubbard dimer, corresponding to 4 qubits. The initial reference state of a spin-up electron occupying site 0 and a spin-down on site 1 ($|\Psi_{\text{ref}} \rangle = |1001\rangle$) can be prepared with Pauli-$X$ gates. The ground state can then be obtained by applying two qubit excitation operators, i.e. unitary operators corresponding to a single-excitation and a double-excitation which act directly on the qubit Hilbert space (hence obviating the need for Pauli-$Z$ gates to maintain fermionic exchange antisymmetry \cite{yordanov20}) $|\Psi_{\text{GS}}(\theta_1, \theta_2) \rangle = e^{\text{i}\theta_1Y_0X_2}e^{\text{i}\theta_2Y_0X_1X_2X_3}|\Psi_{\text{ref}} \rangle$. The doubles excitation portion of this circuit is shown in Fig. \ref{fig:doubles}. 

\begin{center}
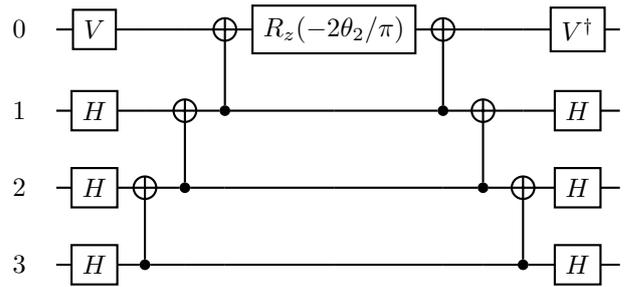
\begin{figure}[ht]
\begin{quantikz}[column sep=0.2cm]
0 &&& \gate{V} & \qw & \qw & \targ{} & \gate{R_z( -2 \theta_2 / \pi )} & \targ{} & \qw & \qw & \gate{V^{\dagger}} & \qw \\
1 &&& \gate{H} & \qw & \targ{} & \ctrl{-1} & \qw & \ctrl{-1} & \targ{} & \qw & \gate{H} & \qw \\
2 &&& \gate{H} & \targ{} & \ctrl{-1} & \qw & \qw & \qw & \ctrl{-1} & \targ{} & \gate{H} & \qw \\
3 &&& \gate{H} & \ctrl{-1} & \qw & \qw & \qw & \qw & \qw & \ctrl{-1} & \gate{H} & \qw \\
\end{quantikz}
\caption{Quantum circuit representation of the double qubit excitation operator $e^{\text{i}\theta_2Y_0X_1X_2X_3}$. The $V$, $V^\dagger$, and $H$ (Hadamard) gates are for rotation into the desired computational basis (such that $H Z H = X$ and $V^{\dagger} Z V = Y$), while the CNOTs and $R_z$ represent the exponentiated Pauli strings corresponding to the qubit excitation. Here 6 CNOTs are needed for the double excitation (2 extra CNOTs are also needed for the singles excitation not shown here, so 8 CNOTs altogether to express the ground state).}
\label{fig:doubles}
\end{figure}
\end{center}

By utilising particle-hole symmetry, as well as total spin symmetry (equal number of spin-$\uparrow$ and spin-$\downarrow$ electrons at half-filling), the number of 2-qubit gates for the double-excitation can be reduced as follows: step 1) treat the double-excitation initially as a single-excitation involving Pauli operators acting on two Hubbard sites instead of four spin orbitals (equivalent to using the molecular spatial orbital index in a chemically aware strategy \cite{khan22}), step 2) specify the qubit excitation between sites as the excitation between qubits representing the same spin on either site (e.g. for a spin-$\uparrow$ to spin-$\uparrow$ excitation in the Hubbard dimer, this corresponds to a qubit excitation involving qubits 0 and 2), step 3) reintroduce spin orbital indexing by applying 2-qubit CNOTs which pair the relevant even-odd indexed qubits. The net result is a series of gates which perform the equivalent action of the double-excitation unitary $e^{\text{i}\theta_2Y_0X_1X_2X_3}$ but with fewer CNOTs. This replaces the double-excitation circuit in Fig. \ref{fig:doubles}. Due to spin symmetry and the lack of $(s, \uparrow) \rightarrow (r, \downarrow$) cross-spin hopping between sites $s$ and $r \neq s$, only one single-excitation unitary is needed. The resulting ansatz circuit is shown in Fig. \ref{fig:ansatz2}. We also note that all measurable circuits in this work are compiled for hardware using the architecture agnostic quantum software compiler $\text{t}|\text{ket}\rangle^\text{TM}$ \cite{tket20}.

\onecolumngrid
\begin{center}
\begin{figure}[ht]
\begin{quantikz}[column sep=0.2cm]
0 &&& \lstick{$\ket{0}$} & \gate{X} & \gate{S^{\dagger}} & \gate{V^{\dagger}} & \targ{} & \qw & \gate{R_z( 2 \theta_2 / \pi )} & \qw & \targ{} & \gate{V} & \gate{S} & \ctrl{3} & \qw & \gate{V} & \qw & \targ{} & \gate{R_z( -2 \theta_1 / \pi )} & \targ{} & \gate{V^{\dagger}} & \qw & \qw \\
1 &&& \lstick{$\ket{0}$} & \qw & \qw & \qw & \qw & \qw & \qw & \qw & \qw & \qw & \targ{} & \qw & \qw & \qw & \qw & \qw & \qw & \qw & \qw & \qw & \qw \\
2 &&& \lstick{$\ket{0}$} & \gate{V} & \qw & \qw & \ctrl{-2} & \gate{H} & \gate{R_z( -2 \theta_2 / \pi )} & \gate{H} & \ctrl{-2} & \gate{V^{\dagger}} & \ctrl{-1} & \qw & \qw & \gate{H} & \qw & \ctrl{-2} & \qw & \ctrl{-2} & \gate{H} & \qw & \qw \\
3 &&& \lstick{$\ket{0}$} & \qw & \qw & \qw & \qw & \qw & \qw & \qw & \qw & \qw & \qw & \targ{} & \qw & \qw & \qw & \qw & \qw & \qw & \qw & \qw & \qw \\
\end{quantikz}
\caption{Parameterised circuit for $e^{\text{i}\theta_1Y_0X_2}e^{\text{i}\theta_2Y_0X_1X_2X_3}|\Psi_{\text{ref}} \rangle$ used to obtain the ground state of the Hubbard dimer, with an optimized 2-body qubit excitation. The $X$ gate on qubit 0 corresponds to Hubbard site 0 occupied by an even-indexed electron. The $S$, $S^\dagger$, $V$, $V^\dagger$, and $H$ gates are for rotation into the desired computational basis. The CNOTs with qubit 0 (2) as targets (controls) bracketing the $R_z$ represent the exponentiated Pauli strings corresponding to the qubit excitations. The 3$^{\text{rd}}$ and 4$^{\text{th}}$ CNOTs translate the action of previous gates to odd-indexed qubits, resulting in a double excitation $e^{\text{i}\theta_2Y_0X_1X_2X_3}$ applied to the initial reference state $|\Psi_{\text{ref}} \rangle = |1001\rangle$, followed by the single excitation $e^{\text{i}\theta_1Y_0X_2}$. Hence in this circuit, 6 CNOTs are needed to represent the ground state.}
\label{fig:ansatz2}
\end{figure}
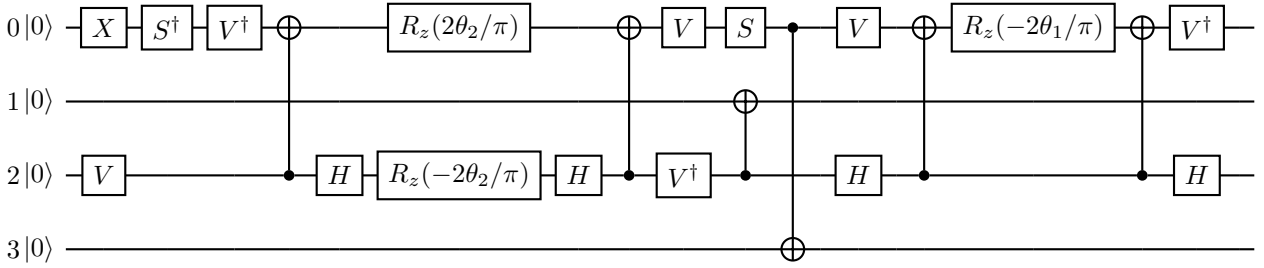
\end{center}
\twocolumngrid

\subsection{Dynamical Mean Field Theory} \label{subsec:dmft}

We also apply our approach to quantum compute the impurity GF within the DMFT solution of the single band Hubbard model on a Bethe lattice \cite{caffarel94}. In the limit of infinite connectivity, the GF of this model can be interpreted as the impurity-site GF of an Anderson model Hamiltonian \cite{georges96, anderson61}

\begin{equation} \label{eqn:h_and}
\begin{split}
\hat{H}_{\text{And}} &= \hat{H}_{\text{imp}} \\ &+ \sum^{N_s + N_b}_{b=N_s} \sum_{\sigma} \epsilon_{b, \sigma} \hat{f}^\dagger_{b, \sigma} \hat{f}^{ }_{b, \sigma} \\&+ \sum^{N_s + N_b}_{b=N_s}  \sum_{\sigma} V_{b, \sigma} \Big(\hat{f}^{\dagger}_\sigma \hat{f}^{ }_{b, \sigma} + \hat{f}^{\dagger}_{b, \sigma} \hat{f}^{ }_\sigma \Big),
\end{split}
\end{equation}
\noindent in which $\hat{H}_{\text{imp}}$ = $\hat{H}_{\text{Hub}}(N_s = 1)$, and baths sites are indexed by $b$. $\epsilon_{b, \sigma}$ and $V_{b, \sigma}$ are variational parameters corresponding to the on-site bath energies and impurity-bath interactions, respectively. Here, we set $\epsilon_{b, \uparrow} = \epsilon_{b, \downarrow}$ and $V_{b, \uparrow} = V_{b, \downarrow}$, and the Hamiltonian is constrained by particle-hole symmetry. We use the same topology for the Anderson impurity model as \cite{kreula16}.

Solving for the eigenspectrum of $\hat{H}_{\text{And}}$ yields the GF of the impurity$+$bath system, the upper $2\times2$ block of which corresponds to the impurity-site GF matrix $\mathbf{G}^{\text{imp}}$ (with elements $G^{\text{imp}}_{i,j}$). Since there is only 1 impurity site and no spin-flipping terms, impurity-related quantities such as $\mathbf{G}^{\text{imp}}$ are $2\times2$ diagonals and reduce to scalar functions of frequency in this case, however we write these as matrices in a basis of impurity spin orbitals $i, j$ for consistency in notation and to emphasise generalisability. We solve for the $G^{\text{imp}}_{i,j}$ elements using the methodology of section \ref{subsec:qgf}.

The dynamical interaction between the impurity and bath is governed by the $U=0$ GF of the Anderson impurity model \cite{caffarel94}, the inverse of which is called hybridisation $\mathbf{\Delta}$, with elements
\begin{equation}
    \Delta_{i,j}(\text{i}\omega_k) = \Bigg( \text{i}\omega_k + \mu - \sum_{\sigma} \sum^{N_\text{bath}}_{b} \frac{V_{b, \sigma}^2}{\text{i}\omega_k - \epsilon_{b, \sigma}} \Bigg)I_{i,j} ,
\end{equation}
\noindent where $I_{i,j}$ is an element of the identity matrix, and $\text{i}\omega_k$ is the $k^{\text{th}}$ Matsubara frequency. The Anderson model can also be related to $\mathbf{G}^{\text{imp}}$ by a self-consistency condition \cite{georges96, caffarel94}
\begin{equation} \label{eqn:deltasc}
    \Delta_{i,j}^{\text{sc}}(\text{i}\omega_k) = \Big( \text{i}\omega_k + \mu \Big)I_{i,j} - \frac{G^{\text{imp}}_{ij}(\text{i}\omega_k)}{2}
\end{equation}
\noindent (with corresponding matrix $\mathbf{\Delta}^{\text{sc}}$). We note that strict self-consistency can only be obtained for $N_{\text{bath}} = \infty$. This limit can be numerically approximated by a finite bath by varying the bath parameters to minimise the cost function

\begin{equation} \label{eqn:dmft_cost}
\begin{split}
    &\mathfrak{C}(\{\epsilon_{b, \sigma}\}, \{V_{b, \sigma}\}) \\
    &= \frac{1}{N_{\omega} + 1} \sum_{k=1}^{N_{\omega}} | \mathbf{\Delta}^{\text{sc}}(\text{i}\omega_k) - \mathbf{\Delta}(\text{i}\omega_k) |_{\text{F}}^2 ,
\end{split}
\end{equation}
\noindent thereby fitting $\mathbf{\Delta}$ to $\mathbf{\Delta}^{\text{sc}}$, where $| \ . \ |_{\text{F}}$ denotes the Frobenius norm. Here, $N_{\omega}$ is the number of fitting frequencies, and we note that all DMFT fitting is performed on a grid of imaginary Matsubara frequencies $\text{i}\omega_k = \frac{\text{i}\pi(2k + 1)}{\upbeta}$ where $\upbeta$ is an inverse temperature which sets a frequency grid cutoff above $\text{i}\omega_k = 0$. In our applications, we run the DMFT algorithm on 1, 2, and 3 bath sites, corresponding to 4, 6, and 8 qubits, respectively. For 1 bath site, bath fitting is performed on 26 imaginary frequencies and $\upbeta$ = 8. While for 2 and 3 bath sites, bath fitting is performed on 64 imaginary frequencies with $\upbeta$ = 16. Numerical minimisation of $\mathfrak{C}$ results in a new set of bath parameters $\{\epsilon_{b, \sigma}\}, \{V_{b, \sigma}\}$, which in turn define a new $\hat{H}_{\text{And}}$, which leads to a new $\mathbf{G}^{\text{imp}}$ via the approach described in section \ref{subsec:qgf}. Eqs. \ref{eqn:h_and} - \ref{eqn:dmft_cost} therefore suggest an iterative algorithm, which can be terminated at a threshold value of the convergence error $\tau$. For $\tau$, we use the Frobenius norm of the change in $\mathbf{\Delta}^{\text{sc}}$ between iterations $m$ and $m+1$, summed over $N_\omega$ imaginary frequencies

\begin{equation} \label{eqn:dmft_tau}
    \tau = \frac{1}{N_{\omega}} \sqrt{\sum_{k=1}^{N_{\omega}} | \mathbf{\Delta}^{\text{sc}}_{m+1}(\text{i}\omega_k) - \mathbf{\Delta}^{\text{sc}}_m(\text{i}\omega_k) |_{\text{F}}^2} \ .
\end{equation}

We assume the half-filled regime, hence $\mu$ is set to $U/2$ which corresponds to 1 electron on the impurity site in the ground state. To find the number of electrons in the bath, the following is performed: after finding the ground state of $\hat{H}_{\text{And}}$ for a given set of bath parameters, the total number of electrons $N_e$ is obtained from the expectation of the total number operator. Bath sites are then filled according to the number of bath electrons =  $N_e - 1$. The resulting occupation state is then used to initialise the Lanczos procedure to solve for the impurity GF using the approach outlined in section \ref{subsec:qgf}. This procedure can be summarised by noting that we consider the Anderson impurity model to be in thermodynamic equilibrium with a reservoir of particles, hence it is treated as a grand canonical ensemble with the number of electrons in the Hubbard impurity set by particle-hole symmetry. In order to find the ground state of $\hat{H}_{\text{And}}$ at each DMFT iteration, the UCCSD \cite{anand22} ansatz is optimized using the VQE \cite{tilly22}.
\section{Results} \label{sec:results}
\subsection{Green's Functions of Hubbard Chains} \label{gf_hubbard_results}

\begin{figure}[H]
    \centering
    \begin{minipage}{8cm}
    \begin{overpic}[width=8cm]{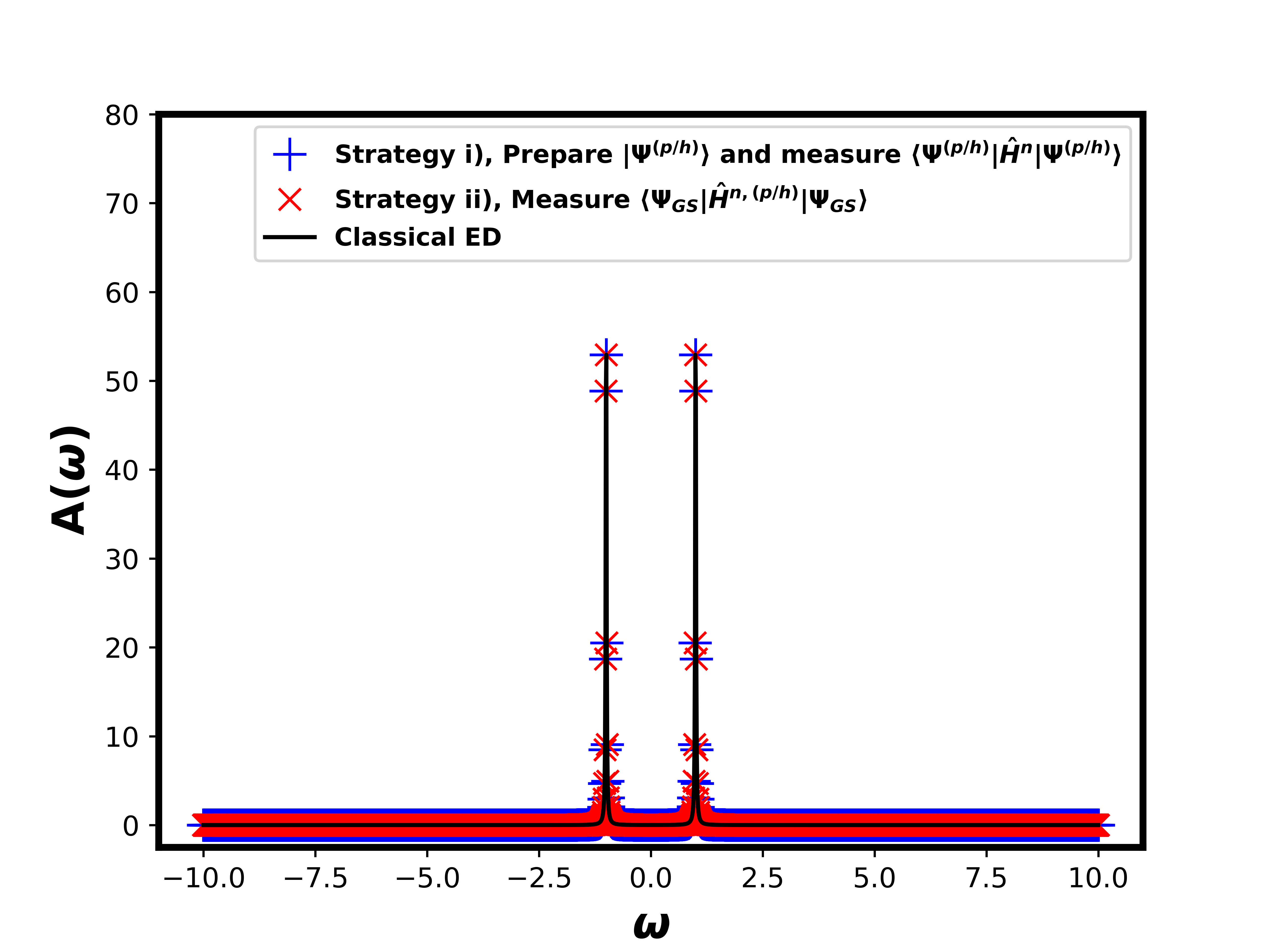}
    \end{overpic}
    \begin{overpic}[width=8cm]{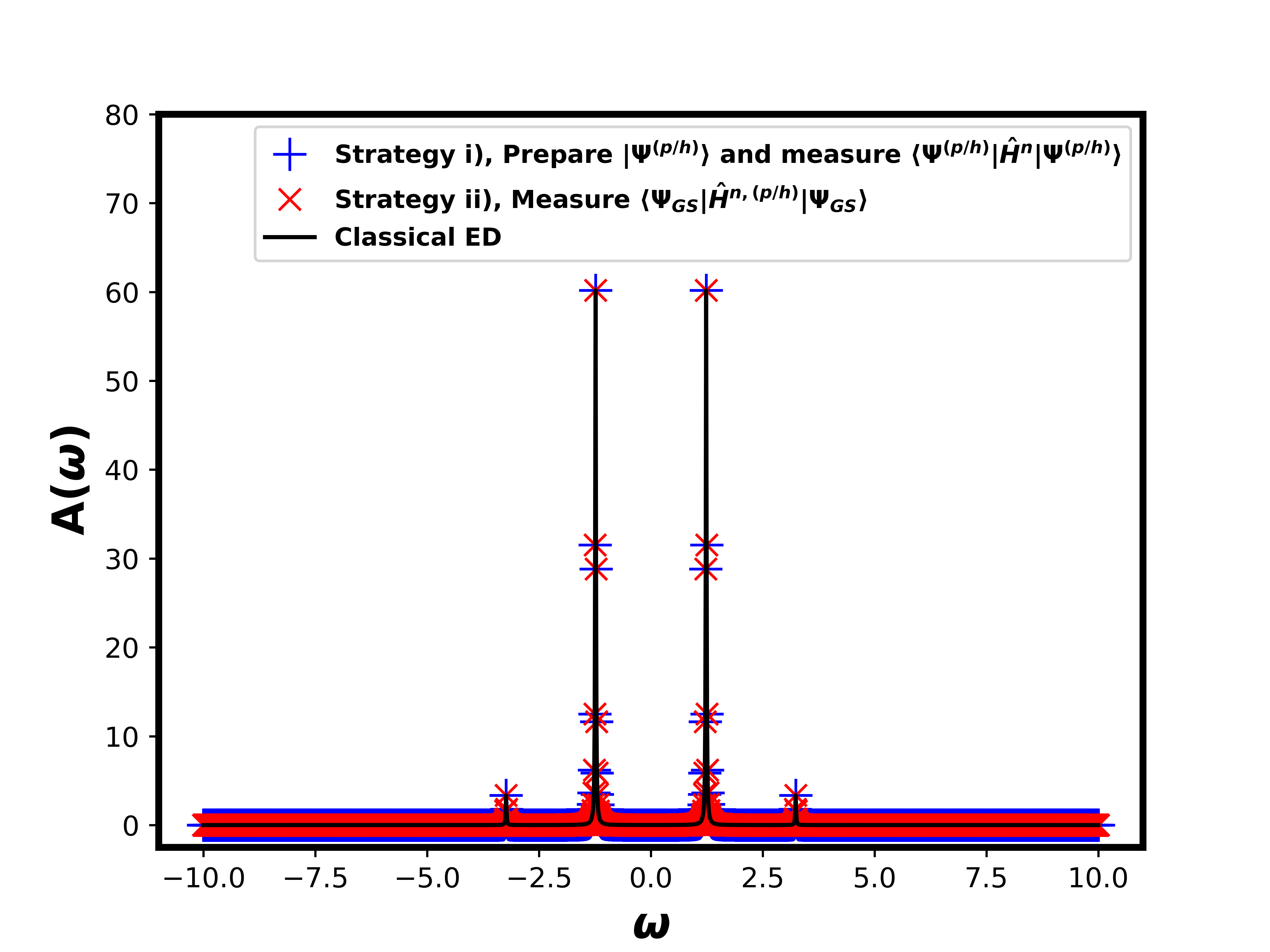}
    \end{overpic}
    \begin{overpic}[width=8cm]{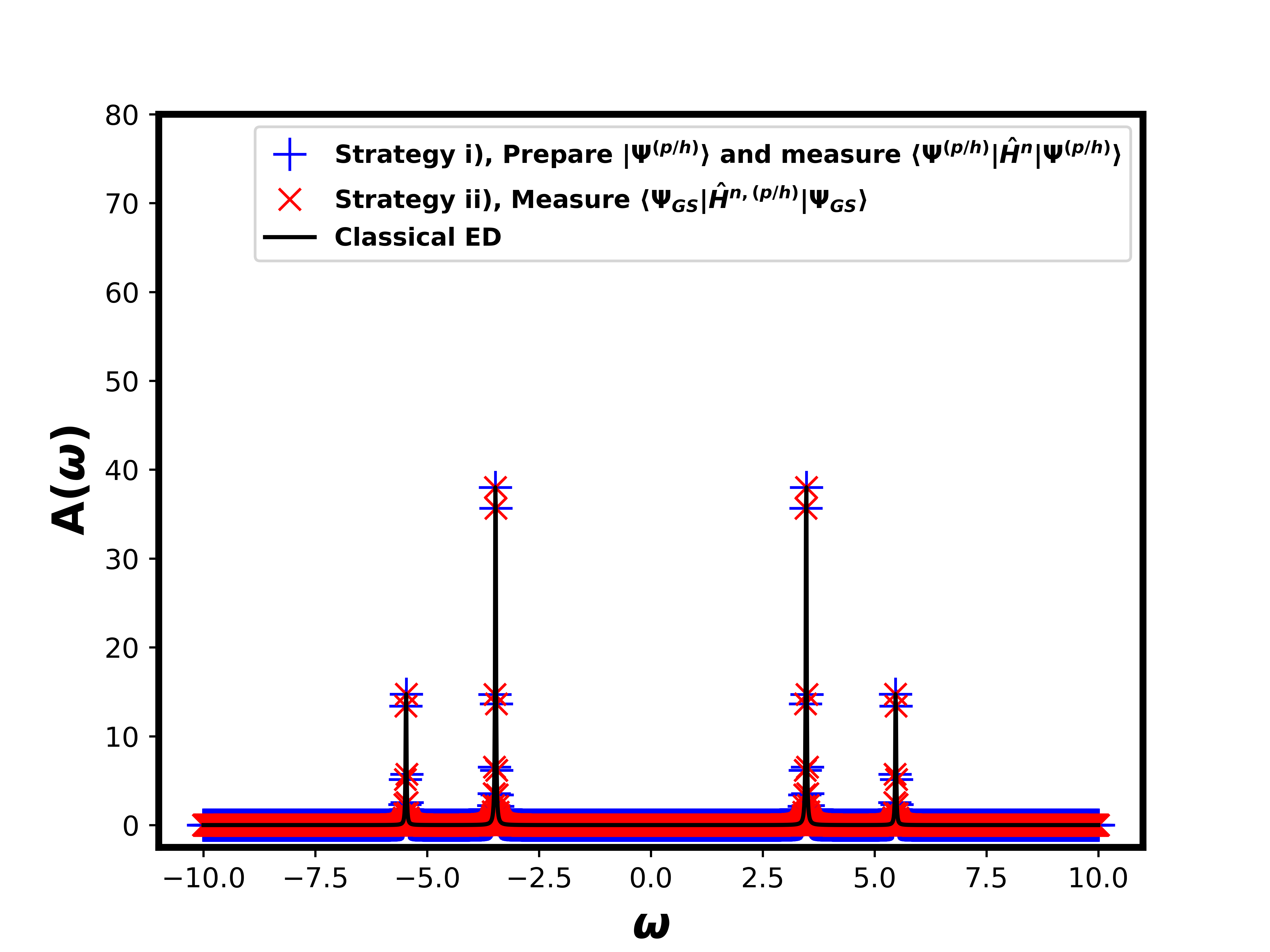}
    \end{overpic}
    \end{minipage}\begin{minipage}{8cm}
    \begin{overpic}[width=8cm]{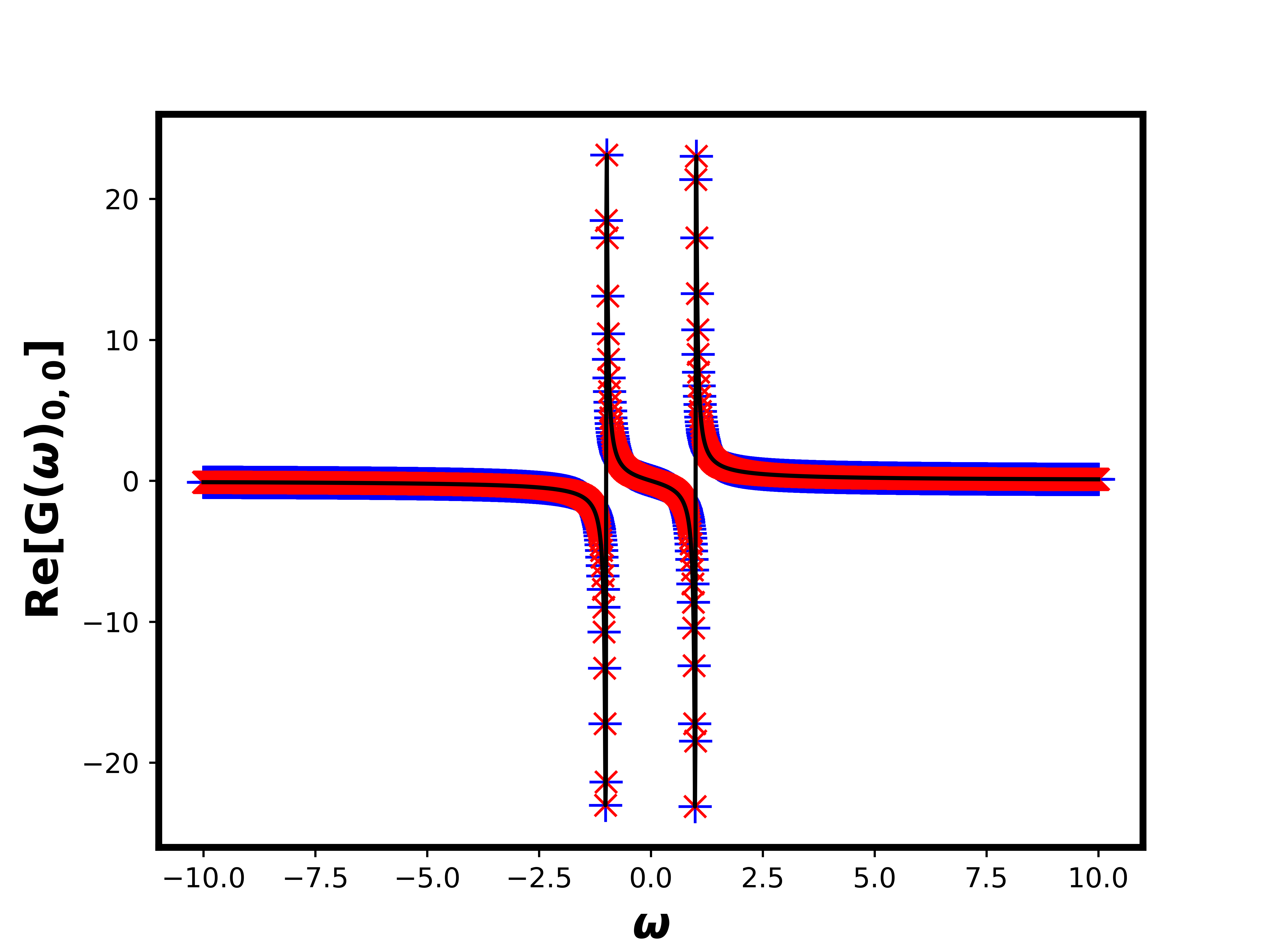}
    \end{overpic}
    \begin{overpic}[width=8cm]{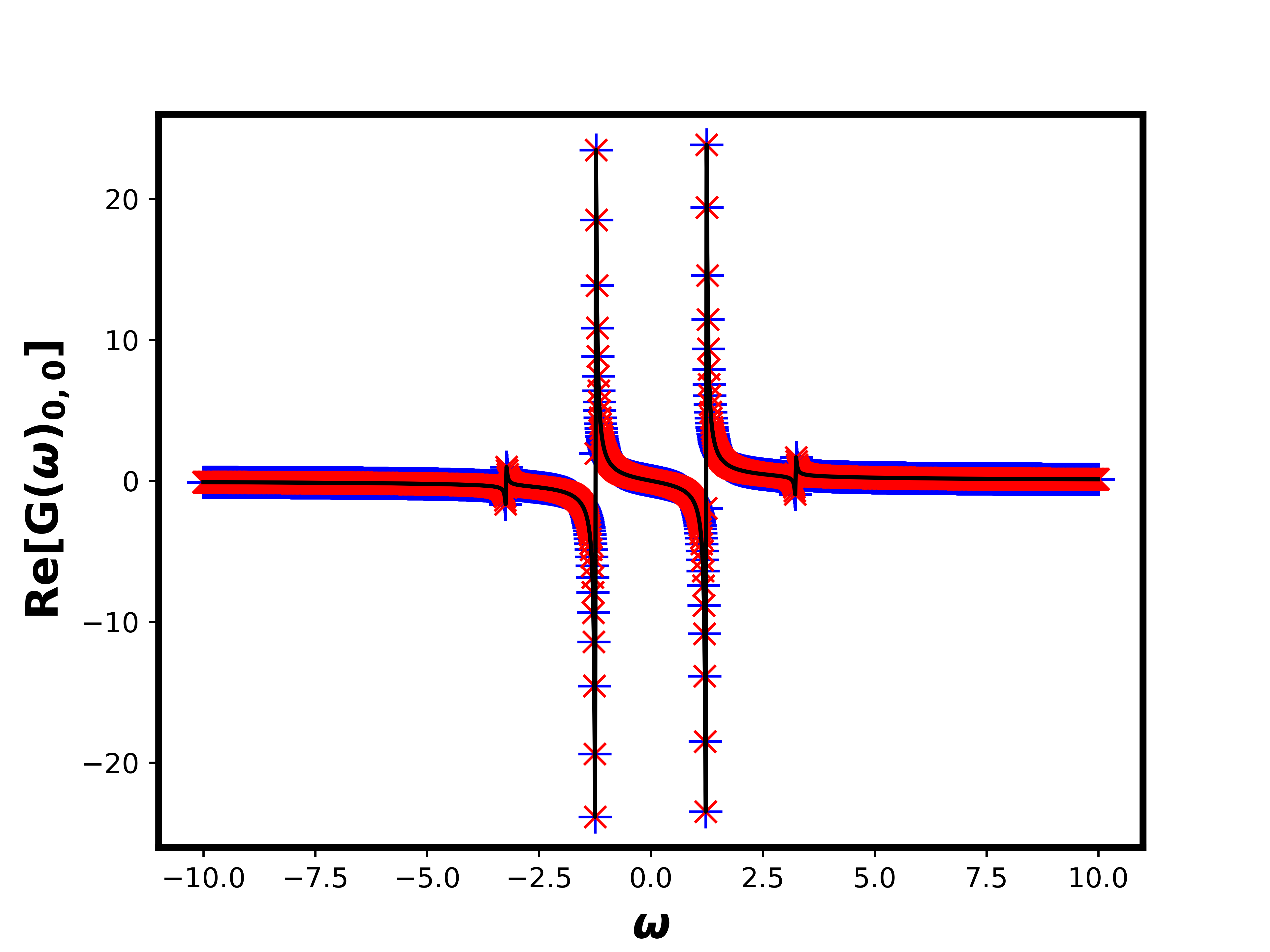}
    \end{overpic}
    \begin{overpic}[width=8cm]{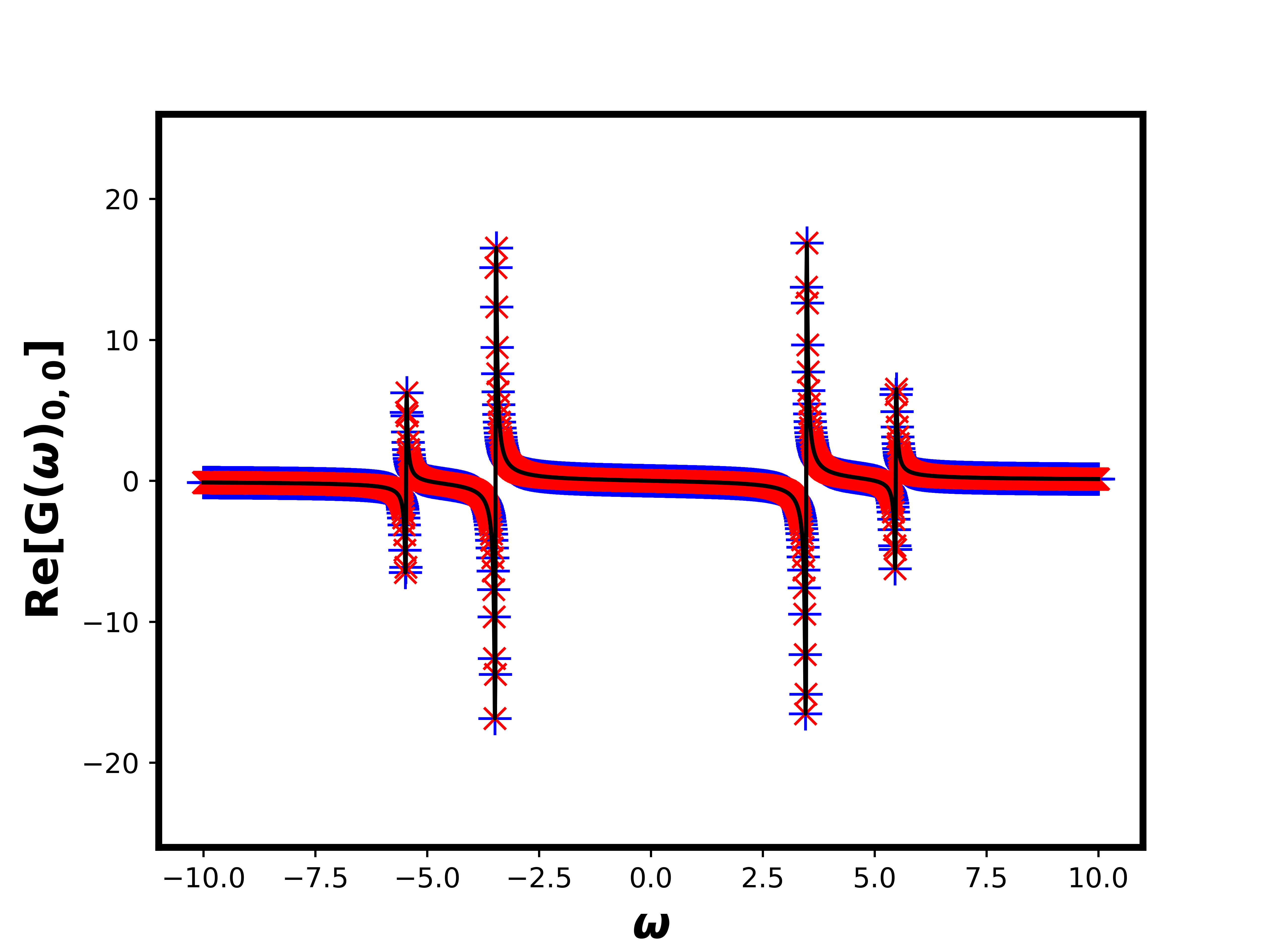}
    \end{overpic}
    \end{minipage}
    \caption{Noiseless simulations of quantum computed Green's function versus $\omega$ (in units of $t$) for the Hubbard dimer when $U = 0$ (top panels), $\abs{\frac{U}{t}} = 2$ (middle panels), and when $\abs{\frac{U}{t}} = 8$ (bottom panels), using two different initialisation strategies for the quantum Lanczos routine: blue `$+$' corresponds to strategy \textit{i)}, while red `$\times$' represents strategy \textit{ii)}. In the legend, spin orbital indexes $ij$ have been omitted and $\hat{H}^{n,(p)} = \hat{f}\hat{H}^{n}\hat{f}^{\dagger}, \hat{H}^{n,(h)} = \hat{f}^{\dagger}\hat{H}^{n}\hat{f}$. Left panels show the spectral function (number of states per unit energy normalised by $\pi$, defined in Eq. \ref{eqn:spectral_fn}), and right panels show the real part of the $G_{0,0}$ element (in units of $\frac{1}{t}$ where we use $t = 1$). We note that in this case the Lanczos routine reproduces the GF obtained from ED.}
\label{fig:gf_dimer_sv}
\end{figure}

\begin{figure}[H]
    \centering
    \begin{minipage}{8cm}
    \begin{overpic}[width=8cm]{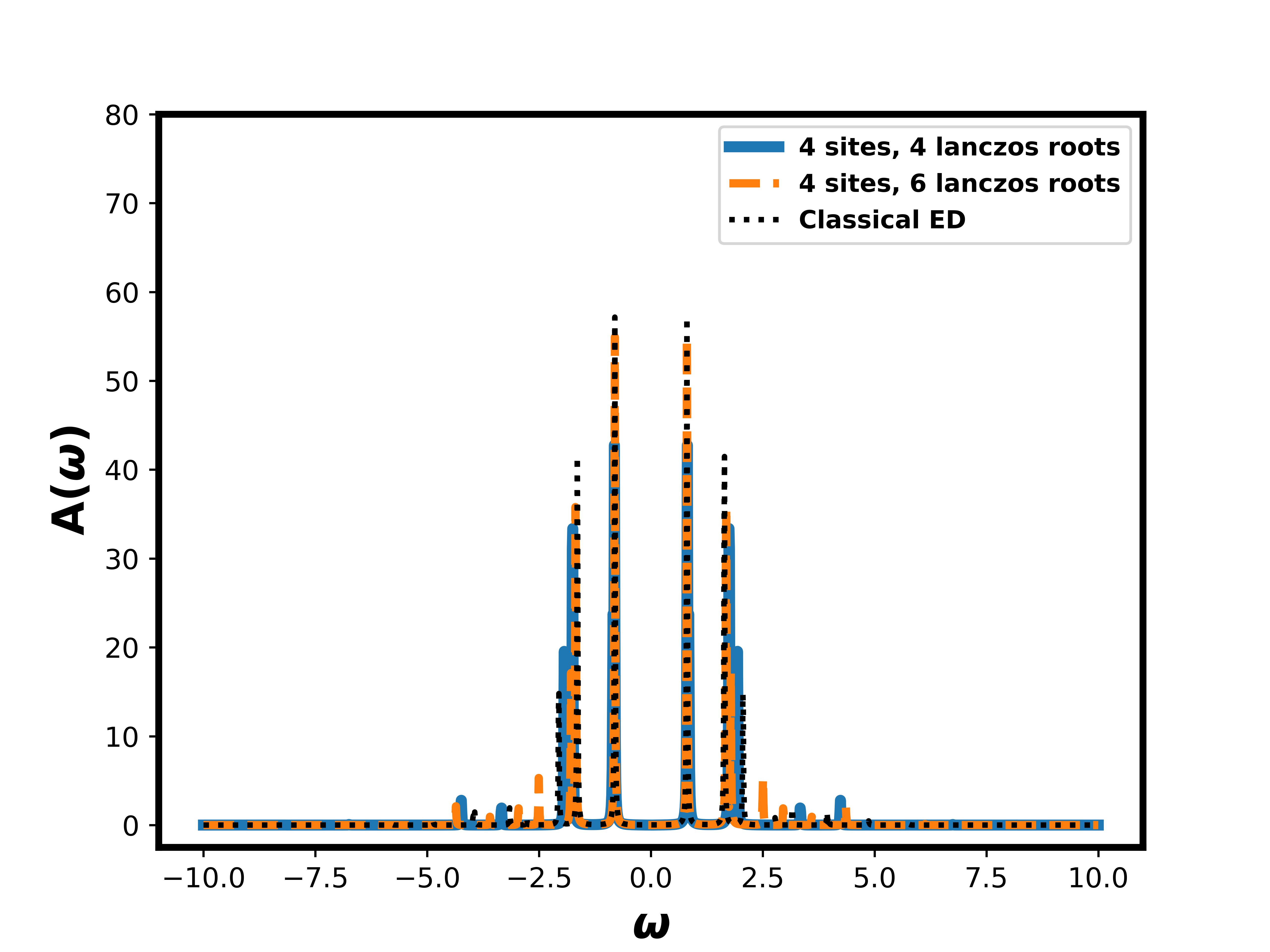}
    \end{overpic}
    \begin{overpic}[width=8cm]{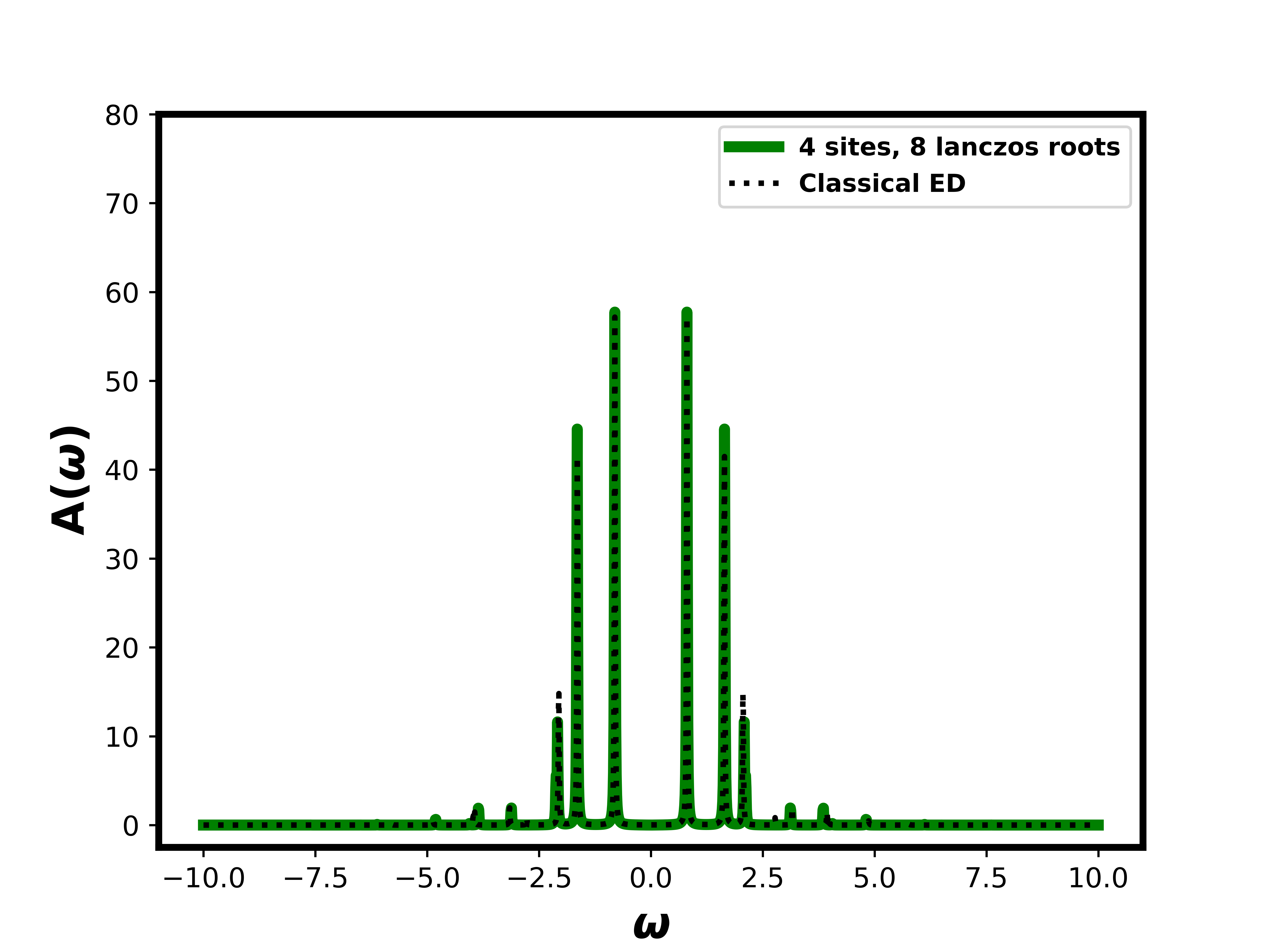}
    \end{overpic}
    \end{minipage}\begin{minipage}{8cm}
    \begin{overpic}[width=8cm]{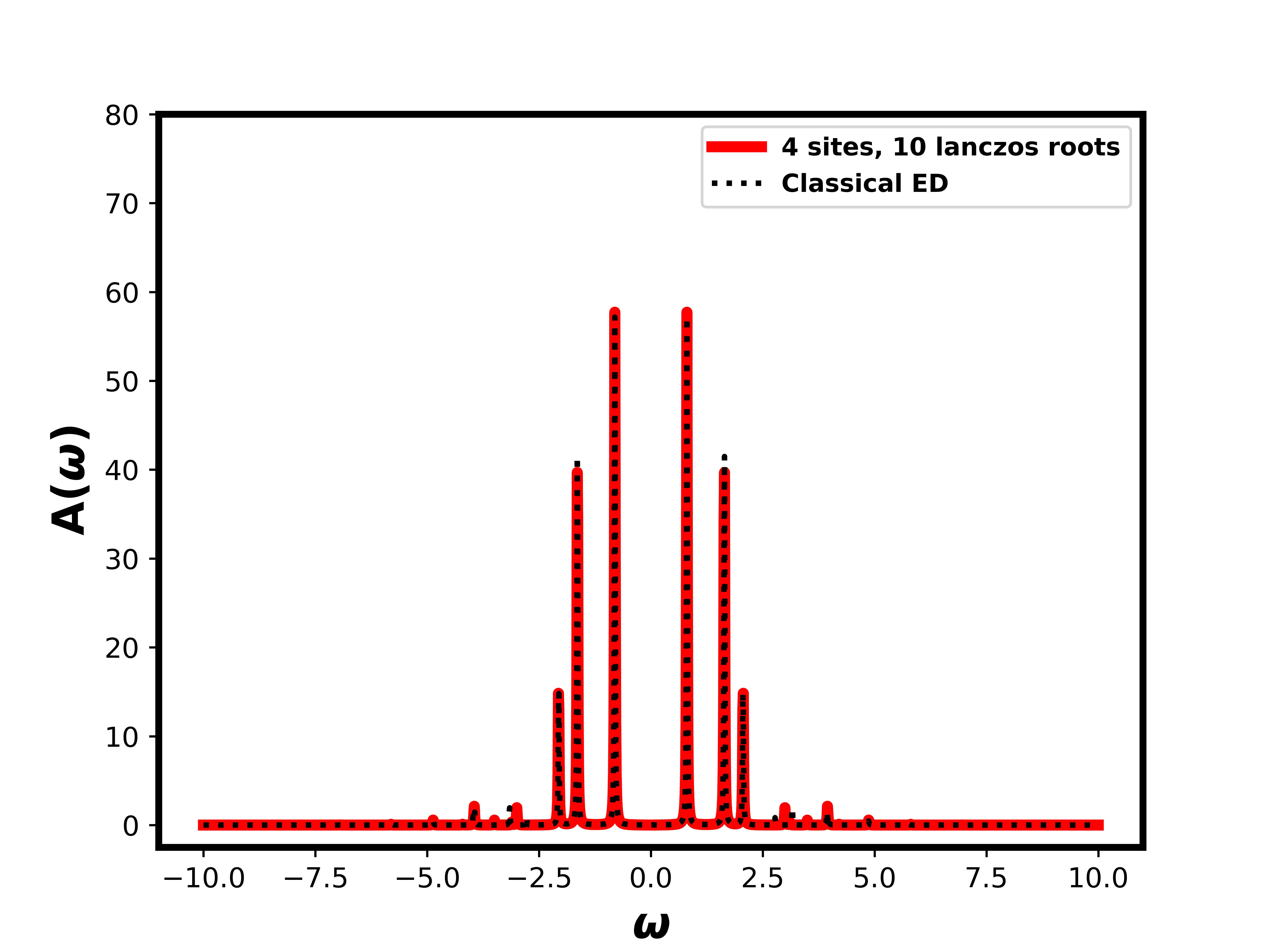}
    \end{overpic}
    \begin{overpic}[width=8cm]{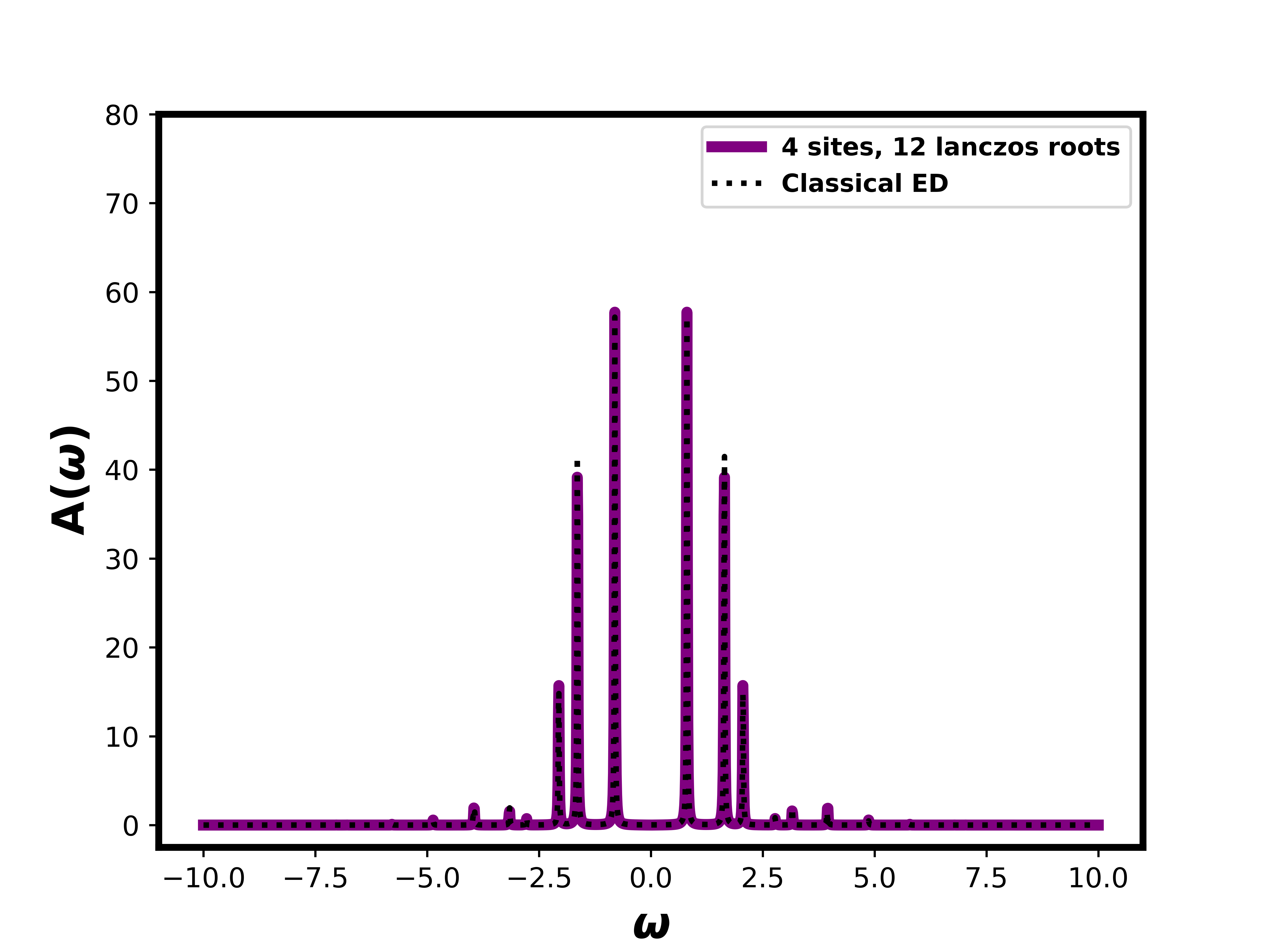}
    \end{overpic}
    \end{minipage}
    \caption{Spectral function of the the 4-site Hubbard model when $\abs{\frac{U}{t}} = 2$, using noiseless simulations of quantum computed Green's functions, using strategy \textit{ii)} for the quantum Lanczos routine. Convergence towards the ED result of the quantum computed spectral function with respect to the number of Lanczos roots (maximum value of $l$) is observed. Peak positions are reasonably converged for $\geq$ 8 Lanczos roots. Dashed black line shows the result from exact diagonalisation. Dimension of spectral function (defined in Eq. \ref{eqn:spectral_fn}) is number of states per unit energy normalised by $\pi$.}
\label{fig:gf_4site_sv}
\end{figure}

In Fig. \ref{fig:gf_dimer_sv} we compare a statevector simulated (noiseless) quantum GF to the classically calculated GF (throughout this work, results labelled `classical ED' correspond to classical exact diagonalisation used to obtain the GF in the Lehmann representation at zero temperature \cite{avella13}), for the non-interacting ($U = 0$), intermediate ($\abs{\frac{U}{t}} = 2$), and strongly correlated ($\abs{\frac{U}{t}} = 8$) regimes of the Hubbard dimer. In Fig. \ref{fig:gf_dimer_sv}, the spectral function and real part of a diagonal element of the GF are plotted. The spectral function (which counts the number of states per energy normalised by $\pi$) is obtained from the imaginary part of the diagonal elements of the GF matrix
\begin{equation}\label{eqn:spectral_fn}
    A(\omega) = -\frac{1}{\pi}\mathfrak{Im}\mathbf{G}(\omega + \text{i}\delta).
\end{equation}
\noindent where $\delta$ is a broadening term, and we set $\delta = 0.01$. Both strategies \textit{i)} and \textit{ii)} (described in section \ref{subsec:qgf}) are used in Fig. \ref{fig:gf_dimer_sv}, showing their equivalence in terms of physical results. In terms of the circuit resources used in either case, interesting differences arise due to the differences in state preparation and operator expectation measurement. In this case, the exact solution is obtained (i.e. the continued fraction GF matches the GF calculated from exact diagonalisation (ED)) for maximum $l = 2$. This corresponds to a maximum Hamiltonian moment of $n=3$, as in Eq. \ref{eqn:alpha_beta}. In Fig. \ref{fig:gf_4site_sv} the spectral function is calculated for a 4-site Hubbard model, with an increasing number of Lanczos roots (i.e. increasing dimensionality of the Krylov space). This shows that at 8 Lanczos roots, reasonable accuracy is obtained in the position of spectral peaks (energies which exhibit poles of the GF), with most of the deviation relative to the exact result exhibited in the weights (rather than positions) of the peaks near $\omega \sim \pm2$ (throughout this paper, $\omega$ is in units of $t$, and we set $t = 1$). 
\onecolumngrid

\twocolumngrid

\subsubsection{Strategy i) Initial Lanczos Vector
Preparation: Hubbard Model} \label{results_lanc_vec_prep}

Following the noiseless VQE optimization of the ground state ansatz (Fig. \ref{fig:ansatz2}), the ground state of the interacting Hubbard dimer exhibits the following expansion in terms of ON basis vectors

\begin{equation} \label{gs_expansion}
\begin{split}
    |\Psi_{\text{GS}}\rangle &= c_{1100} |1100\rangle + c_{1001} |1001\rangle \\
    &+ c_{0110} |0110\rangle + c_{0011} |0011\rangle,
\end{split}
\end{equation}

\noindent where the coefficients depend on $U$ and $t$ Hubbard parameters, with $c_{1100} = c_{0011}, c_{1001} = -c_{0110}$ due to symmetry and fermionic exchange, and for the noninteracting $U=0, t = 1$ case, $|c_{1100}| = |c_{0011}| = |c_{1001}| = |c_{0110}| = 1/2$. Taking the diagonal element of the particle and hole GFs $g_{0,0}^{\text{(p)}}$ and $g_{0,0}^{\text{(h)}}$ as an example, which require preparation of Lanczos vectors $|\Psi^{\text{(p)}}_{00}\rangle = \frac{\hat{f}^{\dagger}_{0}|\Psi_{\text{GS}}\rangle}{\sqrt{n_{00}^{\text{(p)}}}}$ and $|\Psi^{\text{(h)}}_{00}\rangle = \frac{\hat{f}_{0}|\Psi_{\text{GS}}\rangle}{\sqrt{n_{00}^{\text{(h)}}}}$ to measure the $\langle \hat{H}^{n, (\text{p})}_{ii} \rangle$ and $\langle \hat{H}^{n, (\text{h})}_{ii} \rangle$, respectively, the following expansions for these Lanczos vectors are obtained after applying the ladder operators to the ground state

\begin{equation} \label{p_lanc_expansion}
\begin{split}
    |\Psi^{\text{(p)}}_{00}\rangle = \frac{c_{0110}}{\sqrt{n_{00}^{\text{(p)}}}} |1110\rangle + \frac{c_{0011}}{\sqrt{n_{00}^{\text{(p)}}}} |1011\rangle,
\end{split}
\end{equation}

\begin{equation} \label{h_lanc_expansion}
\begin{split}
    |\Psi^{\text{(h)}}_{00}\rangle = \frac{c_{1100}}{\sqrt{n_{00}^{\text{(h)}}}} |0100\rangle + \frac{c_{1001}}{\sqrt{n_{00}^{\text{(h)}}}} |0001\rangle.
\end{split}
\end{equation}

\noindent The quantum circuits representing Eqs. \ref{p_lanc_expansion} and \ref{h_lanc_expansion} (in addition to the off-diagonal terms) can then be prepared using multicontrolled Given's rotations to obtain arbitrary (particle-conserving) linear combinations of basis states \cite{arrazola22}. In strategy \textit{i)}, in contrast to strategy \textit{ii)}, we do not measure the expectation of sandwiched moments with respect to the ground state, but rather take the expectation of the moments $\hat{H}^{n}$ with respect to the Lanczos vectors. Hence, the Pauli strings to be measured for each GF element do not change with the element index of the GF matrix. The Pauli strings take the following form after JW encoding the fermionic Hamiltonian moments (using Hubbard parameters $U=2, t=1$ as an example)

\begin{equation} \label{eqn:p_qham1}
\begin{split}
    &\hat{H}^1 \mapsto \frac{1}{2}\bigg(Z_0Z_1 + Z_2Z_3 - X_0Z_1X_2 - Y_0Z_1Y_2 
    \\&- X_1Z_2X_3 - Y_1Z_2Y_3 \bigg),
\end{split}
\end{equation}

\begin{equation} \label{eqn:p_qham2}
\begin{split}
    &\hat{H}^2 \mapsto \frac{1}{2}\bigg(3I - X_0X_1Y_2Y_3 + X_0Y_1Y_2X_3 + Y_0X_1X_2Y_3 
    \\&- Y_0Y_1X_2X_3 + Z_0Z_1Z_2X_3 - Z_0Z_2 - Z_1Z_3 \bigg),
\end{split}
\end{equation}

\begin{equation} \label{eqn:p_qham3}
\begin{split}
    &\hat{H}^3 \mapsto \frac{1}{2}\bigg(X_0X_1X_2X_3 + X_0Y_1X_2Y_3 - 3X_0Z_1X_2 
    \\ &+ 2X_0X_2Z_3 + Y_0X_1Y_2X_3 + Y_0Y_1Y_2Y_3 - 3Y_0Z_1Y_2 
    \\[7pt] &+ 2Y_0Y_2Z_3  + 2Z_0X_1X_3 + 2Z_0Y_1Y_3 + 2Z_0Z_1 
    \\ &- Z_0Z_3 - 3X_1Z_2X_3 - 3Y_1Z_2Y_3 - Z_1Z_2 + 2Z_2Z_3 \bigg).
\end{split}
\end{equation}

(with $I \equiv I_0I_1I_2I_3$). However, the circuits corresponding to the Lanczos vectors do change with GF matrix element index, and grow rapidly with the size of the system. Considering the scaling of the number of terms in Eq. \ref{gs_expansion} with respect to the number of qubits, the number of terms required to exactly represent the Lanczos vectors (Eqs. \ref{p_lanc_expansion} and \ref{h_lanc_expansion}) scales exponentially. Hence, the feasibility of strategy \textit{i)} largely depends on how the Lanczos vectors are prepared on the quantum circuit. 

A well known issue with the Lanczos method is the degradation of the Lanczos basis due to floating point precision \cite{pavarini11}. Typically, this degradation occurs for large values of $l$, where $\beta_l$ tends to become small and hence numerical noise is amplified when dividing by $\beta_l$ to normalise $|v_l\rangle$. In classical schemes, this is typically treated by re-orthogonalising the current $|v_l\rangle$ to the set of previously calculated vectors $\{|v_{k<l}\rangle\}$. However, quantum noise and the statistical nature of circuit measurements required for the evaluation of Hamiltonian moments can also affect the Lanczos basis and associated Lanczos coefficients calculated from the measured moments. 

\begin{figure}[H]
    \centering
    \begin{minipage}{7cm}
    \begin{overpic}[width=7cm]{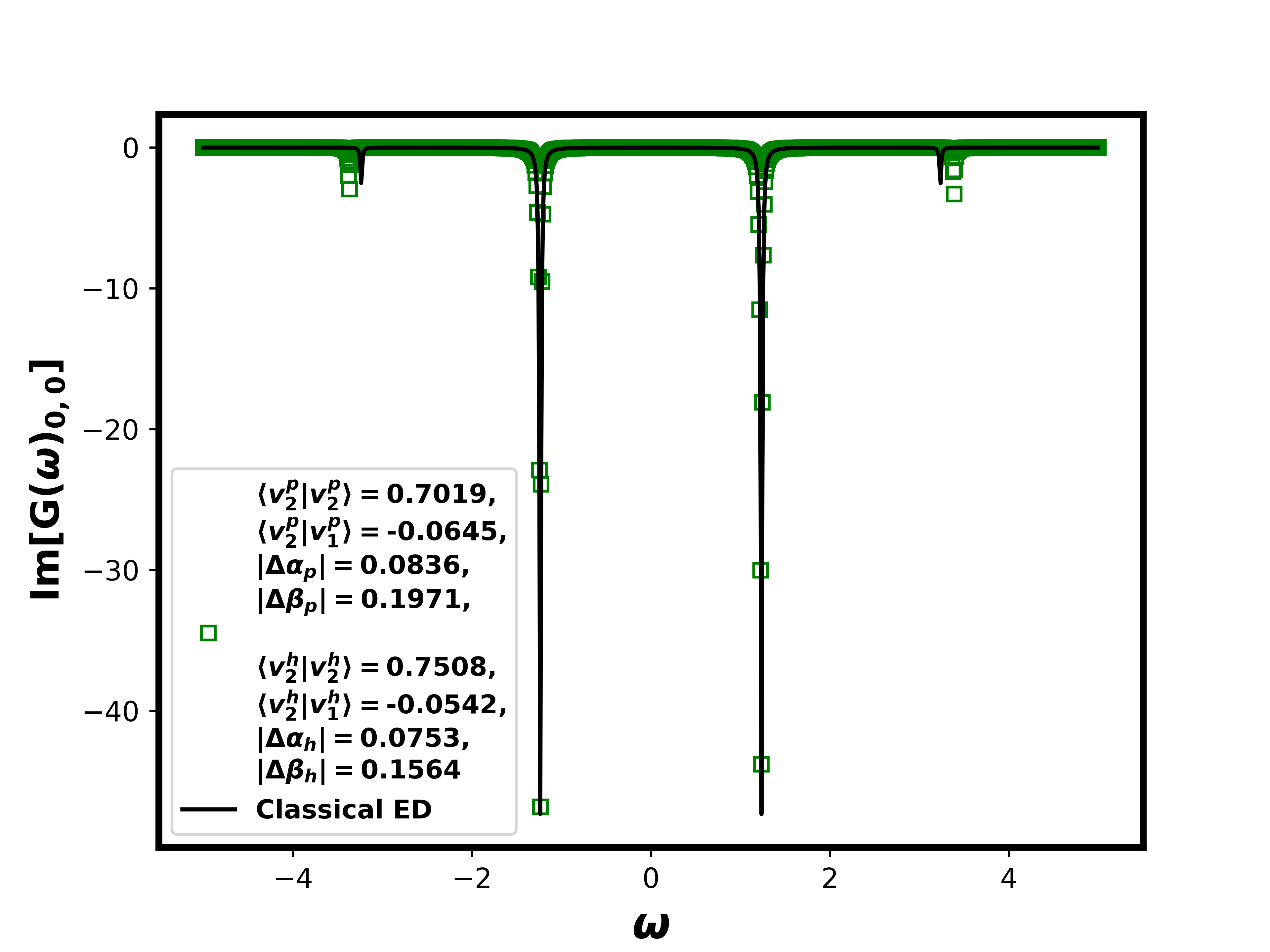}
    \end{overpic}
    \begin{overpic}[width=7cm]{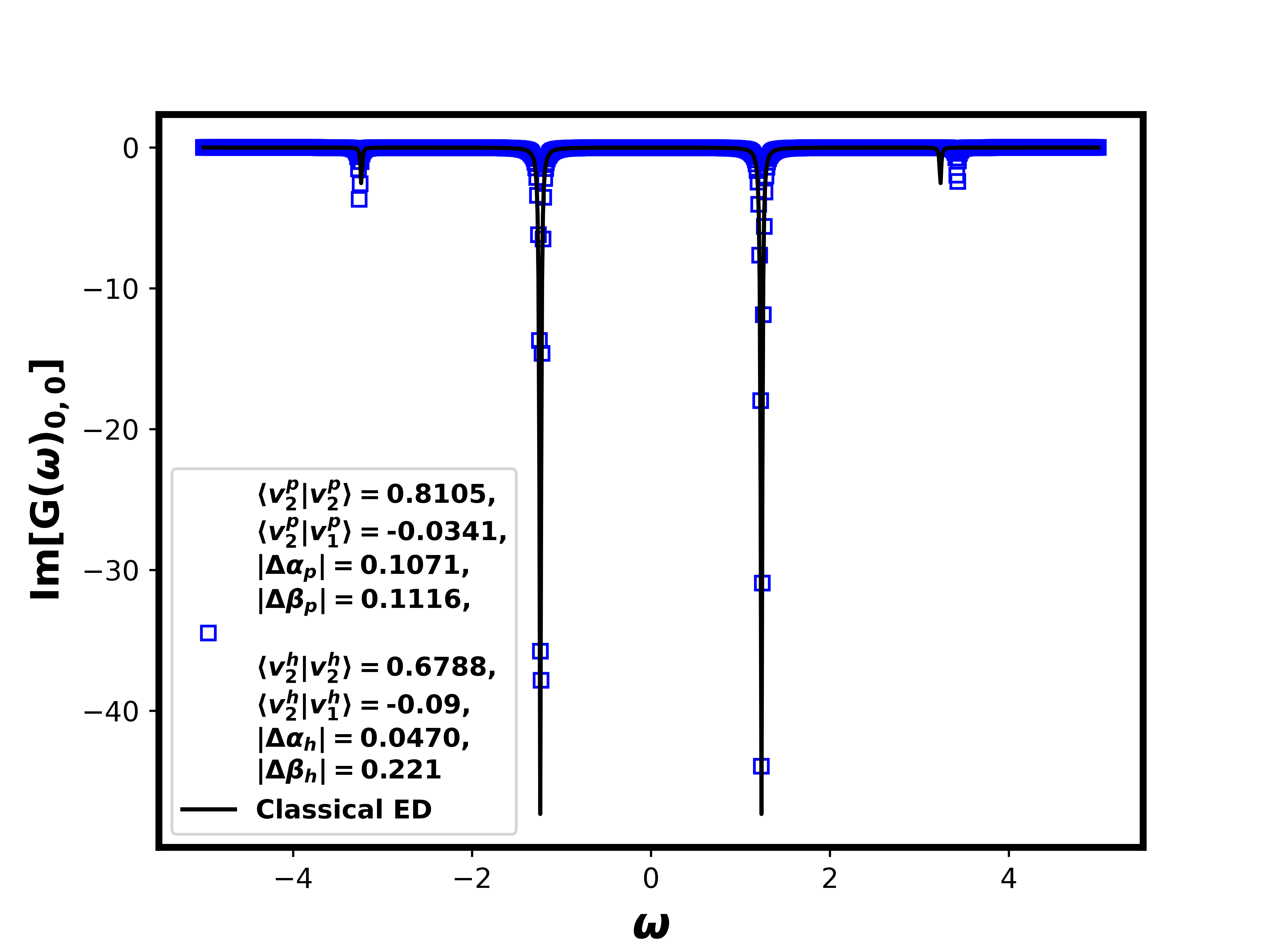}
    \end{overpic}
    \end{minipage}\begin{minipage}{7cm}
    \begin{overpic}[width=7cm]{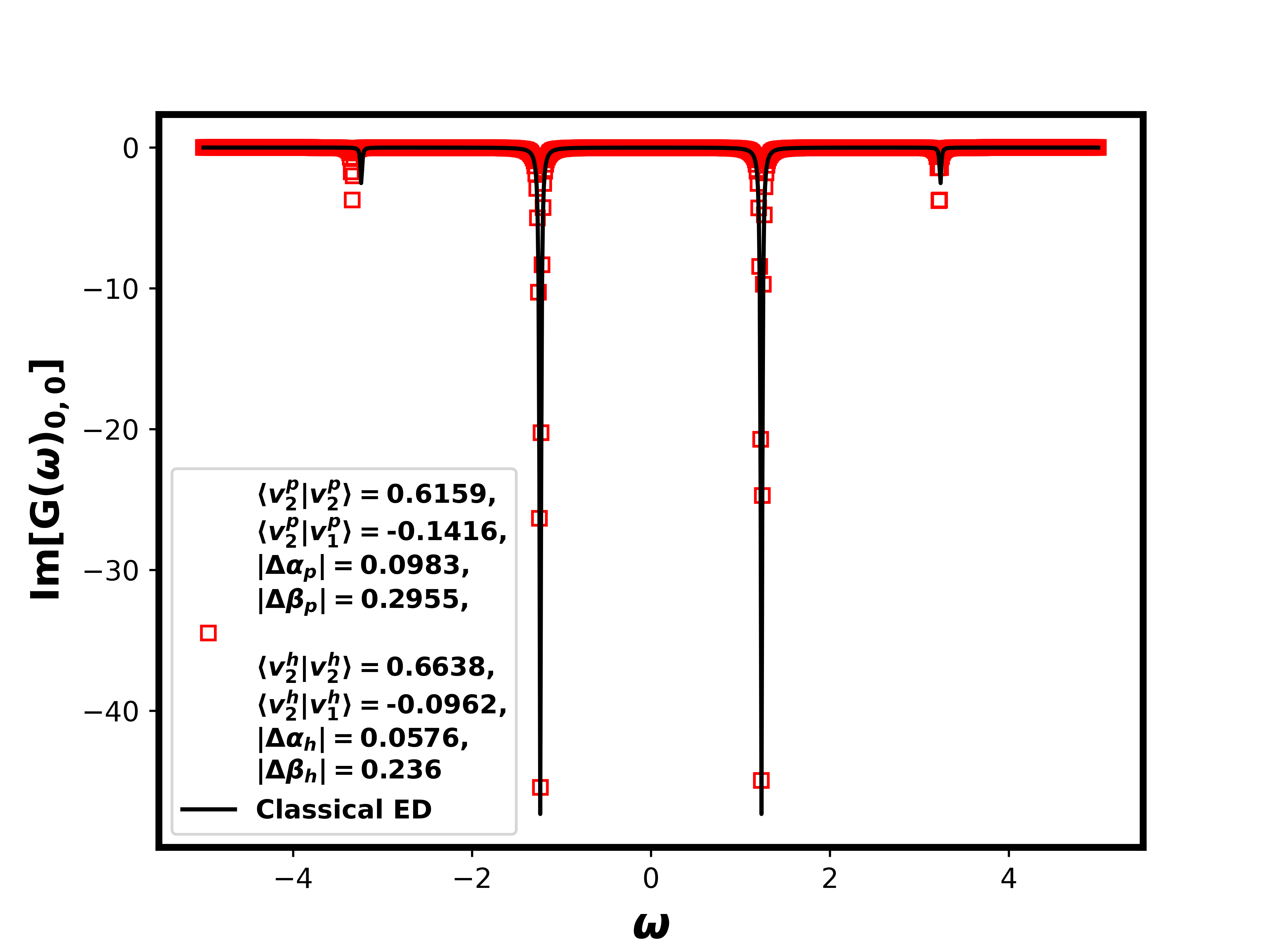}
    \end{overpic}
    \begin{overpic}[width=7cm]{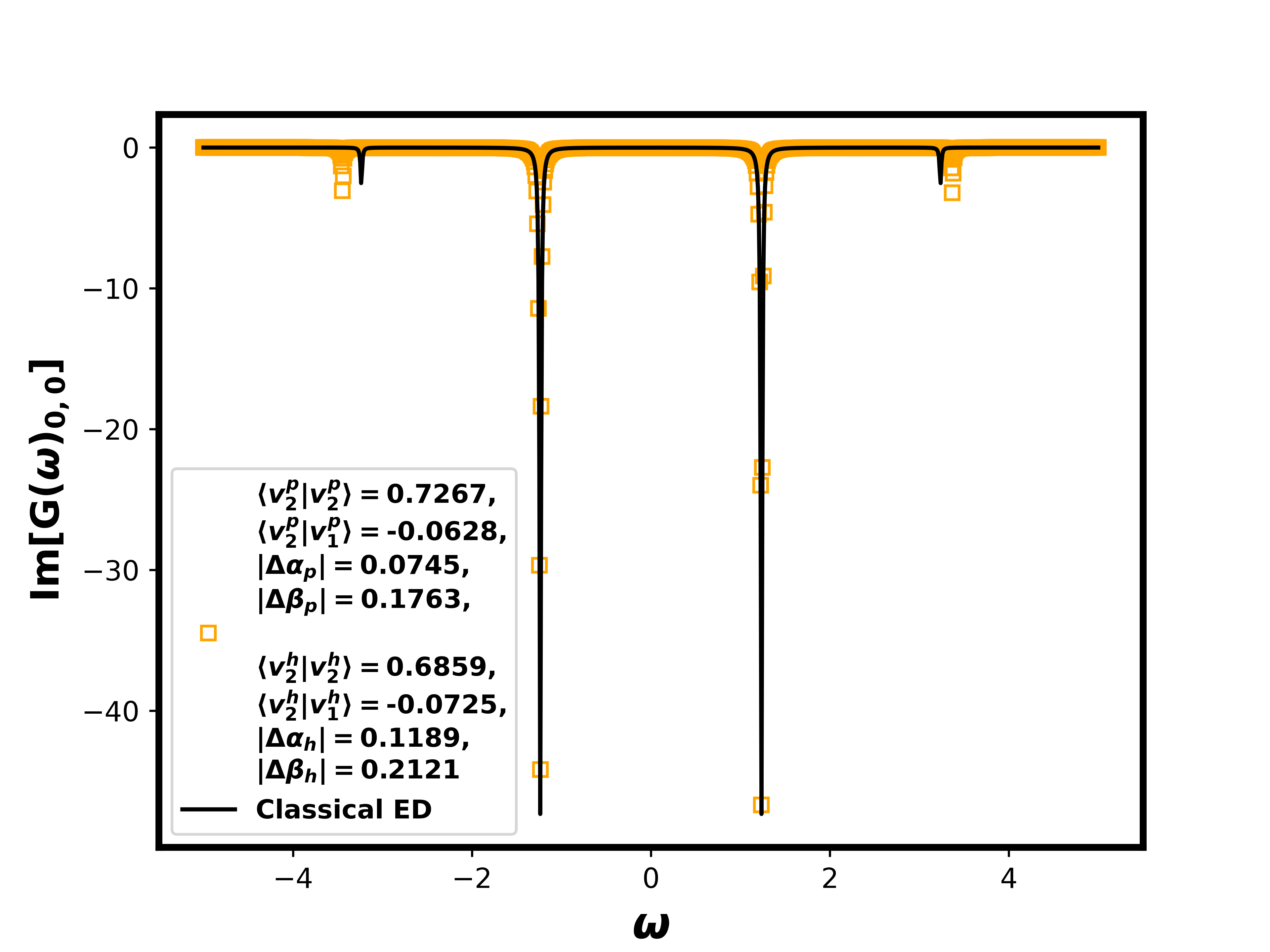}
    \end{overpic}
    \end{minipage}
    \caption{Imaginary part of $G(\omega)_{0,0}$ obtained from the Quantinuum H1-1 emulator, from 4 separate runs. For each run, the associated errors in Lanczos coefficients are shown, along with the norms of the Lanczos vector constructed from the (moment-)measured coefficients.}
    \label{fig:imgf_dimer_coeffs_ovs}
\end{figure}

To study this, the imaginary part of a diagonal element of $\mathbf{G}$ for the Hubbard dimer is obtained from measurements of expectations of Hamiltonian moments corresponding to the particle ($g_{0,0}^{(\text{p})}$) and hole ($g_{0,0}^{(\text{h})}$) contributions to $G_{0,0}$. The results are obtained from the Quantinuum H1-1 emulator, a classical device emulator with a noise model corresponding to the noise profile of the H1-1 device \cite{h11e}. These emulated experiments correspond to one of the 14 measurable circuits to be described in section \ref{hardware_results}, representing the upper diagonal elements of the particle and hole GF matrices, both with 23 (7) total (2-qubit) gates and depth 13. These are plotted alongside the normalised absolute errors in $\alpha_{l}$ and $\beta_{l}$ resulting from measurements, and the diagonal and off-diagonal components of the overlap matrix elements calculated classically from the Lanczos vectors which in turn are built from the measured $\alpha_{l}$ and $\beta_{l}$. Since the ED result for the Hubbard dimer GF is recovered for maximum $l = 2$, only $\alpha_{1}, \alpha_{2}$, and $\beta_{1}$ are used in this case, and we average the error in $\alpha_{l}$. Hence the errors in the Lanczos coefficients are calculated as

\begin{equation}
    |\Delta \alpha^{(\text{p/h})}| = \frac{1}{2} \bigg(\bigg|\frac{\alpha_{1}^{(\text{p/h})} - \alpha_{1,{\text{exact}}}^{(\text{p/h})}}{\alpha_{1,{\text{exact}}}^{(\text{p/h})}}\bigg| + \bigg|\frac{\alpha_{2}^{(\text{p/h})} - \alpha_{2,{\text{exact}}}^{(\text{p/h})}}{\alpha_{2,{\text{exact}}}^{(\text{p/h})}}\bigg|\bigg)
\end{equation}

\begin{equation}
    |\Delta \beta^{(\text{p/h})}| =  \bigg(\bigg|\frac{\beta_{1}^{(\text{p/h})} - \beta_{1,{\text{exact}}}^{(\text{p/h})}}{\beta_{1,{\text{exact}}}^{(\text{p/h})}}\bigg|\bigg)
\end{equation}

\noindent where $\alpha_{l,{\text{exact}}}^{(\text{p/h})}, \beta_{l,{\text{exact}}}^{(\text{p/h})}$ are noiseless ideal values of the Lanczos coefficients. The next Lanczos vector $|v_2\rangle$ is then constructed from these measured Lanczos coefficients, and its norm $\langle v_2|v_2 \rangle$ and overlap with $v_1$ are obtained classically (since the ground state parameters are obtained from an ideal noise free calculation, and the vector norms are calculated classically, the first Lanczos vector $v_1^{(\text{p/h})} = |\Psi^{\text{(p/h)}}_{00}\rangle$ has unity norm by construction). The results of 4 separate emulator runs, corresponding to 4 separate sets of measurements, are shown in Fig. \ref{fig:imgf_dimer_coeffs_ovs}. It can be observed that, for a given set of measurements, the error in $\beta_l$ resulting from the quantum noise correlates to the deviation from unity norm of $v_2$, which is not unexpected since $\beta_l$ governs the normalisation of the Lanczos vectors. This error in normalisation also translates to the deviation of spectral peak heights from the exact result, which gives a qualitative understanding of the effect of quantum noise in $\beta_l$ on the resulting GF: within each plot of Fig.  \ref{fig:imgf_dimer_coeffs_ovs}, a larger $|\Delta \beta^{(\text{p/h})}|$ results in a worse relative peak height (compared to ED) for $g_{0,0}^{(\text{p/h})}$, where superscript "p" ("h") denotes contribution to the positive (negative) frequency spectral peaks.

Also, quantum noise at the circuit level can have a cumulative effect on the orthogonalisation of the Lanczos vectors at higher roots, the study of which requires measurements of corresponding higher moments of the Hamiltonian with respect to the ground state for models larger than the Hubbard dimer. Such measurements require circuits too deep for the current NISQ era, as errors due to gate infidelities and qubit decoherence would accumulate and dominate the measurements of associated Pauli operators. Hence, we leave the effect of quantum noise on the orthogonalisation of Lanczos vectors for higher lying roots as an open question for future studies, although we note the investigation of Vallury \textit{et. al.} \cite{vallury20} into the effect of noise and device errors on the arithmetical operations involved in the assembly of cumulants, in which it was found that infimum estimates of the ground state energy are improved in accuracy and more robust to device noise relative to variational approaches to optimizing $\langle \hat{H}(\boldsymbol{\theta}) \rangle$ \cite{vallury23}. The relation between numerical errors in the Lanczos basis and quantum noise could have interesting implications for error mitigation, which we briefly discuss in the conclusion.

\onecolumngrid

\twocolumngrid

\subsubsection{Strategy ii) Sandwiched Moment Expectation: Hubbard Model} \label{results_sand_moments}

\begin{figure}[H]
    \centering
    \begin{minipage}{7cm}
    \begin{overpic}[width=7cm]{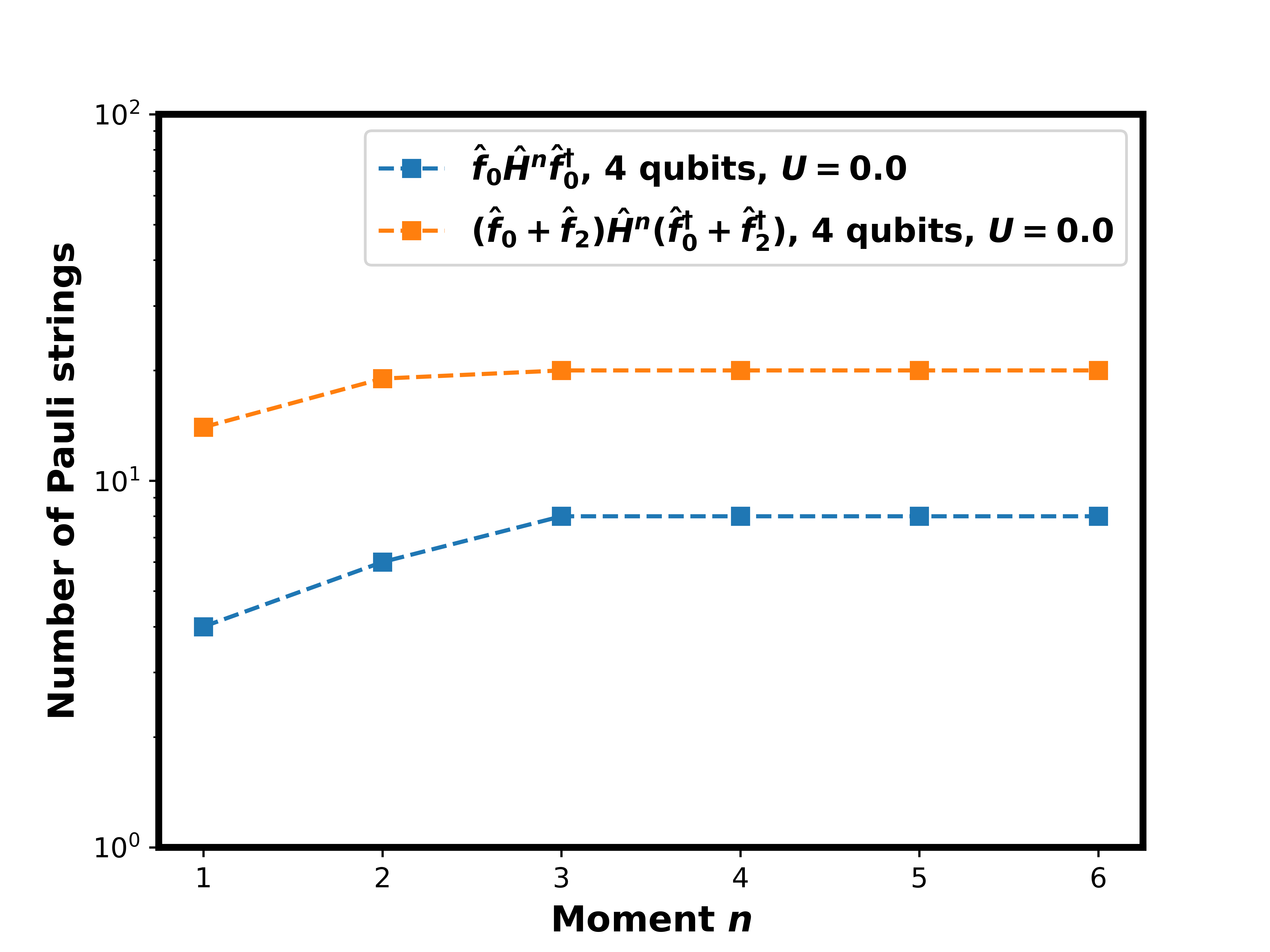}
    \end{overpic}
    \begin{overpic}[width=7cm]{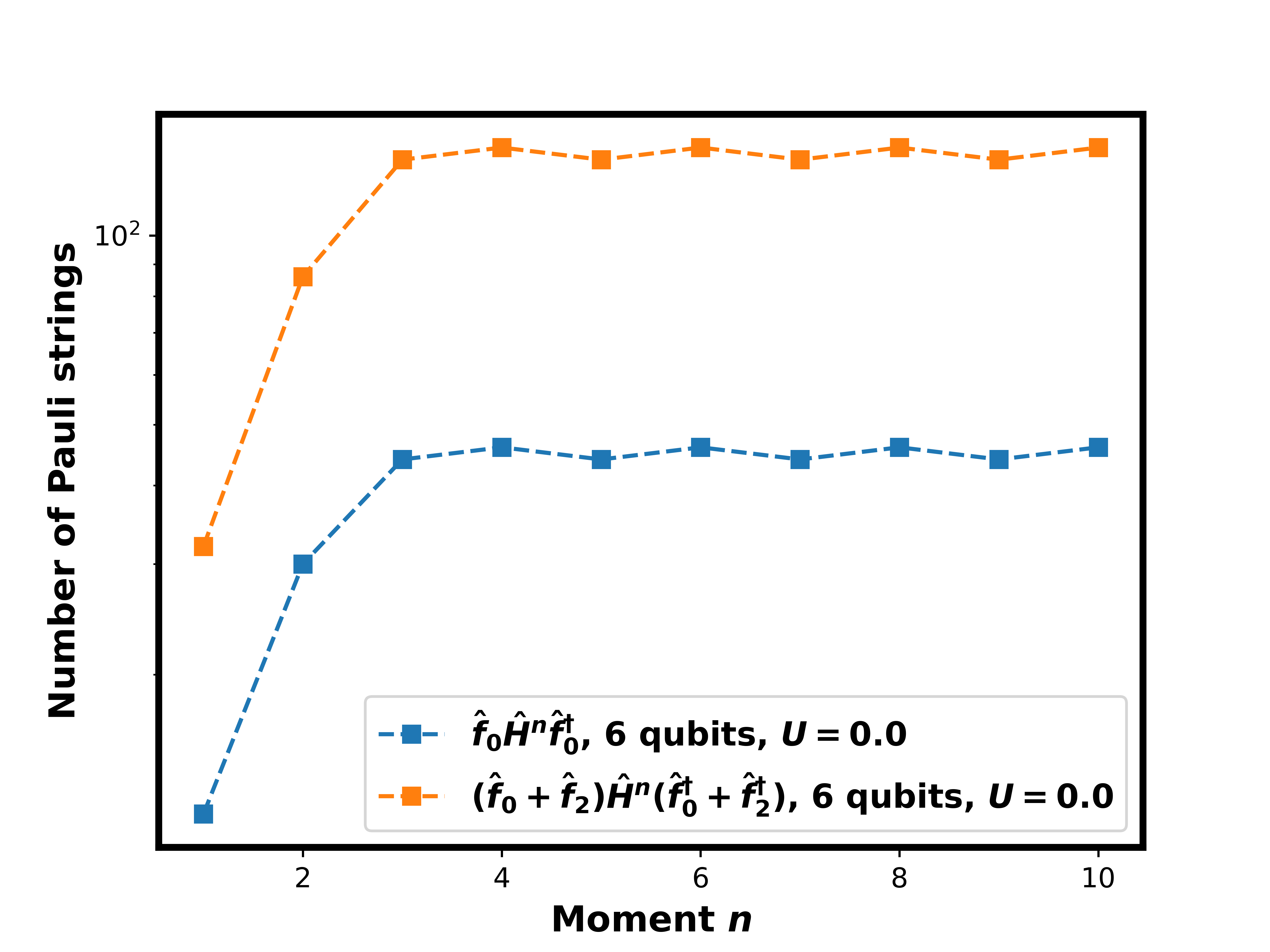}
    \end{overpic}
    \begin{overpic}[width=7cm]{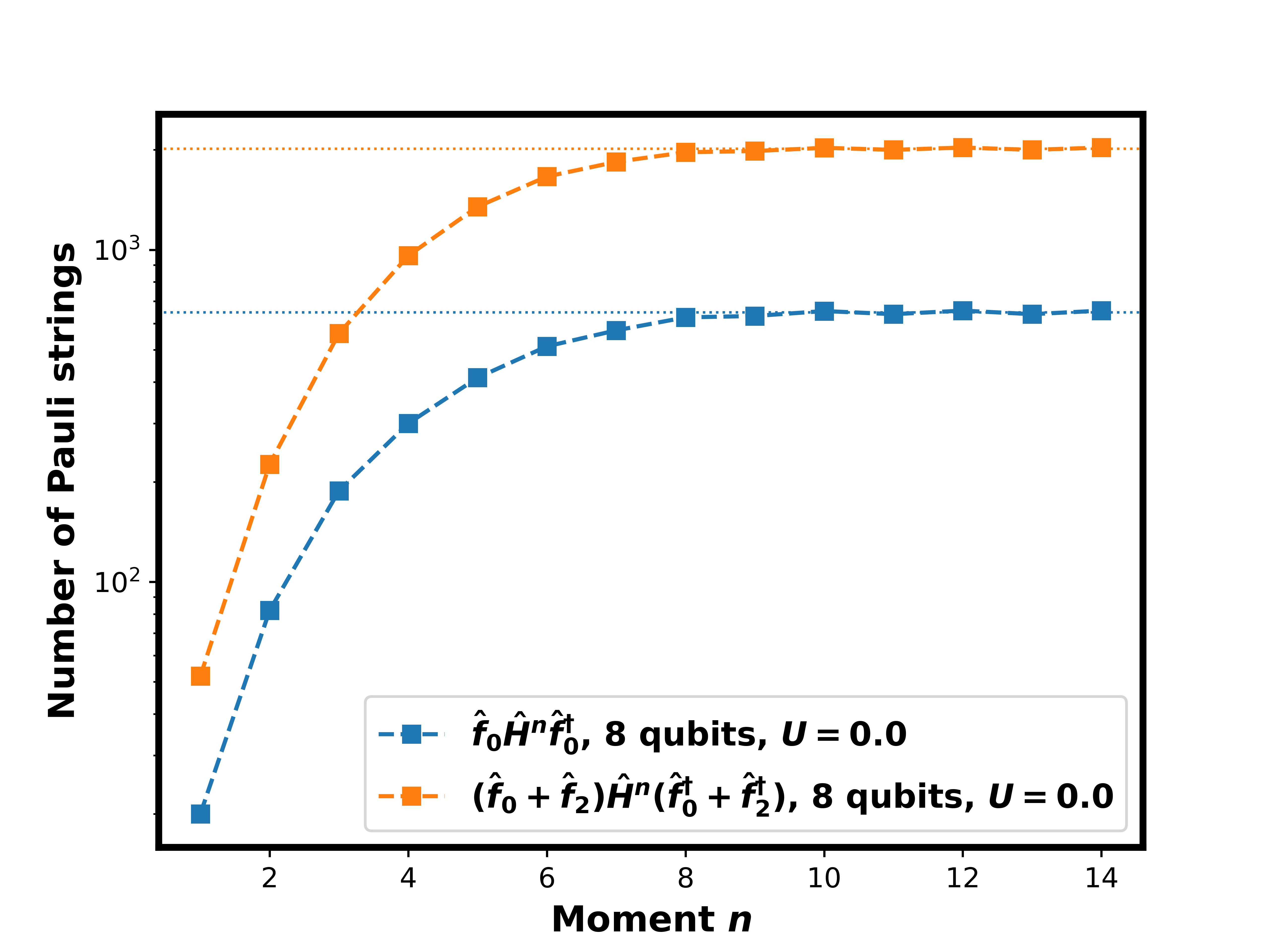}
    \end{overpic}
    \end{minipage}\begin{minipage}{7cm}
    \begin{overpic}[width=7cm]{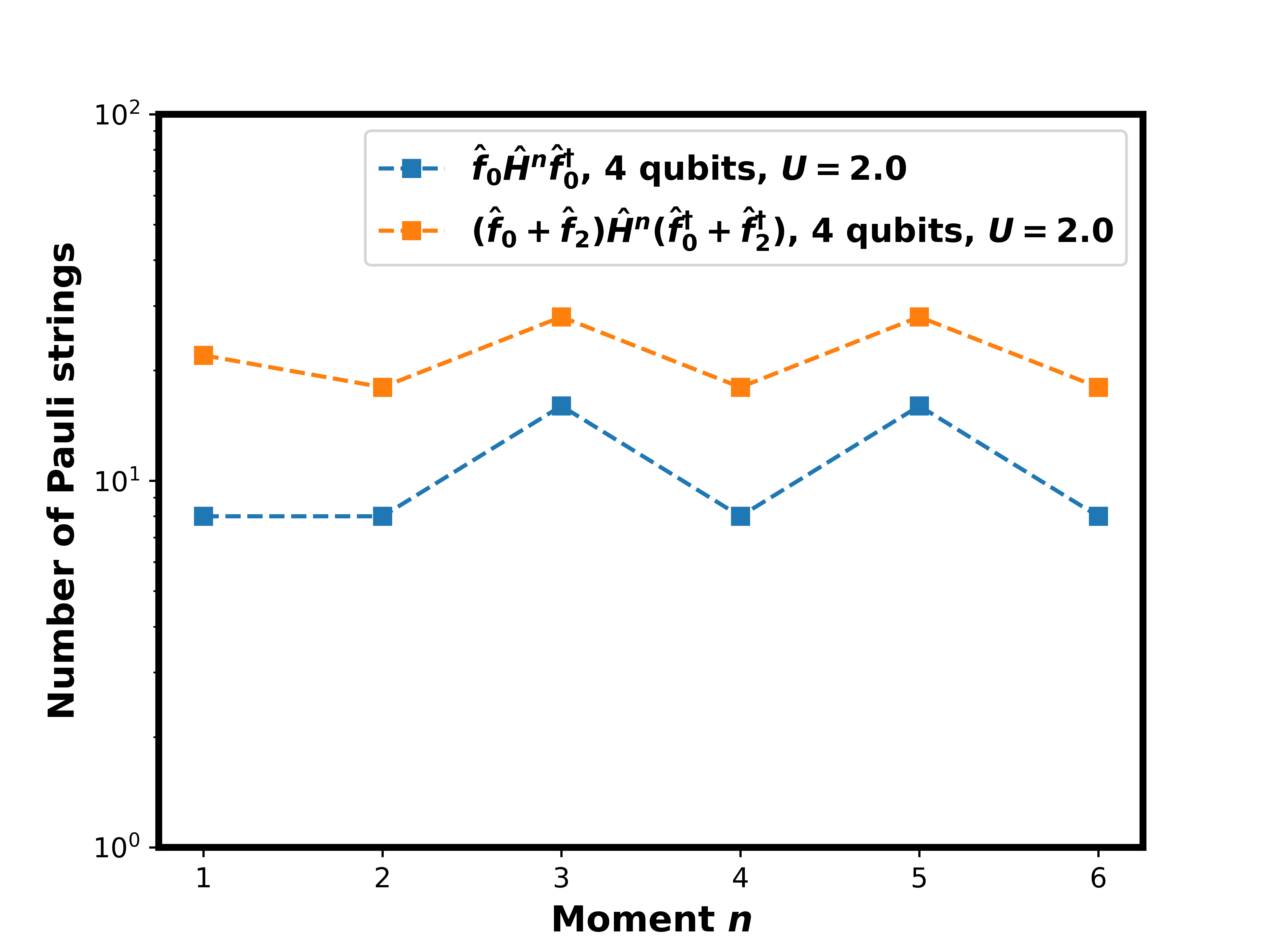}
    \end{overpic}
    \begin{overpic}[width=7cm]{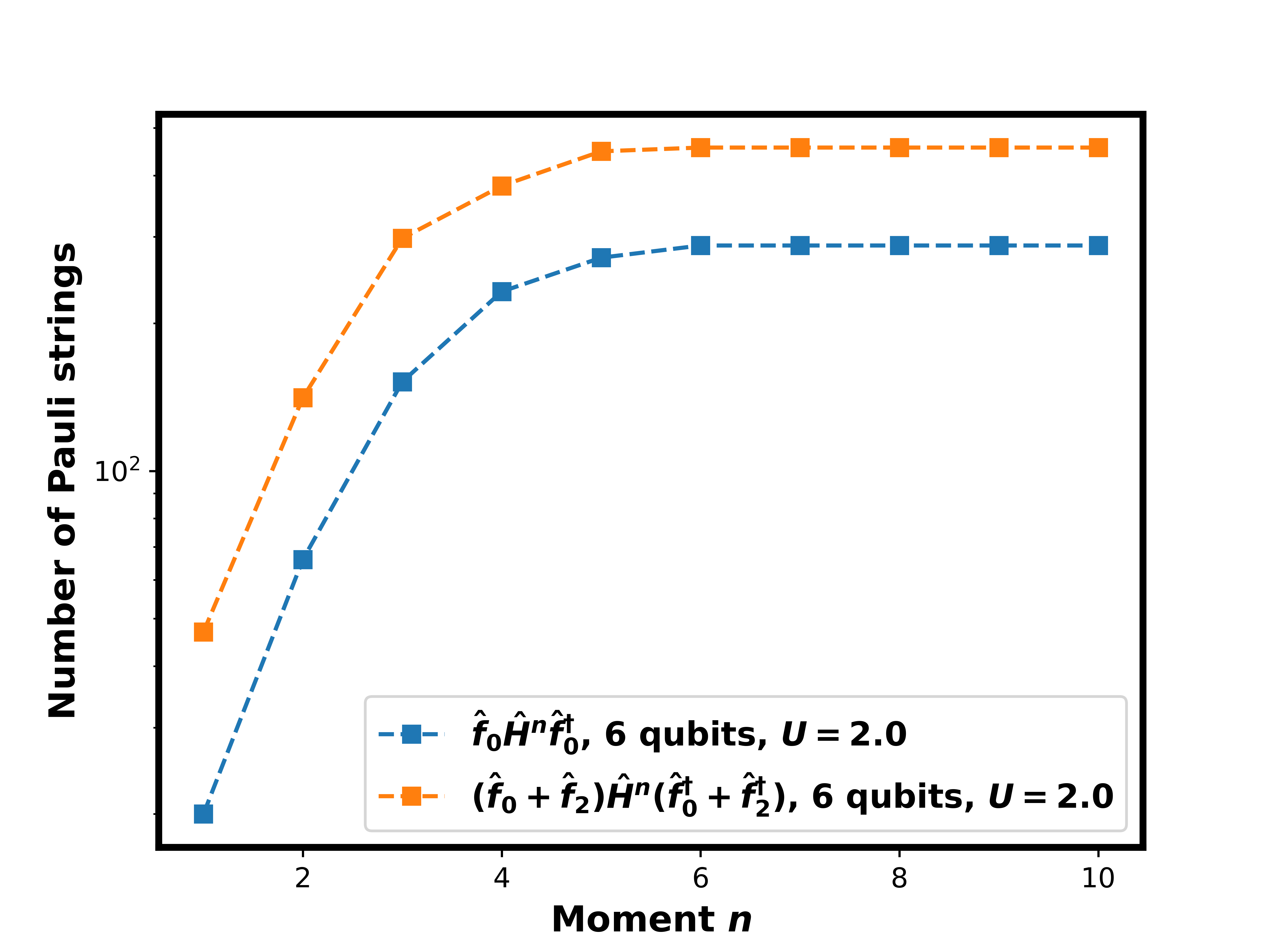}
    \end{overpic}
    \begin{overpic}[width=7cm]{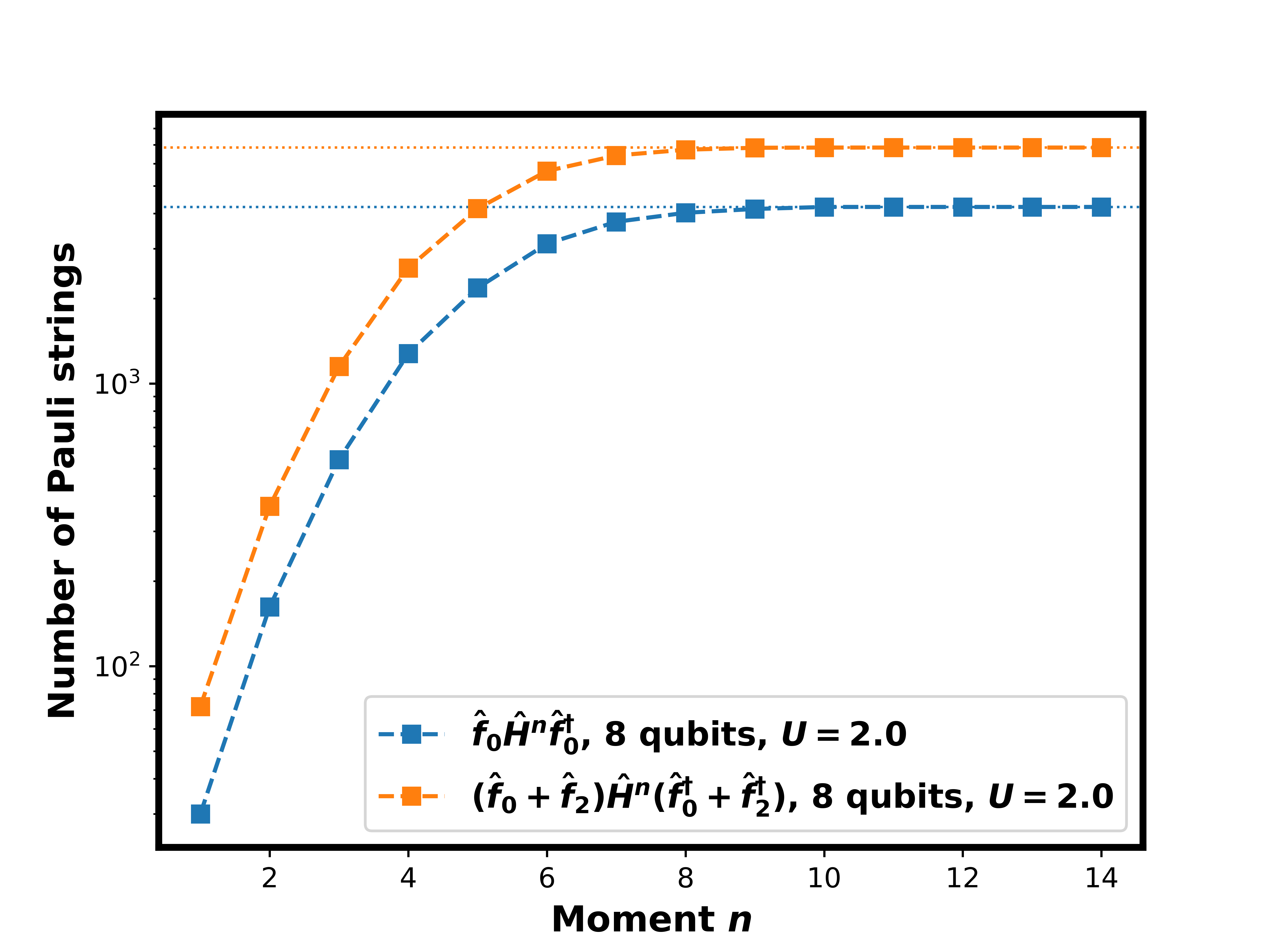}
    \end{overpic}
    \end{minipage}
    \caption{Number of Pauli strings in sandwiched moment operators (for diagonal and off-diagonal elements) versus Hamiltonian moment index $n$, using 4, 6, and 8 qubits, for non-interacting $U=0$ (left panels) and interacting $|\frac{U}{t}|=2$ Hubbard models. For an interacting case of 4 qubits, oscillations in the number of Pauli strings are observed for even and odd $n$; the magnitude of these oscillations rapidly decrease as the system size increases. For larger numbers of qubits, the shape of the curve (semilog plot) is characteristic of polynomial scaling, and indicates a roughly convergent number of Pauli strings as a function of $n$ as the Lanczos procedure approaches the full dimensionality of the Hilbert space (or symmetry-reduced subspace thereof). For 8 qubits, dotted lines are shown to indicate saturated numbers of Pauli strings.}
    \label{fig:npaulis_v_power}
\end{figure}

Using the particle GF with $U=2, t=1$ as an example, the first, second, and third powers of the Hamiltonian, sandwiched between ladder operators, and contributing to the diagonal element $g_{0,0}^{\text{(p)}}$, are mapped to the following sums of Pauli operator strings via JW encoding.

\begin{equation} \label{eqn:p_sand_qham1_00}
\begin{split}
    &\hat{f}_{0} \hat{H}^1 \hat{f}^{\dagger}_{0} \mapsto \bigg(-\frac{1}{4}Z_0X_1Z_2X_3 - \frac{1}{4}Z_0Y_1Z_2Y_3 - \frac{1}{4}Z_0Z_1 \\ &+ \frac{1}{4}Z_0Z_2Z_3 - \frac{1}{4}X_1Z_2X_3- \frac{1}{4}Y_1Z_2Y_3 - \frac{1}{4}Z_1 + \frac{1}{4}Z_2Z_3  \bigg),
\end{split}
\end{equation}
\begin{equation} \label{eqn:p_sand_qham2_00}
\begin{split}
    &\hat{f}_{0} \hat{H}^2 \hat{f}^{\dagger}_{0} \mapsto \bigg(\frac{3}{4}I + \frac{3}{4}Z_0 -\frac{1}{4}Z_0Z_1Z_2Z_3 - \frac{1}{4}Z_0Z_1Z_3 \\ &+ \frac{1}{4}Z_0Z_2 - \frac{1}{4}Z_1Z_2Z_3 - \frac{1}{4}Z_1Z_3 + \frac{1}{4}Z_2 \bigg),
\end{split}
\end{equation}
\begin{equation} \label{eqn:p_sand_qham3_00}
\begin{split}
    &\hat{f}_{0} \hat{H}^3 \hat{f}^{\dagger}_{0} \mapsto \bigg(-\frac{3}{4}Z_0X_1Z_2X_3 - \frac{1}{2}Z_0X_1X_3 - \frac{3}{4}Z_0Y_1Z_2Y_3 \\
    &- \frac{1}{2}Z_0Y_1Y_3 - \frac{1}{2}Z_0Z_1 - \frac{1}{4}Z_0Z_1Z_2 + \frac{1}{2}Z_0Z_2Z_3 + \frac{1}{4}Z_0Z_3 \\ &- \frac{3}{4}X_1Z_2X_3 - \frac{1}{2}X_1X_3 - \frac{3}{4}Y_1Z_2Y_3 - \frac{1}{2}Y_1Y_3 - \frac{1}{2}Z_1 \\ &- \frac{1}{4}Z_1Z_2 + \frac{1}{2}Z_2Z_3 + \frac{1}{4}Z_3 \bigg)
\end{split}
\end{equation}
\noindent At variance to strategy \textit{i)}, in strategy \textit{ii)} the measurable Pauli strings contributing to each GF element change with matrix element index. This is exemplified for 2, 3, and 4 Hubbard sites (4, 6, and 8 qubits, respectively) in Fig. \ref{fig:npaulis_v_power}, in which the number of Pauli strings in $\hat{f}_{0} \hat{H}^n \hat{f}^{\dagger}_{0}$, and in $(\hat{f}_{0} + \hat{f}_{2}) \hat{H}^n (\hat{f}^{\dagger}_{0} + \hat{f}^{\dagger}_{2})$ (contributing to the off-diagonal element $g_{0,2}^{\text{(p)}}$), are plotted as a function of $n$.

Interestingly, the total number of individual Pauli terms does not necessarily increase with the Hamiltonian power. This effect manifests in two ways: \textit{1)} The concatenation of Pauli strings that occurs when taking the power can result in the collapse of products into smaller equivalent strings; this is evident for smaller numbers of qubits where the number of Pauli strings can oscillate for even and odd powers of $\hat{H}$, however the importance of this decreases rapidly as the system size grows (see Fig. \ref{fig:npaulis_v_power}). \textit{2)} In general, the maximum number of Pauli strings for $N$ qubits is $4^N$, however the Hamiltonian will have less terms than this (depending on the interactions of the model) for a given $N$ which will affect the number of individual strings that result from $n$ powers of $\hat{H}$; our results show that the number of Pauli strings for the Hubbard Hamiltonian saturates for values of $n$ large enough for the Lanczos procedure to sufficiently cover the Hilbert (sub)space of the model (in general of lower dimensionality of the full Hilbert space of $N$ qubits due to symmetry), which occurs at large enough values of the Lanczos index $l$ defined in section \ref{subsec:qgf}. 

The resulting manageable number of Pauli terms of small models, in addition to the optimized qubit excitation ansatz circuit for the ground state presented in section \ref{subsec:hubbard} (in this strategy, the circuits corresponding to the \textit{bra} and \textit{ket} of $\langle \hat{H}^{n, (\text{p/h})}_{ii} \rangle$ do not change with GF matrix index, unlike strategy \textit{i)}), facilitate the quantum calculation of the full GF matrix of the Hubbard dimer on Quantinuum's H1 ion-trap machine, the results of which are presented in the next section. 

\onecolumngrid

\twocolumngrid

\subsection{Green's Functions Calculated from Hardware Experiments} \label{hardware_results}

\subsubsection{Hubbard Model} \label{hardware_results_hub}

\begin{figure}[H]
\centering
\includegraphics[width=10cm]{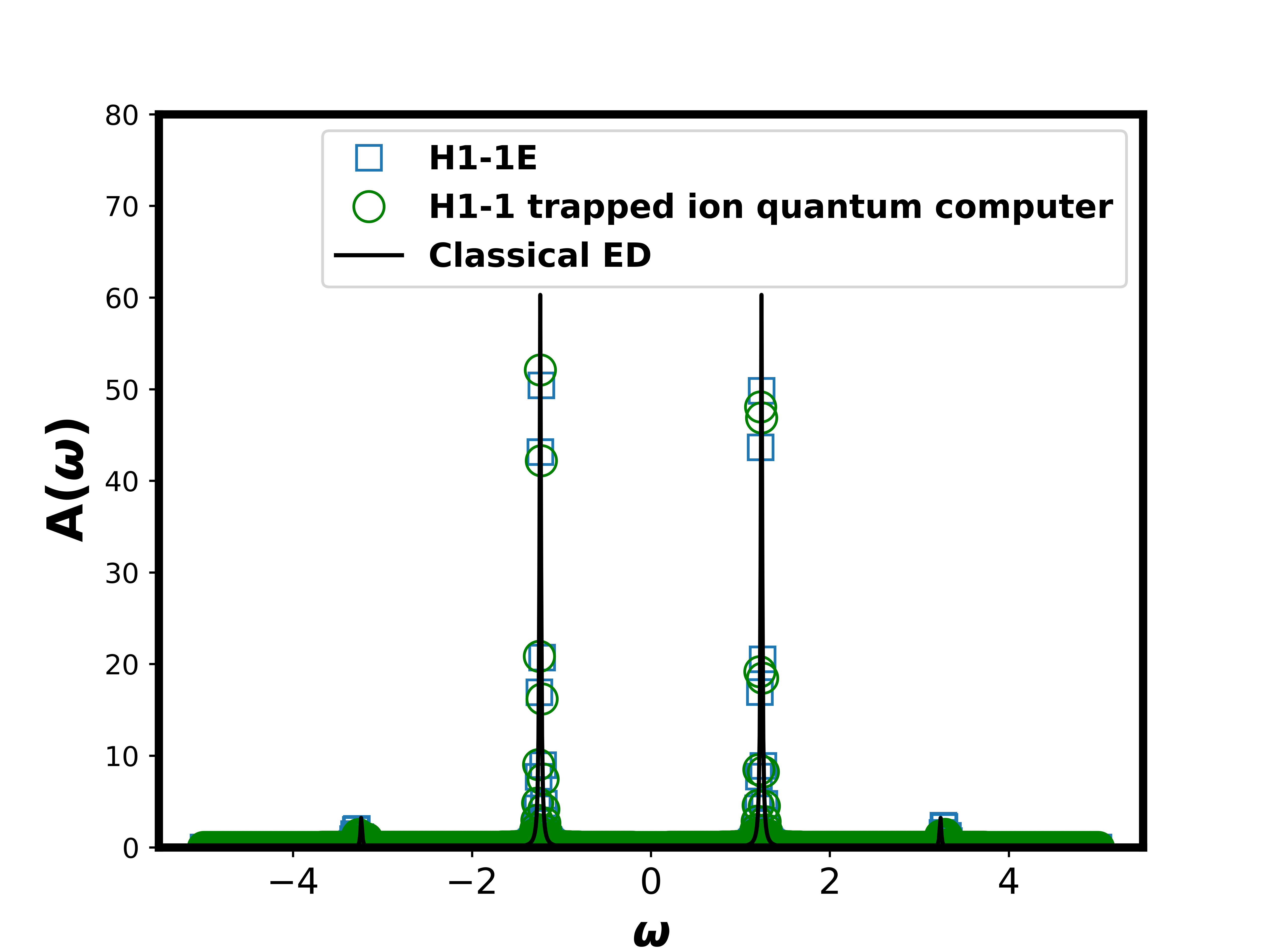}
    \caption{Quantum computed GF obtained from the H1-1 trapped ion quantum computer. Solid black line shows the result from classical ED. Dimension of spectral function (defined in Eq. \ref{eqn:spectral_fn}) is number of states per unit energy normalised by $\pi$.}
    \label{fig:gf_dimer_hardware}
\end{figure}

In Fig. \ref{fig:gf_dimer_hardware}, the spectral function of the Hubbard dimer is obtained from the H1-1 trapped ion quantum computer. In addition, results from the Quantinuum H1-1 emulator (H1-1E) (with its noise model calibrated to the noise profile of H1-1 \cite{h11e}) are also shown. Both the particle and hole GF matrices contain 16 elements, which immediately reduces to 10 elements each due to symmetry about the diagonal \cite{avella13}. This would naively result in 20 measurable circuits for the full GF (10 for both particle and hole GFs): one circuit per matrix element for particle and hole matrices (despite particle-hole symmetry, we explicitly calculate both the particle and hole matrices to more comprehensively test the performance on hardware). Due to certain elements within the particle and hole GF matrices being identical for the Hubbard dimer in this regime, the final total number of measurable circuits for the Hubbard dimer is 14. The total number of (2-qubit ZZ) gates per circuit ranges from 23 to 27 (7 to 9), with depth ranging from 13 to 18. Measurements of each circuit are performed with 8192 shots. Excellent agreement is observed in the spectral peak positions, indicating an accurate description of the excitation energies involved in the particle/hole transitions, when compared to the ideal classical simulation. Previous investigations have demonstrated the robustness of this method to quantum measurement noise and device errors for infimum estimates of the ground state energy \cite{hollenberg96, vallury20, vallury23}, while other recent studies have also shown the robustness to noise of quantum subspace algorithms in general for accessing low lying eigenvalues \cite{epperly22, kirby23}. Our results are consistent with these previous works. The quality of the results is also indicative of low device errors on the H1-1 trapped ion quantum computer, since measurements of Hamiltonian moment expectations were performed without error mitigation. We also note the excellent agreement between H1-1 and its emulator. 

\begin{figure}[H]
\centering
\includegraphics[width=10cm]{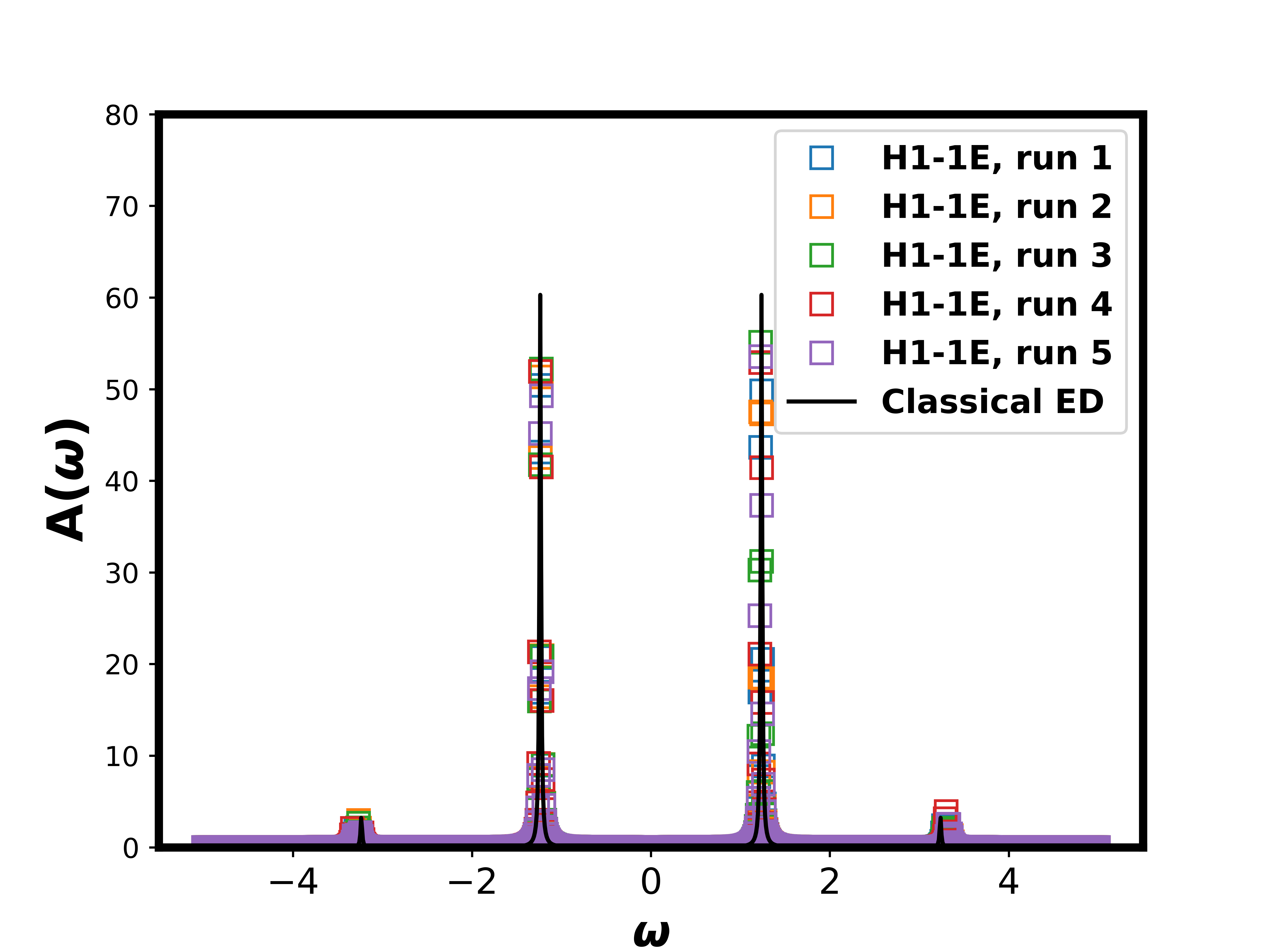}
    \caption{Quantum computed GF obtained from the classical emulator of the H1-1 trapped ion quantum computer, with a noise model calibrated to current hardware data. Solid black line shows the result from classical ED. Dimension of spectral function (defined in Eq. \ref{eqn:spectral_fn}) is number of states per unit energy normalised by $\pi$.}
    \label{fig:gf_dimer_emulator}
\end{figure}

The heights of the two central peaks in Fig. \ref{fig:gf_dimer_hardware} are underestimated with respect to ED by an average (between positive and negative frequency peaks) of approximately $17\%$ for H1-1. Repeated runs on the emulator H1-1E shows this is roughly consistent (Fig. \ref{fig:gf_dimer_emulator}), yielding a mean underestimation of 15$\%$ over the 5 runs for the two central peaks. With regard to the peak positions along the real frequency axis, the hardware results from H1-1 (Fig. \ref{fig:gf_dimer_hardware}) yield an average error relative to ED in the low lying peak frequency (at absolute frequency $|\omega| = 1.236$ in the exact GF) of 0.5\% (averaged over positive and negative frequency peaks), whereas the higher lying peaks (at $|\omega| = 3.238$ in the exact GF) exhibit a frequency-averaged error of 1.3\%. The repeated runs of the emulator (Fig. \ref{fig:gf_dimer_emulator}) yield a roughly similar average error of 0.6\% in the low lying peak positions, and an average higher lying peak position error of 1.5\%. 

We note that the H1-1 device currently exhibits typical one-qubit and two-qubit gate error rates of $4 \times 10^{-5}$ and $2 \times 10^{-3}$, respectively, and a typical state preparation and measurement error rate of $3 \times 10^{-3}$ \cite{h11e}. With this in mind, and to further investigate the impact of noise on our results, we performed bootstrapping of H1-1 hardware measurements to generate ensembles of GFs resampled from the original measured results. Statistical analysis of these bootstrapped samples provide effective error bars on the position and weight of peaks of the spectral function, thereby demonstrating the impact of random variations (e.g. due to device error) on the final output GF along the real frequency axis.

\begin{figure}[H]
\centering
    \begin{minipage}{7cm}
    \begin{overpic}[width=7cm]{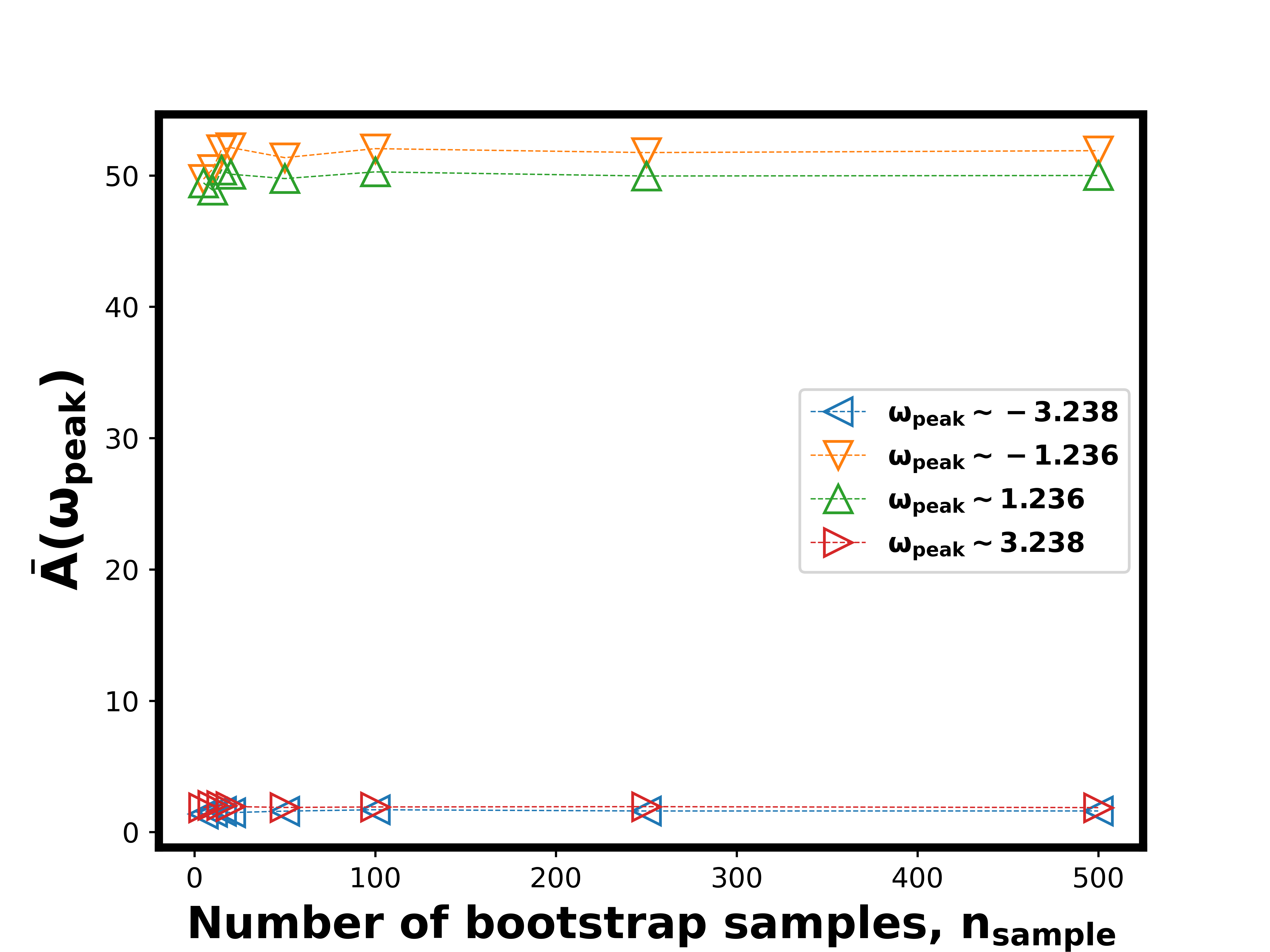}
    \end{overpic}
    \begin{overpic}[width=7cm]{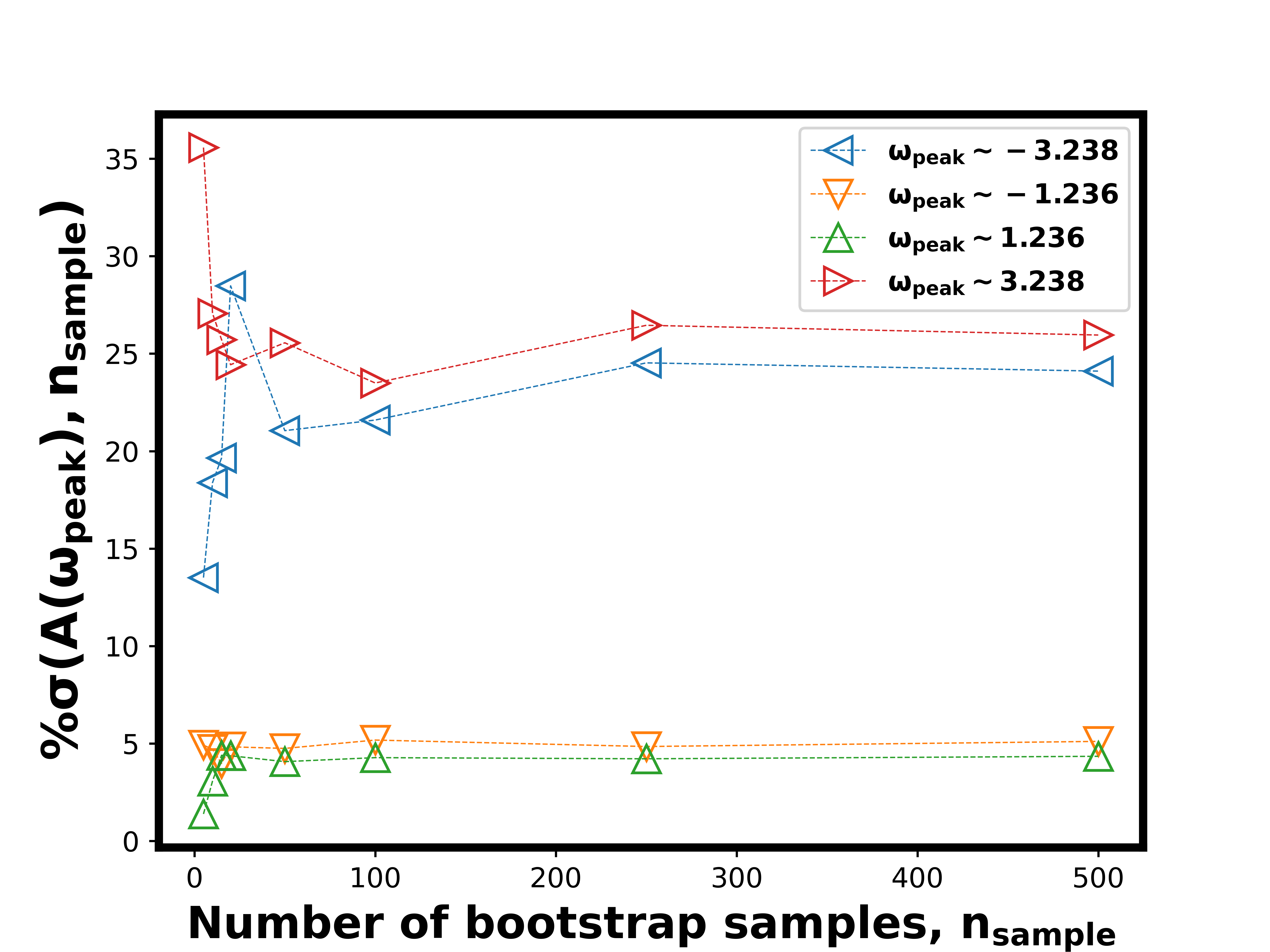}
    \end{overpic}
    \end{minipage}\begin{minipage}{7cm}
    \begin{overpic}[width=7cm]{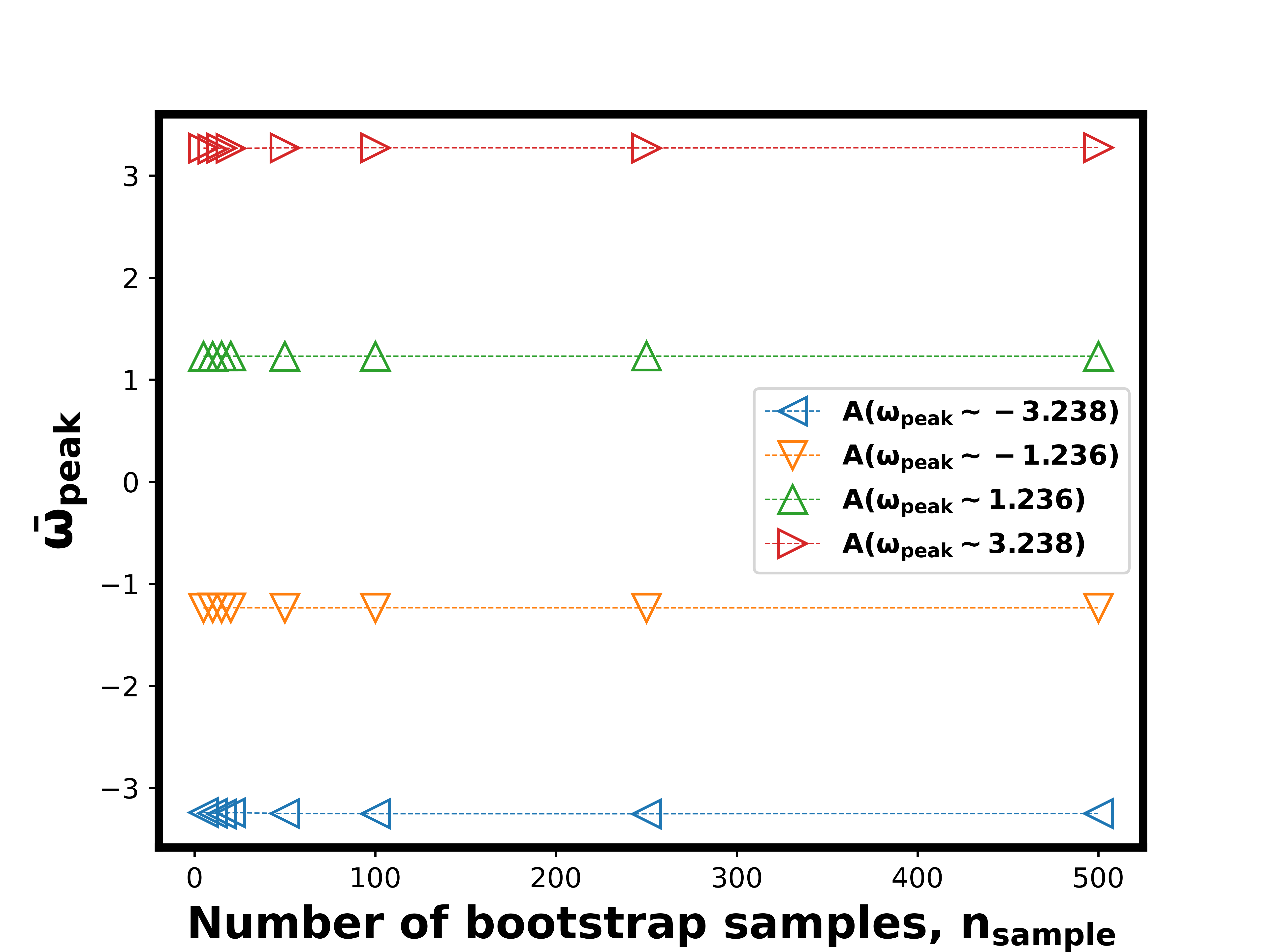}
    \end{overpic}
    \begin{overpic}[width=7cm]{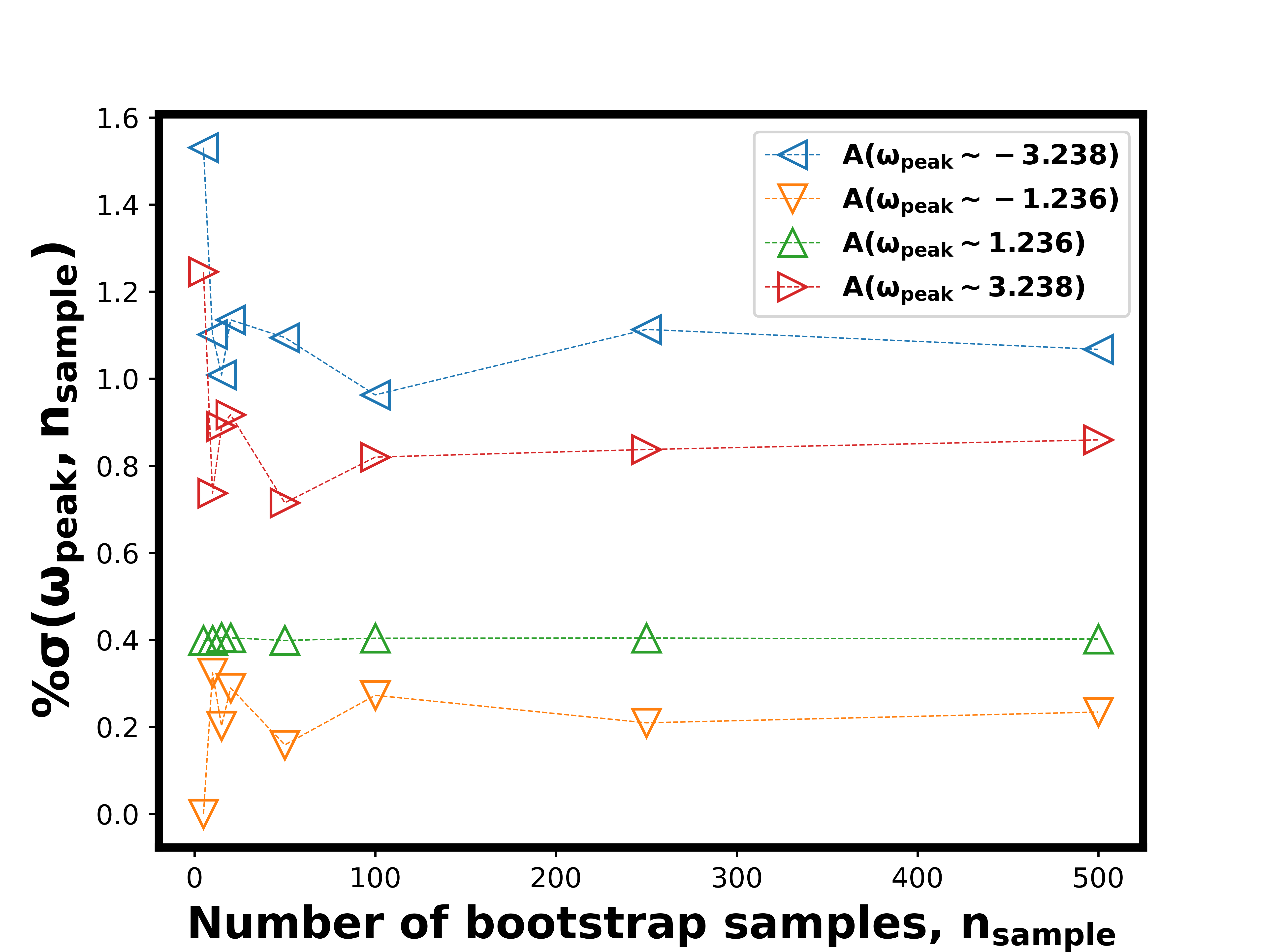}
    \end{overpic}
    \end{minipage}
    \caption{Analysis of bootstrapped GFs derived from measurements obtained on the H1-1 trapped ion quantum computer, showing the variation of results with respect to the number of bootstrap samples. Top left panel shows spectral peak heights averaged over each number of samples ($\bar{A}(\omega_{\text{peak}})$), and bottom left panel shows the peak height standard deviations. Top right panel shows transition frequency (or energy) positions of the 4 peaks of the Hubbard dimer spectral function averaged over the number of samples ($\bar{\omega_{\text{peak}})}$), and bottom right panel shows the peak position standard deviations. All standard deviations $\sigma$ here are expressed as a percentage normalised to the ensemble average for a given number of samples $\%\sigma(X, n_{\text{sample}}) = \sigma(X, n_{\text{sample}}) / \bar{X}(n_{\text{sample}}) \times 100$, where $X=A(\omega_{\text{peak}}), \omega_{\text{peak}}$. Dimension of spectral function (defined in Eq. \ref{eqn:spectral_fn}) is number of states per unit energy normalised by $\pi$.}
    \label{fig:bootstrap_conv}
\end{figure}

\begin{figure}[H]
\centering
    \begin{minipage}{8.5cm}
    \begin{overpic}[width=8.5cm]{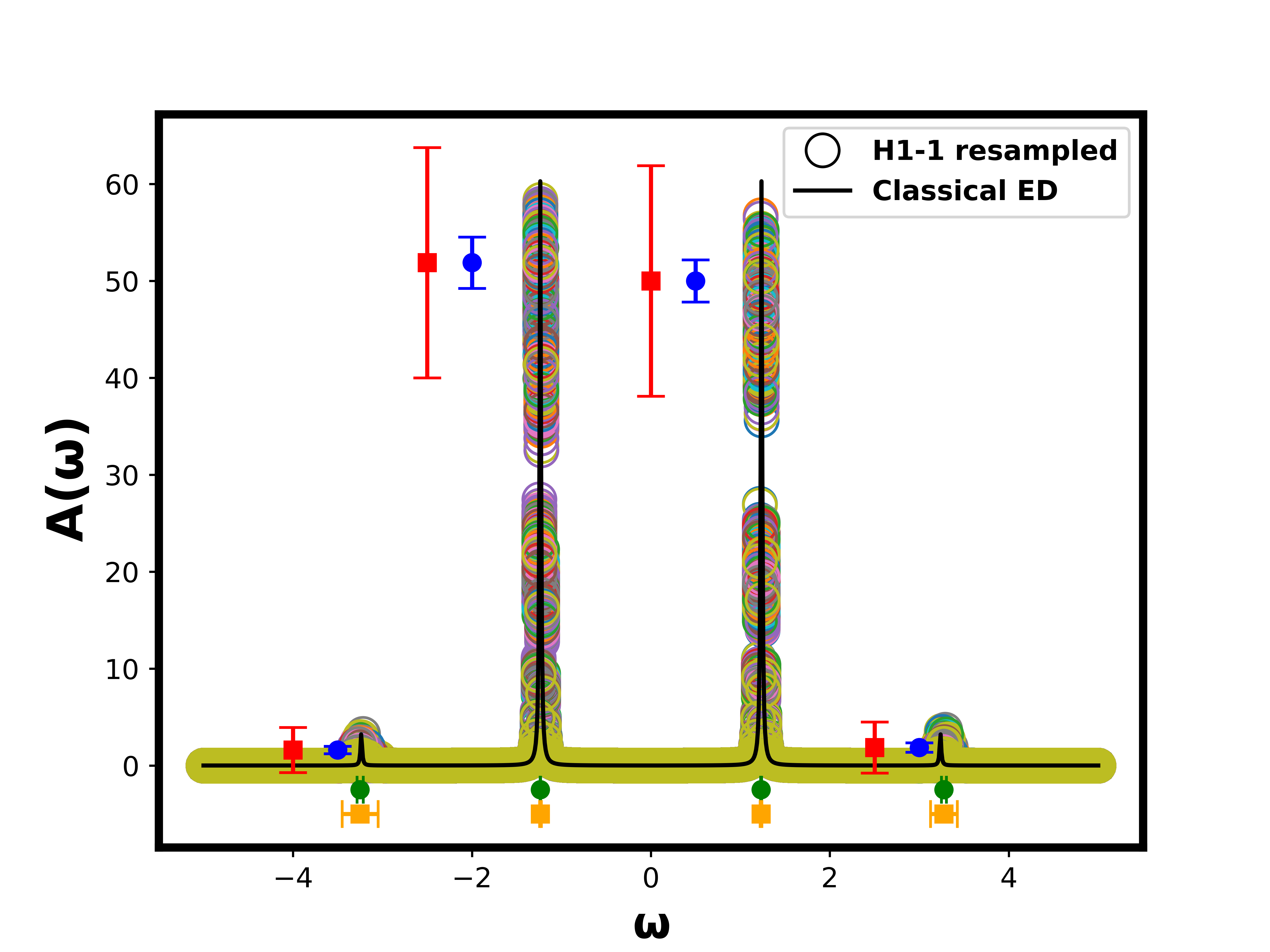}
    \end{overpic}
    \end{minipage}\begin{minipage}{8.5cm}
    \begin{overpic}[width=8.5cm]{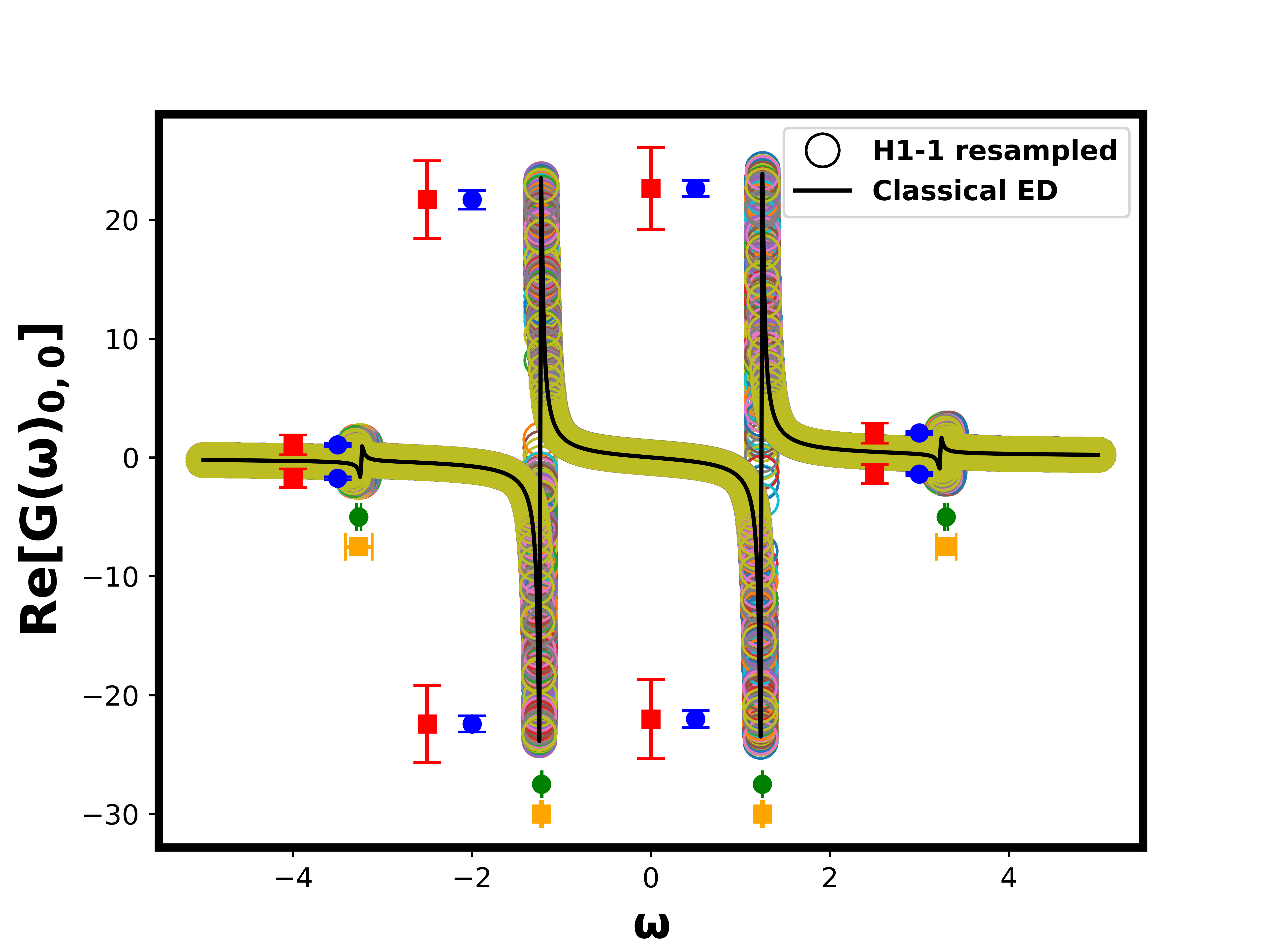}
    \end{overpic}
    \end{minipage}
    \caption{Bootstrapped spectral functions derived from quantum computed GF corresponding to 500 samples, with original measurements obtained from the H1-1 trapped ion quantum computer. On the left panel, blue (red) error bars denote the standard deviations (minimum-maximum spreads) of $A\omega_\text{peak}$, centred on the mean, while green (orange) error bars denote the standard deviations (minimum-maximum spreads) of $\omega_\text{peak}$ values. All error bars are centred on the ensemble mean. The right panel shows the corresponding information for the real part of the $G_{0,0}$ matrix element (in units of $\frac{1}{t}$ where we use $t = 1$). Solid black line shows A($\omega$) from classical ED. In this figure, all standard deviations are in units of the corresponding quantity, i.e. blue bars represent units of $A(\omega)$ while green bars represent units of $\omega$. Dimension of spectral function (defined in Eq. \ref{eqn:spectral_fn}) is number of states per unit energy normalised by $\pi$.}
    \label{fig:bootstrap_500}
\end{figure}

To perform such an analysis of the hardware-calculated GF, the local spectral maxima $A(\omega_\text{peak})$ were sampled at the transition frequencies ($\omega_\text{peak}$, of which there are 4 for the Hubbard dimer, located at frequencies $\omega_\text{peak} = -3.238, -1.236, 1.236, 3.238$ in the exact GF) for each bootstrapped GF. Each of these local maxima were then averaged over the ensemble, with their spread and standard deviations also obtained. A similar procedure was carried out for the positions of spectral peaks along the real frequency axis (the $\omega$ value for which $A(\omega)$ has a peak near $\omega = \omega_\text{peak}$), sampling the frequencies at which the spectrum has local maxima for each of the bootstrapped GFs. For the real part of the GF matrix element (right panel of Fig. \ref{fig:bootstrap_500}), the value of the positive and negative extrema centred at the $\omega_\text{peak}$ values were sampled over the ensemble. Fig. \ref{fig:bootstrap_conv} shows the variation of spectral peak positions and heights with respect to the number of samples, showing reasonable convergence of the statistics at 500 bootstrapped samples. The ensemble of 500 bootstrapped GFs is plotted in Fig. \ref{fig:bootstrap_500}. We observe that the low lying transition energies are well reproduced with respect to the exact result, with very small variation (standard deviation approximately 0.2\% - 0.4\% normalised to the ensemble average at 500 samples as shown in Fig. \ref{fig:bootstrap_conv}) between bootstrapped samples. For the higher lying spectral peaks ($\omega_\text{peak} \approx \pm 3.238$), standard deviations of the transition energy are larger (approximately 0.8\% - 1\% at 500 samples in Fig. \ref{fig:bootstrap_conv}) which is unsurprising given the diminished accuracy of higher eigenspectra for truncated Krylov spaces, and is consistent with the results shown in Fig. \ref{fig:imgf_dimer_coeffs_ovs}.

In section \ref{results_lanc_vec_prep} we discussed how noise in the measured moments translate to errors in the normalisation of Lanczos vectors, which in turn translates to errors in the peak heights. Here we see that these errors exceed (in proportion) errors in the peak position frequencies (this is reflected in the larger average error relative to ED for the peak heights, in addition to the larger spread of spectral weights, compared to the transition frequencies). Hence variations due to device error seem to have a larger impact on spectral weights compared to errors in the values of transition frequencies. However spectral weights of low and high lying peaks are qualitatively in proportion with respect to exact diagonalisation: the dominance of low lying peaks over higher lying excitations is maintained even in the presence of quantum noise. This combined with the relatively accurate peak positions empirically demonstrates that our quantum Lanczos approach can reproduce important frequency dependent features of the GF in the presence of noise on the H1-1 quantum computer.

Whether the observed errors are large enough to prevent application of this quantum GF approach to more elaborate simulations, such as quantum embedding within DMFT, remains an interesting question. In the next section, we address this issue and demonstrate the usefulness of this technique for DMFT.

\onecolumngrid


\twocolumngrid

\subsubsection{DMFT} \label{hardware_results_dmft}

\begin{figure}[ht]
\centering
\includegraphics[width=7.0cm]{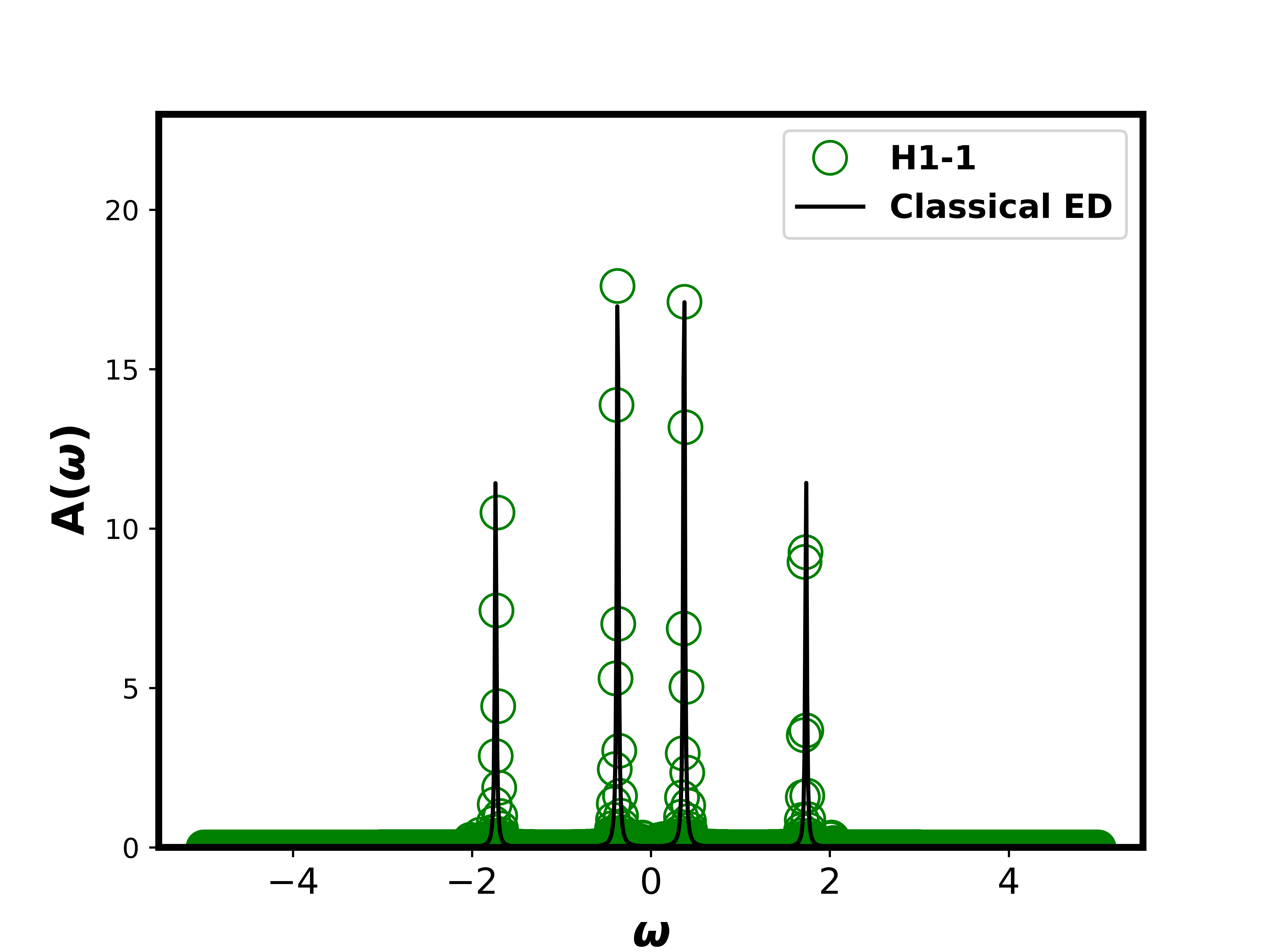}
    \caption{Quantum computed impurity GF, in which the H1-1 trapped ion quantum computer is used to compute the impurity GF at the final DMFT iteration, following classical impurity GF computations for the previous iterations. Solid black line shows the final impurity GF obtained from classical ED. Dimension of spectral function (defined in Eq. \ref{eqn:spectral_fn}) is number of states per unit energy normalised by $\pi$.}
    \label{fig:gf_imp_dmft_final}
\end{figure}

The quantum computed moments approach is also used to obtain the GF of the impurity site in a DMFT algorithm. Running the DMFT procedure with the impurity GF (corresponding to 1 bath site) obtained from quantum computed moments at the ideal, noiseless statevector level reproduces the ED result (in which the impurity GF is calculated using classical exact diagonalisation at each DMFT iteration) exactly, corresponding to the black line in Fig. \ref{fig:gf_imp_dmft_final}. Also in Fig. \ref{fig:gf_imp_dmft_final} the impurity GF corresponding to 1 bath site is quantum computed at the final DMFT step in which the Hamiltonian corresponds to converged bath parameters (where convergence was obtained using ideal noiseless simulations of the quantum GF at each DMFT iteration), using measurements of expectations of Hamiltonian moments performed on the H1-1 trapped ion quantum computer (green circles correspond to real hardware \cite{h11e} in Fig. \ref{fig:gf_imp_dmft_final}). In this case 2 Lanczos roots were used, which reproduces ED (consistent with the Hubbard dimer). For 1 bath site, 14 circuits were measured (corresponding to 14 impurity+bath GF matrix elements, similar to the Hubbard dimer), ranging from 58 (22) to 67 (25) gates (2-qubit gates), with depths ranging from 44 to 51. The larger number of gates compared to the Hubbard dimer is a result of the reduced symmetry of the impurity-bath Hamiltonian for DMFT. Each circuit is measured with 8192 shots. 

Excellent agreement is observed between the hardware, emulator, and noiseless simulations, and we comment here on the observed robustness to error. At 8192 shots per circuit the absolute error in the expectation values of Hamiltonian moments due to shot noise (i.e. only from measurement sampling) is approximately 0.0062 for the first and second power, and 0.0127 for the third power (2 Lanczos roots requires up to the third power of the Hamiltonian). However, the GF matrix elements involve continued fractions of polynomials of Pauli strings which can lead to partial error cancellation when grouping Pauli terms into commuting sets: once the Pauli commuting sets are found, the measurement sampling algorithm \cite{inquanto_docs} can associate certain Pauli operator expectations with both $\alpha_l$ and $\beta_l$, hence errors in those Pauli expectations can cancel when evaluating the $\alpha_l$ and $\beta_l$ terms. We note that this is an interesting effect related to the combination of Krylov basis construction using Pauli-encoded Hamiltonian moments along with elements of the resulting tridiagonal matrix to build a Green’s functions via continued fractions (the latter being the novelty of this work).

\begin{figure}[ht]
\centering
\includegraphics[width=7cm]{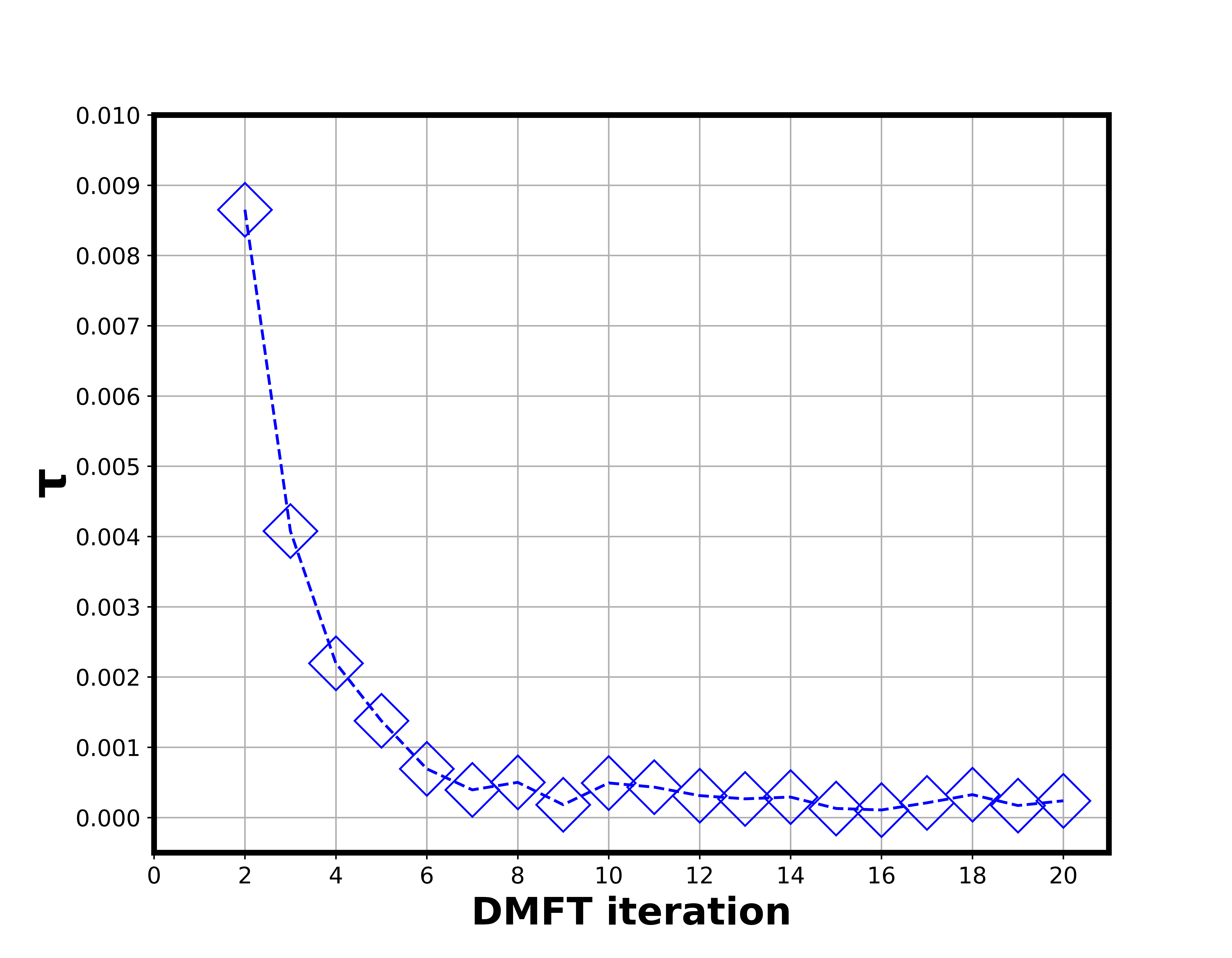}
    \caption{Error in $\mathbf{\Delta}^{\text{sc}}$, as defined in Eq. \ref{eqn:dmft_tau} (which sets the threshold for DMFT convergence) at each DMFT iteration, in which the H1-1E emulator is used to quantum compute the impurity GF at each iteration.}
    \label{fig:tau_vs_iter}
\end{figure}

\begin{figure}[ht]
\centering
\includegraphics[width=7.0cm]{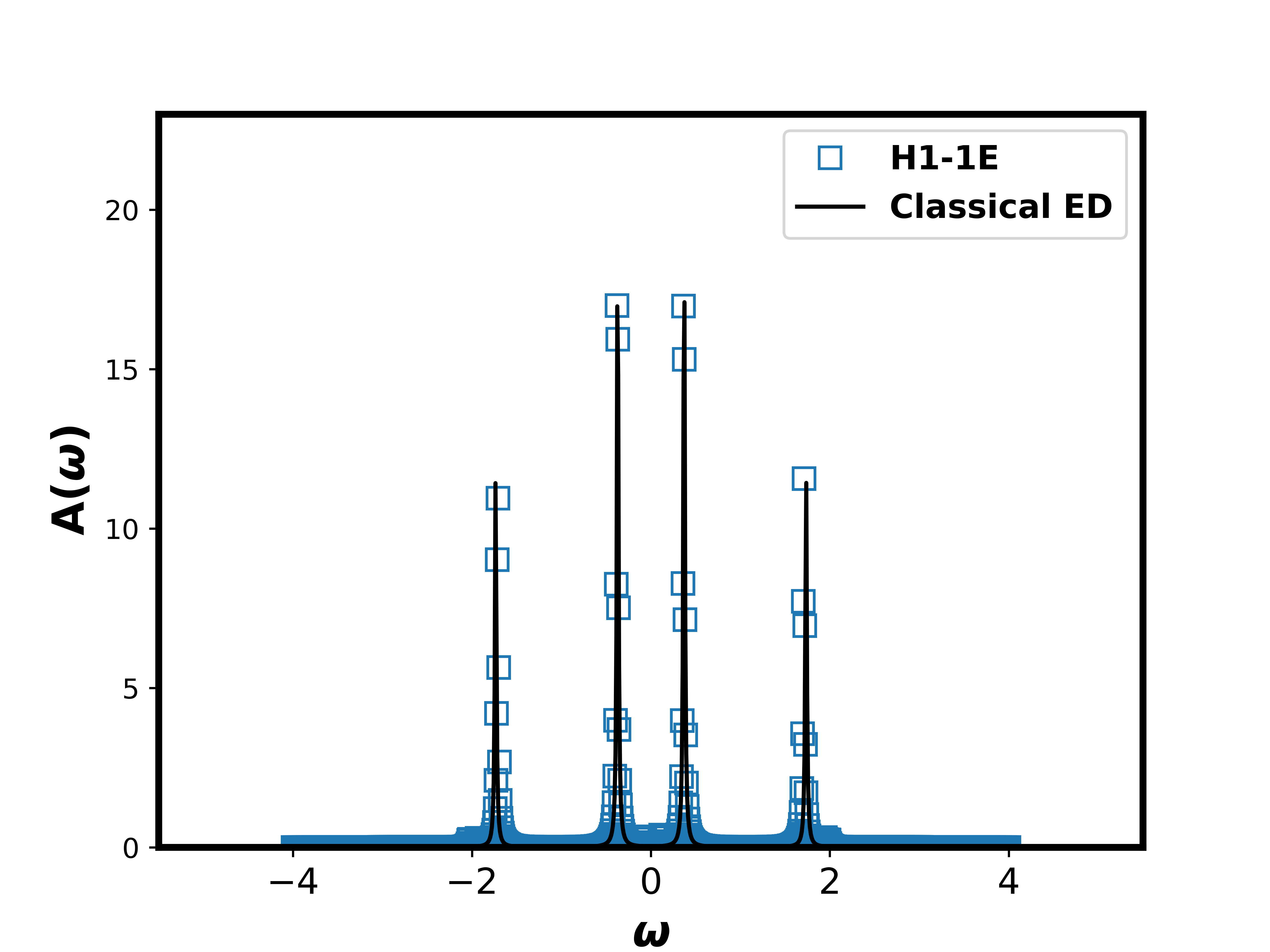}
    \caption{Quantum computed impurity GF after 20 DMFT iterations, in which the emulator of the H1-1 trapped ion quantum computer is used to compute the impurity GF at each iteration. Black line shows the final impurity GF obtained from classical ED. Dimension of spectral function (defined in Eq. \ref{eqn:spectral_fn}) is number of states per unit energy normalised by $\pi$.}
    \label{fig:gf_imp_dmft}
\end{figure}

\begin{figure}[H]
    \centering
    \begin{minipage}{5.5cm}
    \begin{overpic}[width=5.5cm]{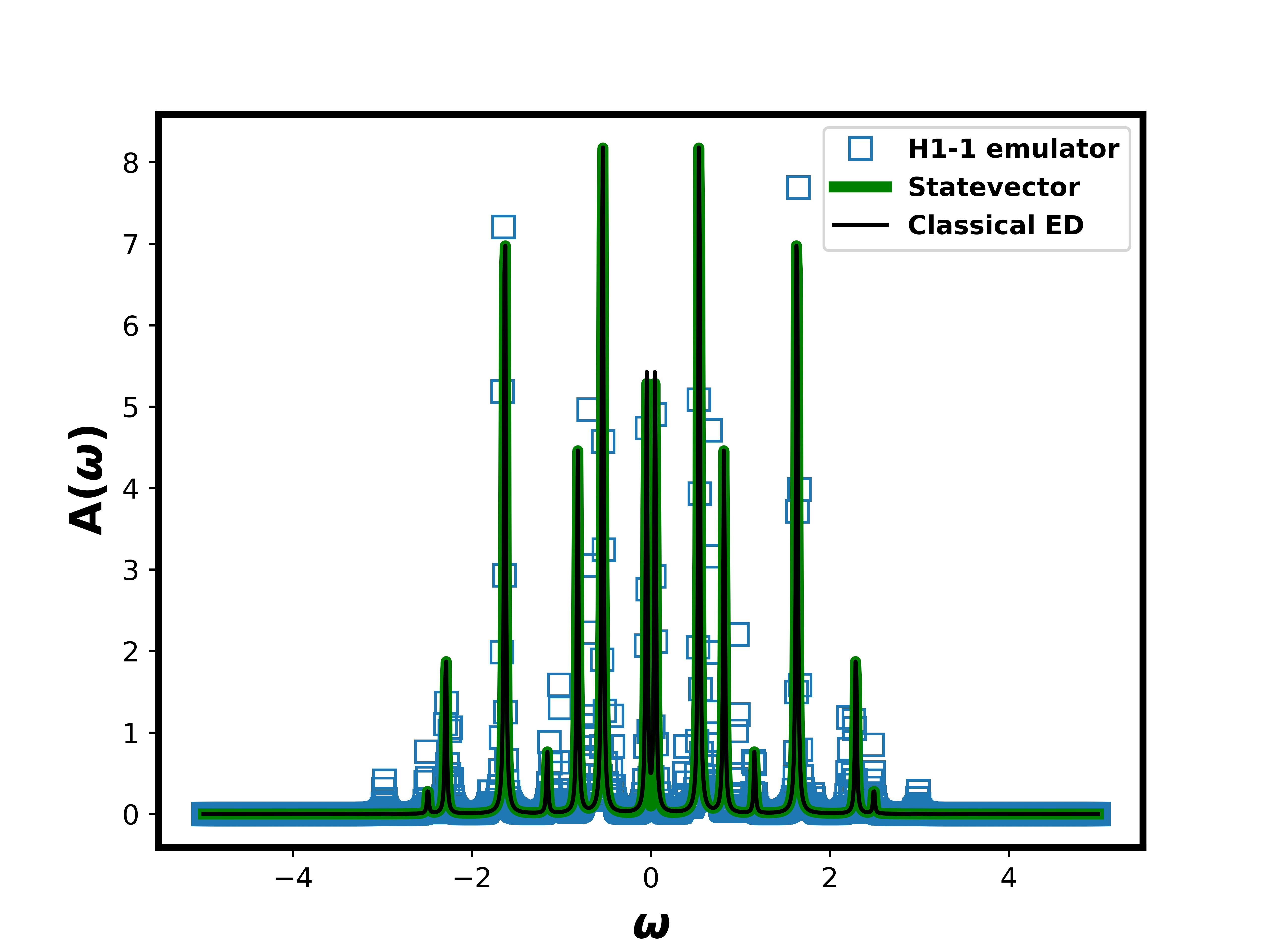}
    \end{overpic}
    \begin{overpic}[width=5.5cm]{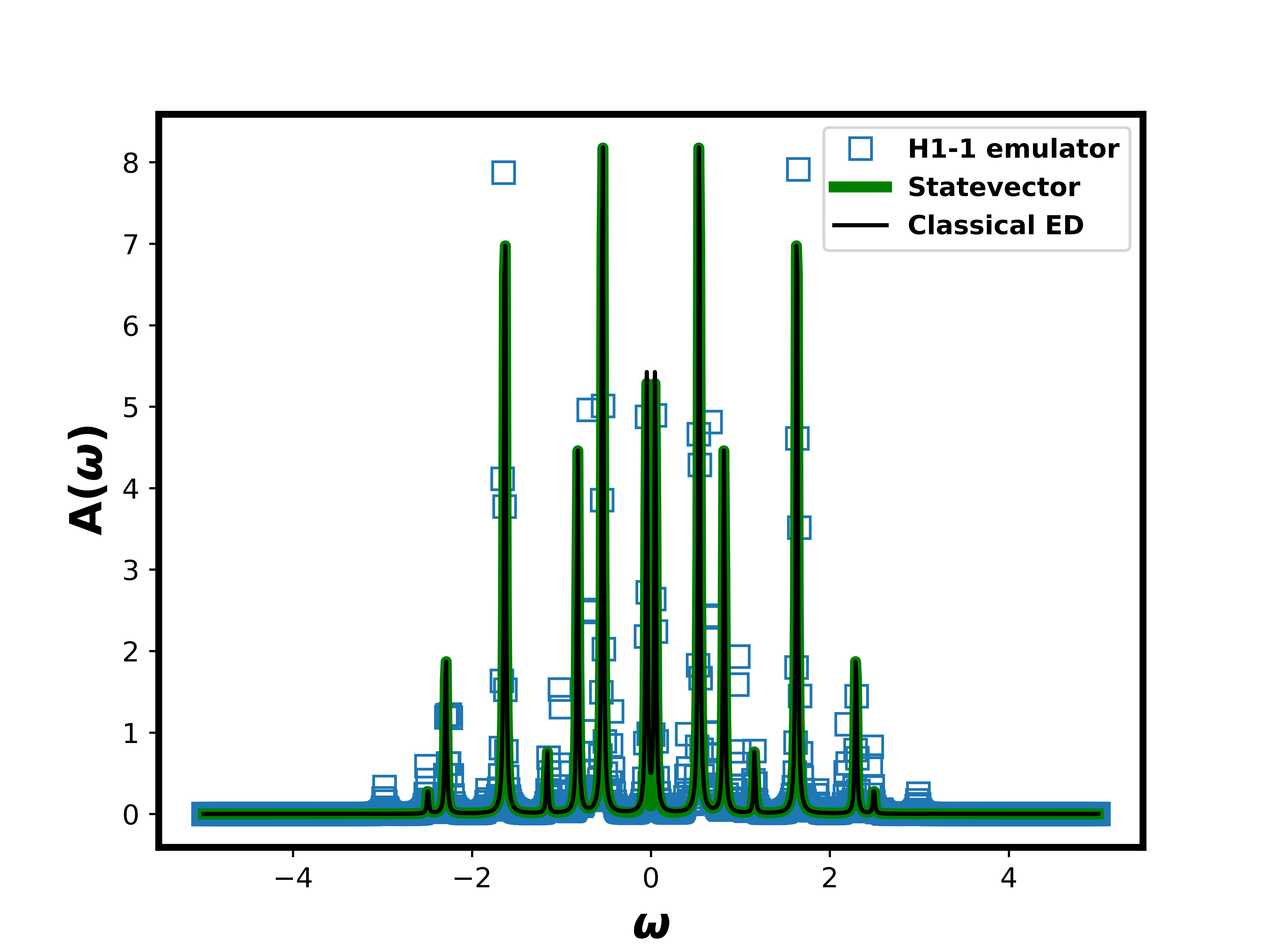}
    \end{overpic}
    \begin{overpic}[width=5.5cm]{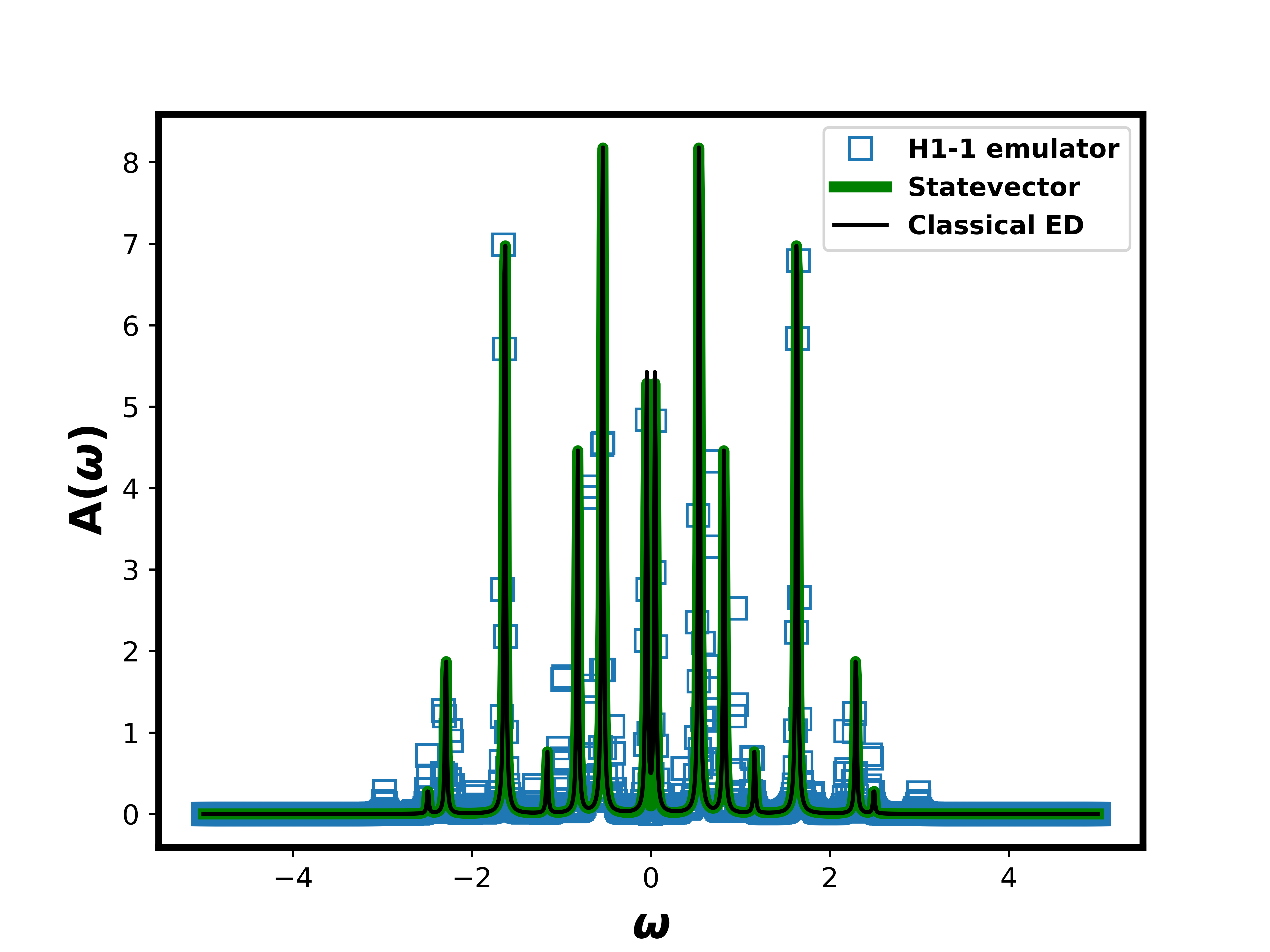}
    \end{overpic}
    \end{minipage}
    \caption{Quantum computed impurity GF of DMFT with 1 impurity and 2 bath sites, in which the emulator of the H1-1 trapped ion quantum computer is used to compute the impurity GF at the final DMFT iteration, following statevector computations for the previous iterations. Solid green line shows the converged impurity GF from ideal noiseless quantum computed moments. Solid black line shows the converged impurity GF when classical ED is used at each DMFT iteration. The emulator H1-1E is applied in 3 separate calculations of the quantum computed moments impurity GF (using bath parameters obtained from the noiseless run), to simulate 3 separate hardware runs and assess variation of results. Dimension of spectral function (defined in Eq. \ref{eqn:spectral_fn}) is number of states per unit energy normalised by $\pi$.}
    \label{fig:gf_imp_dmft_2bath}
\end{figure}

\begin{figure}[H]
\centering
\includegraphics[width=6.3cm]{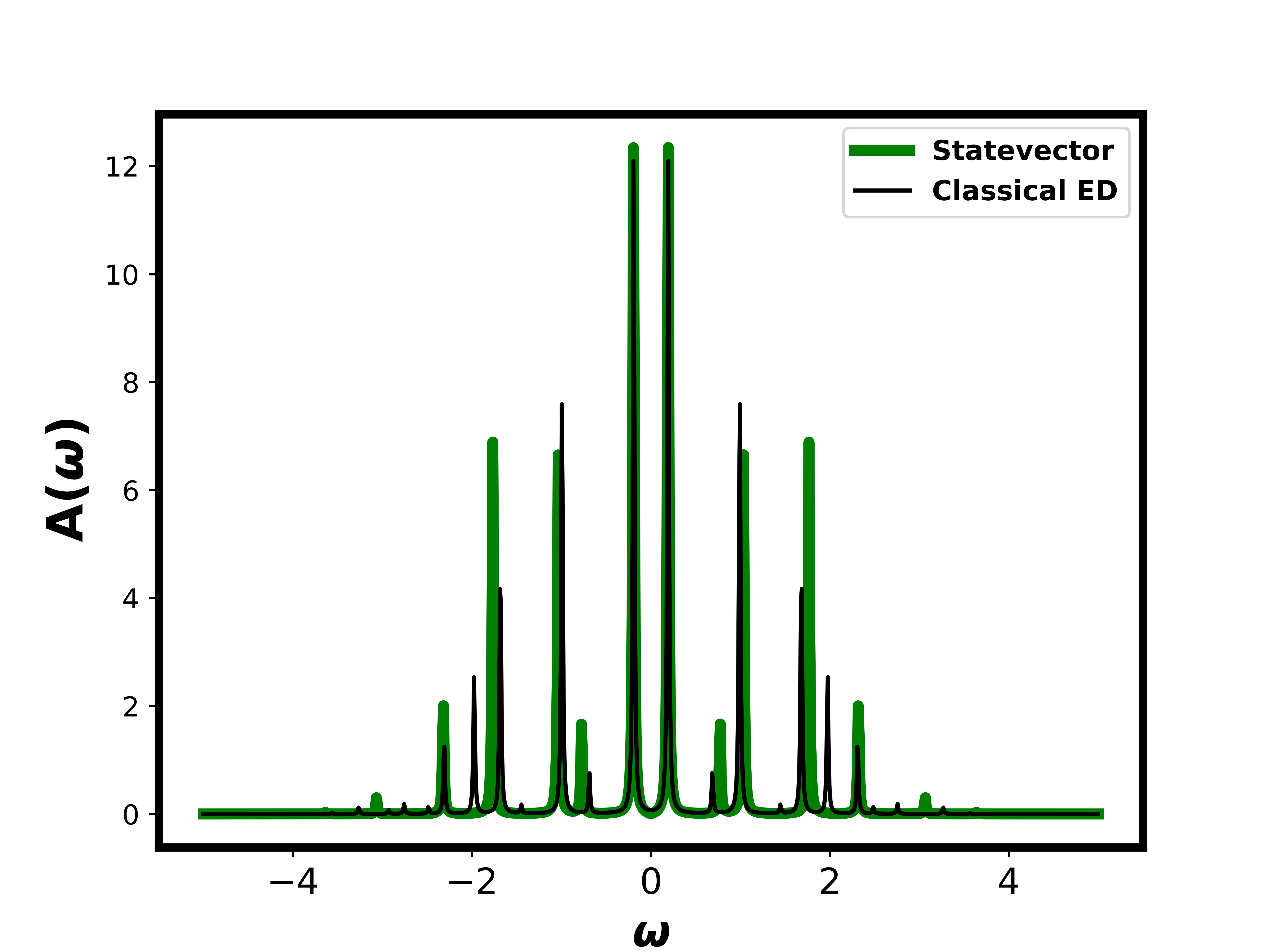}
    \caption{Quantum computed impurity GF for 3 bath sites, calculated to full convergence of the DMFT algorithm. Solid green line shows the converged impurity GF from ideal quantum computed moments. Solid black line shows the converged impurity GF when classical ED is used at each DMFT iteration. Dimension of spectral function (defined in Eq. \ref{eqn:spectral_fn}) is number of states per unit energy normalised by $\pi$.}
    \label{fig:gf_imp_dmft_3bath}
\end{figure}

To investigate the impact of noise on DMFT convergence of the impurity GF, the final impurity GF obtained from the quantum computed moments method was calculated using a simple noise model and compared to the ideal noiseless result. To this end, bath parameters were first obtained by converging the impurity GF at the ideal statevector level to $\tau = 0$. Using these bath parameters, the impurity GF was then re-computed at the circuit level using a simulated noiseless backend \cite{qiskit} with 8192 shots per circuit, and with a range of values of a noise parameter $\lambda$. Focusing on two-qubit gate errors, our simplified noise model introduces a depolarising channel to all two-qubit gates in the circuit, where each two-qubit gate error can be expressed by the error channel $E$ as

\begin{equation} \label{eq:error}
    E(\rho) = (1 - \lambda)\rho + \lambda\text{Tr}[\rho]\frac{I}{4}
\end{equation}

\noindent which maps a corresponding state $\rho$ to its linear combination with a maximally mixed state. This depolarising channel yields a corresponding error in the impurity GF. To quantify the latter, we re-use the threshold for DMFT convergence, $\tau$, defined in Eq. \ref{eqn:dmft_tau}. We calculate the additional error in the impurity problem due to depolarising noise as the value of $\tau$ resulting from comparing the ideal noiseless (statevector) converged $\mathbf{\Delta}^{\text{sc}}$ (related to the impurity GF by the Anderson model self-consistency condition, see Eq. \ref{eqn:deltasc}), $\mathbf{\Delta}^{\text{sc}}_{\text{ideal}}$, to its value obtained in the presence of depolarising error channel $E$ (with the same ground state and bath parameters), labelled as $\mathbf{\Delta}^{\text{sc}}_{\text{noise}}$.
\begin{equation} \label{eqn:dsc_error}
    \tau_{E} = \frac{1}{N_{\omega}} \sqrt{\sum_{k=1}^{N_{\omega}} | \mathbf{\Delta}^{\text{sc}}_{\text{noise}}(\text{i}\omega_k) - \mathbf{\Delta}^{\text{sc}}_{\text{ideal}}(\text{i}\omega_k) |_{\text{F}}^2} \ .
\end{equation}
\begin{figure}[ht]
\centering
\includegraphics[width=7cm]{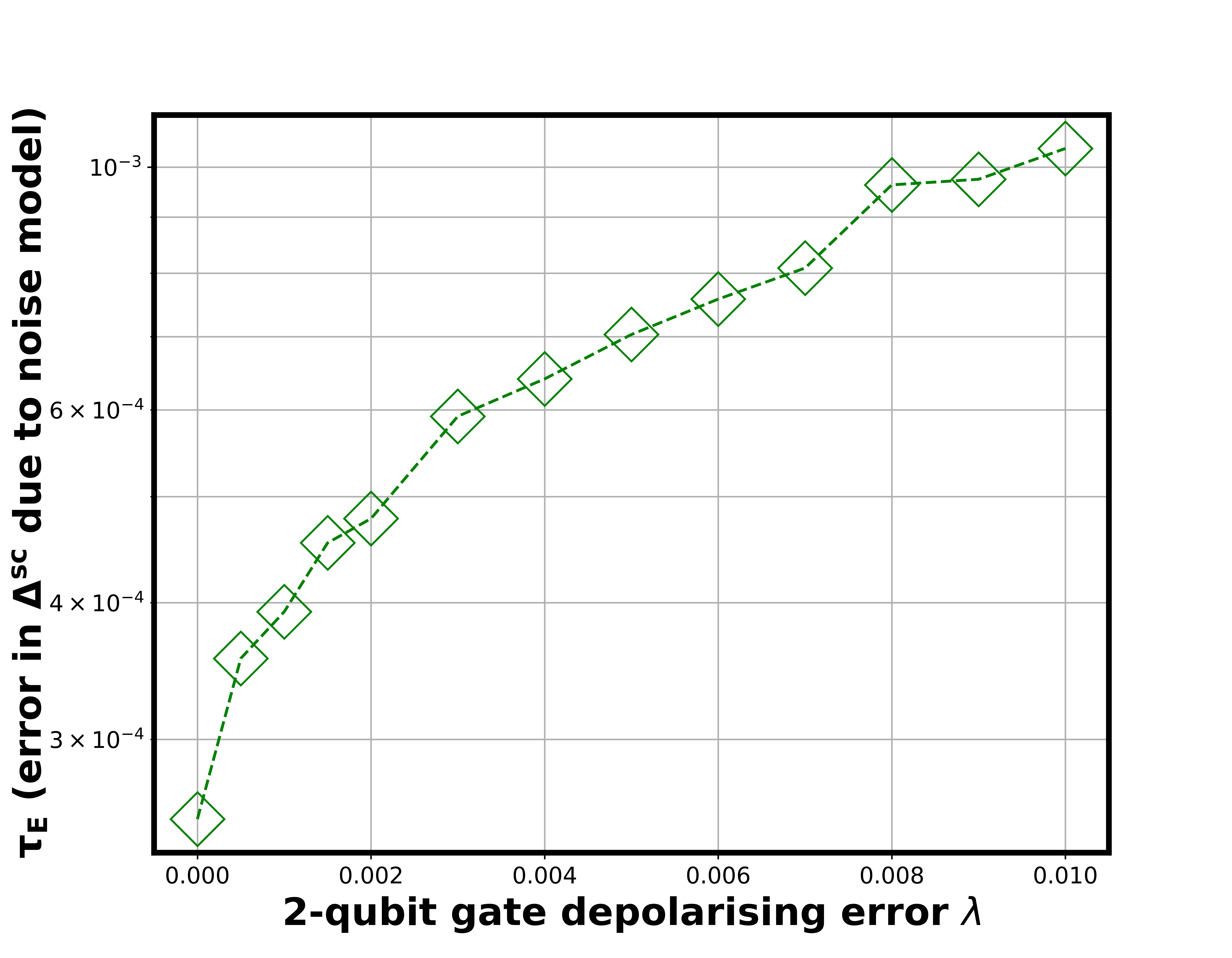}
    \caption{Error in $\mathbf{\Delta}^{\text{sc}}$, as defined in Eq. \ref{eqn:dsc_error}, versus two-qubit gate depolarising error. At $\lambda=0$ corresponding to zero noise, $\tau_E$ = 0.000254 which is close to the value of $\tau$ obtained from the H1-1 emulator after 20 DMFT iterations.}
    \label{fig:dsc_error}
\end{figure}
The error in $\mathbf{\Delta}^{\text{sc}}$ as a function of $\lambda$ is plotted in Fig. \ref{fig:dsc_error}. A non-zero $\tau_E$ at $\lambda=0$ is due to finite sampling of the measurement distributions. We observe that in this range of $\lambda$, $\tau_E$ increases monotonically with $\lambda$, and for small $\lambda$ ($\lesssim0.01$) reasonably small errors in the impurity GF matrix on the order of ~$10^{-3} - 10^{-4}$ are obtained. We also note that for this impurity GF errors of this magnitude do not change the number of GF poles and maintain the dominance of the two central peaks of the spectral function, as was observed in Fig. \ref{fig:gf_imp_dmft_final}. While not directly equivalent to the two-qubit gate infidelity of the H1-1 device ($2 \times 10^{-3}$ \cite{h11e}), we expect two-qubit error rates on H1-1 to correspond to $\lambda$ well below 0.01, which could translate to the ability to converge the DMFT procedure to correspondingly small errors in the impurity GF. As a step further, and to test this hypothesis, a hardware emulator in which the noise model is calibrated to H1-1 was also used to quantum compute the GF at each iteration of DMFT, for 20 iterations. The value of $\tau$ at each DMFT iteration is plotted in Fig. \ref{fig:tau_vs_iter}. Starting with random bath parameters, the initial error in $\mathbf{\Delta}^{\text{sc}}$ is $\tau = 0.00865$. After 20 iterations this error reduced to $\tau = 0.000237$, a value sufficiently low as to obtain a reasonably accurate solution to the impurity problem. We also notice that this value of $\tau$ is slightly less than the $\lambda=0$ value of $\tau_E$ shown in Fig. \ref{fig:dsc_error} (0.000254), indicating that the impurity GF error between DMFT iterations near convergence performed on the H1-1 emulator does not exceed the error (relative to the ideal noiseless DMFT result) introduced by measurement sampling errors (without noise), for this case. The resulting spectral function is  shown in Fig. \ref{fig:gf_imp_dmft}. Hence the error due to quantum noise from the H1-1 emulator was sufficiently low to run the DMFT algorithm applied to 1 impurity site and 1 bath site for multiple iterations towards convergence. 
We also compute the impurity GF corresponding to converged bath parameters for 2 bath sites in the DMFT algorithm, in which 8 Lanczos roots are calculated. This is estimated to be reasonably accurate for 2 and 3 bath sites, in accordance with Fig. \ref{fig:gf_4site_sv} in which it was shown that 8 Lanczos roots provide a good approximation to exact diagonalisation for 4 Hubbard sites. Results for 2 bath sites are shown in Fig. \ref{fig:gf_imp_dmft_2bath}. Initially, the DMFT algorithm was run in an ideal fashion, in which the impurity GF was calculated using a noiseless classical (statevector) simulation of the quantum computed moments. In this setting, full DMFT convergence ($\tau=0$) was achieved after 14 DMFT iterations. This is shown by the green line in Fig. \ref{fig:gf_imp_dmft_2bath}. For comparison, a DMFT run was also performed in which classical ED was used to obtain the impurity GF at each DMFT iteration (again achieving convergence after 14 iterations), shown by the black line in Fig. \ref{fig:gf_imp_dmft_2bath}. This compares well with the noiseless approach for 8 Lanczos roots. Following this, the H1-1 emulator (H1-1E) \cite{h11e} was used to quantum compute the impurity GF using the statevector-converged bath parameters (blue squares in Fig. \ref{fig:gf_imp_dmft_2bath}). For 2 bath sites, 65 circuits were measured (emulated hardware) at the final DMFT iteration to obtain the impurity+bath GF, ranging from 193 (81) to 204 (86) total gates (2-qubit gates), with depths of 212 to 214. The spectral function is shown in Fig. \ref{fig:gf_imp_dmft_2bath}. In this case 3 separate applications of the quantum method are applied to the emulator H1-1E, in order to simulate 3 separate hardware runs and assess the variation of results. We observe that the energy positions of low lying spectral peaks generally agree well with the exact result, and are also stable against statistical variations in measurements. Higher lying spectral peaks differ in position from the exact result more significantly, while peak heights (associated with normalisation of Lanczos vectors) are also less accurate and are more sensitive to variations due to measurement statistics. 
Finally, ideal simulations of this approach, in which the Hamiltonian moment expectations are evaluated in an noiseless statevector fashion at all DMFT iterations, are performed for 3 bath sites. As in the previous case, we also ran a variant of the DMFT algorithm in which the impurity GF is obtained from classical ED at each DMFT iteration. In both the quantum (statevector) and classical ED runs for 3 bath sites, the impurity GF was iterated to convergence ($\tau=0$) after 14 DMFT iterations. The resulting spectral functions are shown in Fig. \ref{fig:gf_imp_dmft_3bath}. Disagreements arise in the total number of spectral peaks, and in the weight of low energy poles (consistent with observations made for the spectral function of the 4-site Hubbard model, see section \ref{gf_hubbard_results} and Fig. \ref{fig:gf_4site_sv}). In this case, disagreements with classical ED are due to the continued fraction representation of the GF in Lanczos method; While the Lehmann representation of the GF used in the classical ED is exact for all temperatures T including the limit $\text{T} \rightarrow 0$, this is not the case for the Lanczos calculated GF. In the latter, the exponential factor containing the inverse temperature $\upbeta$ is discarded. Hence, despite the common practice, it is strictly an approximation to use a finite $\upbeta$ in the fitting of the Lanczos-calculated impurity GF. However, this practice is justified by the common observation that the discrepancies between the Lanczos and ED impurity GFs vanish at sufficiently high $\upbeta$ (sufficiently low T). We note that full DMFT convergence is achieved for the quantum Lanczos computed impurity GF, as well as for the classical ED GF, hence the DMFT algorithm finds slightly different solutions corresponding to the differences arising from approximations in the Lanczos computed GF. We choose a relatively low value of $\upbeta$ for quicker convergence of the impurity GF, however we expect the differences between Lanczos and ED impurity GFs to vanish as $\upbeta \rightarrow \infty$. Hence, discrepancies here are not due to errors from quantum noise or measurements, and these results show that this quantum approach in principle works for multiple bath sites in DMFT.
\section{Conclusions} \label{sec:conclusion}

In this paper, a quantum computational approach has been presented to calculate the single particle Green's function in a spin orbital basis, using a cumulant expansion of the Lanczos procedure involving quantum computed moments. This is implemented in the InQuanto package \cite{inquanto_web, inquanto_medium}, and is an extension of a recent work which presented the quantum computed moments approach to obtain infimum estimates of the ground state energy \cite{hollenberg96, vallury20, vallury23}. In our work, we utilise quantum computed moments to obtain the Lanczos coefficients, which in turn can be used to obtain the Green's function in the continued fraction representation. 

Following a brief outline of the method, we show that our implementation allows for multiple strategies to initialise and run the Lanczos procedure; we present two separate strategies, one involving the explicit preparation of the first Lanczos vector on a quantum circuit (strategy \textit{i)}), the other involving measurements of Pauli terms representing the Hamiltonian moments sandwiched between ladder operators indexed by the corresponding Green's function matrix element (strategy \textit{ii)}). While strategy \textit{i)} allows for a flexibility in the choice of representation of the Lanczos vector, we focus on strategy \textit{ii)} for application of the method on trapped ion quantum computers due to the "NISQ-friendly" number of measurable Pauli terms for small sized Hubbard models. 

Using our approach, we computed the GF matrix of the Hubbard dimer on the H1-1 trapped ion quantum computer, showing excellent agreement with the ideal noiseless result in terms of spectral peak positions. We also note the good agreement between H1-1 and H1-1E in Fig. \ref{fig:gf_dimer_hardware}, which also indicates the accurate approximation of hardware results when applying the emulator to larger models in section \ref{subsec:dmft}. Following the GF of the Hubbard Hamiltonian, we then apply our approach to obtain the impurity GF in a DMFT algorithm with up to 3 bath sites. Hardware results again show good agreement with the ideal noiseless result when applied to the final DMFT iteration, and emulated hardware results indicate that errors in the GF due to quantum noise do not prevent convergence of the DMFT algorithm when the GF is quantum computed at all DMFT iterations. 

We note that our approach does not require ancillary qubits (only the ground state or Lanczos vector state circuits, and the Hamiltonian as a sum of Pauli operators). We also emphasise that no error mitigation has been applied to the hardware or emulator results in this paper. This is in contrast to a recent interesting work which utilised error mitigation to calculate linear response properties of molecules, and which would require ancillas for non-symmetric cases \cite{huang22}. In addition, another recent work also investigated a Lanczos recursion method to obtain the GF from quantum computers \cite{baker21}. In the latter, a state-preserving quantum counting algorithm is used to find the Lanczos coefficients. This quantum counting algorithm \cite{baker21} relies on ancillary qubits and a QPE block, which likely render this approach difficult for near-term applications. This is in contrast to our approach which utilizes a cumulant expansion of the Lanczos coefficients in terms of the Hamiltonian moments, rather than a counting algorithm which requires QPE.

We also studied the scaling of the number of measurable Pauli terms with respect to the Hamiltonian moment index, for the sandwiched moment operators required for strategy \textit{ii)}, indicating polynomial scaling and a saturation for high-lying moments. Finally, we mention our investigation of errors in the Lanczos basis and corresponding coefficients arising from quantum noise, and how these can impact elements of the GF matrix. As mentioned in section \ref{results_lanc_vec_prep}, an accurate description of the relation between quantum noise and errors in the Lanczos basis could lead to very useful techniques to correct for these errors. For example, knowledge of this relation could be used to design error mitigation protocols in which noisy measurements of Hamiltonian moments, or the resulting errors in the Lanczos coefficients, could be corrected/mitigated by known calibration data of a particular device. This could potentially widen the application domain of quantum computed Green's functions by allowing for larger system sizes, represented by larger, more error prone circuits. We consider this a fruitful direction for future work. 

\section*{Acknowledgements} \label{sec:acknowledgements}

We acknowledge Yu-ya Ohnishi for the productive discussions throughout this work. We also gratefully acknowledge Georgia Prokopiou and Ramil Nigmatullin for reading the manuscript and providing useful suggestions. We thank Andrew Tranter for helpful discussions on commuting sets of measurable Pauli terms. Finally, we thank Cono Di Paola for useful comments and suggestions during the implementation of the method. 

\bibliography{references.bib}

\begin{thebibliography}{10}

\bibitem{fetter03}
A.L. Fetter and J.D. Walecka.
\newblock ``Quantum theory of many-particle systems''.
\newblock Dover Books on Physics. Dover Publications. ~(2003).

\bibitem{golze19}
Dorothea Golze, Marc Dvorak, and Patrick Rinke.
\newblock ``The {GW} compendium: A practical guide to theoretical photoemission spectroscopy''.
\newblock \href{https://dx.doi.org/10.3389/fchem.2019.00377}{Frontiers in Chemistry {}{\bf 7}}~(2019).

\bibitem{senechal00}
D.~S\'en\'echal, D.~Perez, and M.~Pioro-Ladri\`ere.
\newblock ``Spectral weight of the hubbard model through cluster perturbation theory''.
\newblock \href{https://dx.doi.org/10.1103/PhysRevLett.84.522}{Phys. Rev. Lett. {\bf 84}, 522--525}~(2000).

\bibitem{senechal02}
David S\'en\'echal, Danny Perez, and Dany Plouffe.
\newblock ``Cluster perturbation theory for hubbard models''.
\newblock \href{https://dx.doi.org/10.1103/PhysRevB.66.075129}{Phys. Rev. B {\bf 66}, 075129}~(2002).

\bibitem{georges96}
Antoine Georges, Gabriel Kotliar, Werner Krauth, and Marcelo~J. Rozenberg.
\newblock ``Dynamical mean-field theory of strongly correlated fermion systems and the limit of infinite dimensions''.
\newblock \href{https://dx.doi.org/10.1103/RevModPhys.68.13}{Rev. Mod. Phys. {\bf 68}, 13--125}~(1996).

\bibitem{caffarel94}
Michel Caffarel and Werner Krauth.
\newblock ``Exact diagonalization approach to correlated fermions in infinite dimensions: Mott transition and superconductivity''.
\newblock \href{https://dx.doi.org/10.1103/PhysRevLett.72.1545}{Phys. Rev. Lett. {\bf 72}, 1545--1548}~(1994).

\bibitem{avella13}
A.~Avella and F.~Mancini.
\newblock ``Strongly correlated systems: Numerical methods''.
\newblock \href{https://dx.doi.org/https://doi.org/10.1007/978-3-642-21831-6}{Springer Series in Solid-State Sciences}. Springer Berlin Heidelberg. ~(2013).

\bibitem{coveney23}
Christopher J.~N. Coveney and David~P. Tew.
\newblock ``A regularized second-order correlation method from green's function theory''.
\newblock \href{https://dx.doi.org/10.1021/acs.jctc.3c00246}{Journal of Chemical Theory and Computation {\bf 19}, 3915--3928}~(2023).

\bibitem{marie23}
Antoine Marie, Abdallah Ammar, and Pierre-François Loos.
\newblock ``The ${GW}$ approximation: A quantum chemistry perspective''~(2023).
\newblock  url:~\url{https://doi.org/10.48550/arXiv.2311.05351}.

\bibitem{teke19}
Nakul~K. Teke, Fabijan Pavošević, Chong Peng, and Edward~F. Valeev.
\newblock ``{Explicitly correlated renormalized second-order Green’s function for accurate ionization potentials of closed-shell molecules}''.
\newblock \href{https://dx.doi.org/10.1063/1.5090983}{The Journal of Chemical Physics {\bf 150}, 214103}~(2019).

\bibitem{jamet_kvqa21}
Francois Jamet, Abhishek Agarwal, Carla Lupo, Dan~E. Browne, Cedric Weber, and Ivan Rungger.
\newblock ``Krylov variational quantum algorithm for first principles materials simulations''~(2021).
\newblock  url:~\url{https://doi.org/10.48550/arXiv.2105.13298}.

\bibitem{tilly22}
Jules Tilly, Hongxiang Chen, Shuxiang Cao, Dario Picozzi, Kanav Setia, Ying Li, Edward Grant, Leonard Wossnig, Ivan Rungger, George~H. Booth, and Jonathan Tennyson.
\newblock ``The variational quantum eigensolver: A review of methods and best practices''.
\newblock \href{https://dx.doi.org/https://doi.org/10.1016/j.physrep.2022.08.003}{Physics Reports {\bf 986}, 1--128}~(2022).

\bibitem{kanasugi23}
Shota Kanasugi, Shoichiro Tsutsui, Yuya~O. Nakagawa, Kazunori Maruyama, Hirotaka Oshima, and Shintaro Sato.
\newblock ``Computation of green{\textquotesingle}s function by local variational quantum compilation''.
\newblock \href{https://dx.doi.org/10.1103/physrevresearch.5.033070}{Physical Review Research {}{\bf 5}}~(2023).

\bibitem{jamet_qse22}
Francois Jamet, Abhishek Agarwal, and Ivan Rungger.
\newblock ``Quantum subspace expansion algorithm for green’s functions''~(2022).
\newblock  url:~\url{https://doi.org/10.48550/arXiv.2205.00094}.

\bibitem{libbi22}
Francesco Libbi, Jacopo Rizzo, Francesco Tacchino, Nicola Marzari, and Ivano Tavernelli.
\newblock ``Effective calculation of the green's function in the time domain on near-term quantum processors''.
\newblock \href{https://dx.doi.org/10.1103/PhysRevResearch.4.043038}{Phys. Rev. Res. {\bf 4}, 043038}~(2022).

\bibitem{delre22}
Lorenzo~Del Re, Brian Rost, Michael Foss-Feig, A.~F. Kemper, and J.~K. Freericks.
\newblock ``Robust measurements of n-point correlation functions of driven-dissipative quantum systems on a digital quantum computer''~(2022).
\newblock  url:~\url{https://doi.org/10.48550/arXiv.2204.12400}.

\bibitem{sakurai22}
Rihito Sakurai, Wataru Mizukami, and Hiroshi Shinaoka.
\newblock ``Hybrid quantum-classical algorithm for computing imaginary-time correlation functions''.
\newblock \href{https://dx.doi.org/10.1103/PhysRevResearch.4.023219}{Phys. Rev. Res. {\bf 4}, 023219}~(2022).

\bibitem{dhawan23}
Diksha Dhawan, Dominika Zgid, and Mario Motta.
\newblock ``Quantum algorithm for imaginary-time green's functions''~(2023).
\newblock  url:~\url{https://doi.org/10.48550/arXiv.2309.09914}.

\bibitem{kosugi20}
Taichi Kosugi and Yu-ichiro Matsushita.
\newblock ``Construction of green's functions on a quantum computer: Quasiparticle spectra of molecules''.
\newblock \href{https://dx.doi.org/10.1103/PhysRevA.101.012330}{Phys. Rev. A {\bf 101}, 012330}~(2020).

\bibitem{wang23}
Samson Wang, Sam McArdle, and Mario Berta.
\newblock ``Qubit-efficient randomized quantum algorithms for linear algebra''.
\newblock \href{https://dx.doi.org/10.1103/PRXQuantum.5.020324}{PRX Quantum {\bf 5}, 020324}~(2024).

\bibitem{sun23}
Shi-Ning Sun, Brian Marinelli, Jin~Ming Koh, Yosep Kim, Long~B. Nguyen, Larry Chen, John~Mark Kreikebaum, David~I. Santiago, Irfan Siddiqi, and Austin~J. Minnich.
\newblock ``Quantum computation of frequency-domain molecular response properties using a three-qubit i{T}offoli gate''~(2023).
\newblock  url:~\url{https://doi.org/10.48550/arXiv.2302.04271}.

\bibitem{Endo20}
Suguru Endo, Iori Kurata, and Yuya~O. Nakagawa.
\newblock ``Calculation of the green's function on near-term quantum computers''.
\newblock \href{https://dx.doi.org/10.1103/PhysRevResearch.2.033281}{Phys. Rev. Res. {\bf 2}, 033281}~(2020).

\bibitem{huang22}
Kaixuan Huang, Xiaoxia Cai, Hao Li, Zi-Yong Ge, Ruijuan Hou, Hekang Li, Tong Liu, Yunhao Shi, Chitong Chen, Dongning Zheng, Kai Xu, Zhi-Bo Liu, Zhendong Li, Heng Fan, and Wei-Hai Fang.
\newblock ``Variational quantum computation of molecular linear response properties on a superconducting quantum processor''.
\newblock \href{https://dx.doi.org/10.1021/acs.jpclett.2c02381}{The Journal of Physical Chemistry Letters {\bf 13}, 9114--9121}~(2022).

\bibitem{rungger19}
I.~Rungger, N.~Fitzpatrick, H.~Chen, C.~H. Alderete, H.~Apel, A.~Cowtan, A.~Patterson, D.~Munoz Ramo, Y.~Zhu, N.~H. Nguyen, E.~Grant, S.~Chretien, L.~Wossnig, N.~M. Linke, and R.~Duncan.
\newblock ``Dynamical mean field theory algorithm and experiment on quantum computers''~(2019).
\newblock  url:~\url{https://doi.org/10.48550/arXiv.1910.04735}.

\bibitem{keen20}
Trevor Keen, Thomas Maier, Steven Johnston, and Pavel Lougovski.
\newblock ``Quantum-classical simulation of two-site dynamical mean-field theory on noisy quantum hardware''.
\newblock \href{https://dx.doi.org/10.1088/2058-9565/ab7d4c}{Quantum Science and Technology {\bf 5}, 035001}~(2020).

\bibitem{jaderberg20}
B~Jaderberg, A~Agarwal, K~Leonhardt, M~Kiffner, and D~Jaksch.
\newblock ``Minimum hardware requirements for hybrid quantum{\textendash}classical {DMFT}''.
\newblock \href{https://dx.doi.org/10.1088/2058-9565/ab972b}{Quantum Science and Technology {\bf 5}, 034015}~(2020).

\bibitem{kreula16}
J.~M. Kreula, S.~R. Clark, and D.~Jaksch.
\newblock ``Non-linear quantum-classical scheme to simulate non-equilibrium strongly correlated fermionic many-body dynamics''.
\newblock \href{https://dx.doi.org/10.1038/srep32940}{Scientific Reports {\bf 6}, 32940}~(2016).

\bibitem{bauer16}
Bela Bauer, Dave Wecker, Andrew~J. Millis, Matthew~B. Hastings, and Matthias Troyer.
\newblock ``Hybrid quantum-classical approach to correlated materials''.
\newblock \href{https://dx.doi.org/10.1103/PhysRevX.6.031045}{Phys. Rev. X {\bf 6}, 031045}~(2016).

\bibitem{PhysRevA.104.032405}
Hongxiang Chen, Max Nusspickel, Jules Tilly, and George~H. Booth.
\newblock ``Variational quantum eigensolver for dynamic correlation functions''.
\newblock \href{https://dx.doi.org/10.1103/PhysRevA.104.032405}{Phys. Rev. A {\bf 104}, 032405}~(2021).

\bibitem{gomes23}
Niladri Gomes, David~B. Williams-Young, and Wibe~A. de~Jong.
\newblock ``Computing the many-body green's function with adaptive variational quantum dynamics''.
\newblock \href{https://dx.doi.org/10.1021/acs.jctc.3c00150}{Journal of Chemical Theory and Computation {\bf 19}, 3313--3323}~(2023).

\bibitem{besserve22}
P.~Besserve and T.~Ayral.
\newblock ``Unraveling correlated material properties with noisy quantum computers: Natural orbitalized variational quantum eigensolving of extended impurity models within a slave-boson approach''.
\newblock \href{https://dx.doi.org/10.1103/PhysRevB.105.115108}{Phys. Rev. B {\bf 105}, 115108}~(2022).

\bibitem{backes23}
Steffen Backes, Yuta Murakami, Shiro Sakai, and Ryotaro Arita.
\newblock ``Dynamical mean-field theory for the hubbard-holstein model on a quantum device''.
\newblock \href{https://dx.doi.org/10.1103/PhysRevB.107.165155}{Phys. Rev. B {\bf 107}, 165155}~(2023).

\bibitem{cruz23}
Diogo Cruz and Duarte Magano.
\newblock ``Superresolution of green's functions on noisy quantum computers''.
\newblock \href{https://dx.doi.org/10.1103/PhysRevA.108.012618}{Phys. Rev. A {\bf 108}, 012618}~(2023).

\bibitem{steckmann23}
Thomas Steckmann, Trevor Keen, Efekan K\"okc\"u, Alexander~F. Kemper, Eugene~F. Dumitrescu, and Yan Wang.
\newblock ``Mapping the metal-insulator phase diagram by algebraically fast-forwarding dynamics on a cloud quantum computer''.
\newblock \href{https://dx.doi.org/10.1103/PhysRevResearch.5.023198}{Phys. Rev. Res. {\bf 5}, 023198}~(2023).

\bibitem{vallury20}
Harish~J. Vallury, Michael~A. Jones, Charles~D. Hill, and Lloyd C.~L. Hollenberg.
\newblock ``Quantum computed moments correction to variational estimates''.
\newblock \href{https://dx.doi.org/10.22331/q-2020-12-15-373}{{Quantum} {\bf 4}, 373}~(2020).

\bibitem{vallury23}
Harish~J. Vallury, Michael~A. Jones, Gregory A.~L. White, Floyd~M. Creevey, Charles~D. Hill, and Lloyd C.~L. Hollenberg.
\newblock ``Noise-robust ground state energy estimates from deep quantum circuits''.
\newblock \href{https://dx.doi.org/10.22331/q-2023-09-11-1109}{{Quantum} {\bf 7}, 1109}~(2023).

\bibitem{cade20}
Chris Cade, Lana Mineh, Ashley Montanaro, and Stasja Stanisic.
\newblock ``Strategies for solving the fermi-hubbard model on near-term quantum computers''.
\newblock \href{https://dx.doi.org/10.1103/PhysRevB.102.235122}{Phys. Rev. B {\bf 102}, 235122}~(2020).

\bibitem{stanisic22}
Stasja Stanisic, Jan~Lukas Bosse, Filippo~Maria Gambetta, Raul~A. Santos, Wojciech Mruczkiewicz, Thomas~E. O'Brien, Eric Ostby, and Ashley Montanaro.
\newblock ``Observing ground-state properties of the fermi-hubbard model using a scalable algorithm on a quantum computer''.
\newblock \href{https://dx.doi.org/10.1038/s41467-022-33335-4}{Nature Communications {\bf 13}, 5743}~(2022).

\bibitem{poulin09}
David Poulin and Pawel Wocjan.
\newblock ``Preparing ground states of quantum many-body systems on a quantum computer''.
\newblock \href{https://dx.doi.org/10.1103/physrevlett.102.130503}{Physical Review Letters {}{\bf 102}}~(2009).

\bibitem{ge19}
Yimin Ge, Jordi Tura, and J.~Ignacio Cirac.
\newblock ``{Faster ground state preparation and high-precision ground energy estimation with fewer qubits}''.
\newblock \href{https://dx.doi.org/10.1063/1.5027484}{Journal of Mathematical Physics {\bf 60}, 022202}~(2019).

\bibitem{lin20}
Lin Lin and Yu~Tong.
\newblock ``Near-optimal ground state preparation''.
\newblock \href{https://dx.doi.org/10.22331/q-2020-12-14-372}{Quantum {\bf 4}, 372}~(2020).

\bibitem{inquanto_web}
Andrew Tranter, Cono Di~Paola, David Mu\~{n}oz Ramo, David~Zsolt Manrique, Duncan Gowland, Evgeny Plekhanov, Gabriel Greene-Diniz, Georgia Christopoulou, Georgia Prokopiou, Harry D~J Keen, Iakov Polyak, Irfan~T Khan, Jerzy Pilipczuk, Josh J~M Kirsopp, Kentaro Yamamoto, Maria Tudorovskaya, Michal Krompiec, Michelle Sze, and Nathan Fitzpatrick.
\newblock ``\href{https://www.quantinuum.com/products/inquanto}{In{Q}uanto: Quantum {C}omputational {C}hemistry}''~(2022).

\bibitem{inquanto_medium}
Andrew Tranter, Cono Di~Paola, David Mu\~{n}oz Ramo, David~Zsolt Manrique, Duncan Gowland, Evgeny Plekhanov, Gabriel Greene-Diniz, Georgia Christopoulou, Georgia Prokopiou, Harry D~J Keen, Iakov Polyak, Irfan~T Khan, Jerzy Pilipczuk, Josh J~M Kirsopp, Kentaro Yamamoto, Maria Tudorovskaya, Michal Krompiec, Michelle Sze, and Nathan Fitzpatrick.
\newblock ``\href{https://medium.com/cambridge-quantum-computing/introduction-to-the-inquanto-computational-chemistry-platform-for-quantum-computers-4fced08d66cc}{Introduction to the {I}n{Q}uanto {C}omputational {C}hemistry {P}latform {F}or {Q}uantum {C}omputers}''~(2022).

\bibitem{molmer99}
Klaus M\o{}lmer and Anders S\o{}rensen.
\newblock ``Multiparticle entanglement of hot trapped ions''.
\newblock \href{https://dx.doi.org/10.1103/PhysRevLett.82.1835}{Phys. Rev. Lett. {\bf 82}, 1835--1838}~(1999).

\bibitem{h11e}
Quantinuum {H}1-1. \url{https://www.quantinuum.com/}, {F}ebruary - {M}arch, 2023.

\bibitem{hollenberg96}
Lloyd C.~L. Hollenberg and N.~S. Witte.
\newblock ``Analytic solution for the ground-state energy of the extensive many-body problem''.
\newblock \href{https://dx.doi.org/10.1103/PhysRevB.54.16309}{Phys. Rev. B {\bf 54}, 16309--16312}~(1996).

\bibitem{lanczos50}
Cornelius Lanczos.
\newblock ``An iteration method for the solution of the eigenvalue problem of linear differential and integral operators''.
\newblock \href{https://dx.doi.org/10.6028/jres.045.026}{Journal of research of the National Bureau of Standards {\bf 45}, 255--282}~(1950).

\bibitem{hollenberg93}
Lloyd C.~L. Hollenberg.
\newblock ``Plaquette expansion in lattice hamiltonian models''.
\newblock \href{https://dx.doi.org/10.1103/PhysRevD.47.1640}{Phys. Rev. D {\bf 47}, 1640--1644}~(1993).

\bibitem{hollenberg94}
Lloyd C.~L. Hollenberg and N.~S. Witte.
\newblock ``General nonperturbative estimate of the energy density of lattice hamiltonians''.
\newblock \href{https://dx.doi.org/10.1103/PhysRevD.50.3382}{Phys. Rev. D {\bf 50}, 3382--3386}~(1994).

\bibitem{andrews23}
Bartholomew Andrews and Gunnar Möller.
\newblock ``Self-similarity of spectral response functions for fractional quantum hall states''.
\newblock \href{https://dx.doi.org/10.1098/rspa.2023.0021}{Proceedings of the Royal Society A: Mathematical, Physical and Engineering Sciences {\bf 479}, 20230021}~(2023).

\bibitem{pavarini11}
Eva Pavarini, Erik Koch, Alexander Lichtenstein, and Dieter Vollhardt, editors.
\newblock ``{T}he {LDA}+{DMFT} approach to strongly correlated materials''.
\newblock Volume~1.
\newblock Forschungszenrum Jülich GmbH. ~(2011).
\newblock  url:~\url{http://hdl.handle.net/2128/7348}.

\bibitem{jordan28}
P.~Jordan and E.~Wigner.
\newblock ``{\"U}ber das paulische {\"a}quivalenzverbot''.
\newblock \href{https://dx.doi.org/10.1007/BF01331938}{Zeitschrift f{\"u}r Physik {\bf 47}, 631--651}~(1928).

\bibitem{tazi23}
Lila~Cadi Tazi and Alex J.~W. Thom.
\newblock ``Folded spectrum vqe : A quantum computing method for the calculation of molecular excited states''~(2023).
\newblock  url:~\url{https://arxiv.org/abs/2305.04783}.

\bibitem{verteletskyi20}
Vladyslav Verteletskyi, Tzu-Ching Yen, and Artur~F. Izmaylov.
\newblock ``{Measurement optimization in the variational quantum eigensolver using a minimum clique cover}''.
\newblock \href{https://dx.doi.org/10.1063/1.5141458}{The Journal of Chemical Physics {\bf 152}, 124114}~(2020).

\bibitem{arrazola22}
Juan~Miguel Arrazola, Olivia Di~Matteo, Nicol{\'{a}}s Quesada, Soran Jahangiri, Alain Delgado, and Nathan Killoran.
\newblock ``Universal quantum circuits for quantum chemistry''.
\newblock \href{https://dx.doi.org/10.22331/q-2022-06-20-742}{{Quantum} {\bf 6}, 742}~(2022).

\bibitem{mcardle19}
Sam McArdle, Tyson Jones, Suguru Endo, Ying Li, Simon~C. Benjamin, and Xiao Yuan.
\newblock ``Variational ansatz-based quantum simulation of imaginary time evolution''.
\newblock \href{https://dx.doi.org/10.1038/s41534-019-0187-2}{npj Quantum Information {\bf 5}, 75}~(2019).

\bibitem{motta20}
Mario Motta, Chong Sun, Adrian T.~K. Tan, Matthew~J. O'Rourke, Erika Ye, Austin~J. Minnich, Fernando G. S.~L. Brand{\~a}o, and Garnet Kin-Lic Chan.
\newblock ``Determining eigenstates and thermal states on a quantum computer using quantum imaginary time evolution''.
\newblock \href{https://dx.doi.org/10.1038/s41567-019-0704-4}{Nature Physics {\bf 16}, 205--210}~(2020).

\bibitem{anand22}
Abhinav Anand, Philipp Schleich, Sumner Alperin-Lea, Phillip W.~K. Jensen, Sukin Sim, Manuel Díaz-Tinoco, Jakob~S. Kottmann, Matthias Degroote, Artur~F. Izmaylov, and Alán Aspuru-Guzik.
\newblock ``A quantum computing view on unitary coupled cluster theory''.
\newblock \href{https://dx.doi.org/10.1039/D1CS00932J}{Chem. Soc. Rev. {\bf 51}, 1659--1684}~(2022).

\bibitem{kandala17}
Abhinav Kandala, Antonio Mezzacapo, Kristan Temme, Maika Takita, Markus Brink, Jerry~M. Chow, and Jay~M. Gambetta.
\newblock ``Hardware-efficient variational quantum eigensolver for small molecules and quantum magnets''.
\newblock \href{https://dx.doi.org/10.1038/nature23879}{Nature {\bf 549}, 242--246}~(2017).

\bibitem{grimsley19}
Harper~R. Grimsley, Sophia~E. Economou, Edwin Barnes, and Nicholas~J. Mayhall.
\newblock ``An adaptive variational algorithm for exact molecular simulations on a quantum computer''.
\newblock \href{https://dx.doi.org/10.1038/s41467-019-10988-2}{Nature Communications {\bf 10}, 3007}~(2019).

\bibitem{yordanov21}
Yordan~S. Yordanov, V.~Armaos, Crispin H.~W. Barnes, and David R.~M. Arvidsson-Shukur.
\newblock ``Qubit-excitation-based adaptive variational quantum eigensolver''.
\newblock \href{https://dx.doi.org/10.1038/s42005-021-00730-0}{Communications Physics {\bf 4}, 228}~(2021).

\bibitem{gomes21}
Niladri Gomes, Anirban Mukherjee, Feng Zhang, Thomas Iadecola, Cai-Zhuang Wang, Kai-Ming Ho, Peter~P. Orth, and Yong-Xin Yao.
\newblock ``Adaptive variational quantum imaginary time evolution approach for ground state preparation''.
\newblock \href{https://dx.doi.org/https://doi.org/10.1002/qute.202100114}{Advanced Quantum Technologies {\bf 4}, 2100114}~(2021).

\bibitem{khan22}
I.~T. Khan, M.~Tudorovskaya, J.~J.~M. Kirsopp, D.~Muñoz Ramo, P.~W. Warrier, D.~K. Papanastasiou, and R.~Singh.
\newblock ``Chemically aware unitary coupled cluster with ab initio calculations on system model h1: A refrigerant chemicals application''~(2022).
\newblock  url:~\url{https://arxiv.org/abs/2210.14834}.

\bibitem{yordanov20}
Yordan~S. Yordanov, David R.~M. Arvidsson-Shukur, and Crispin H.~W. Barnes.
\newblock ``Efficient quantum circuits for quantum computational chemistry''.
\newblock \href{https://dx.doi.org/10.1103/PhysRevA.102.062612}{Phys. Rev. A {\bf 102}, 062612}~(2020).

\bibitem{tket20}
Seyon Sivarajah, Silas Dilkes, Alexander Cowtan, Will Simmons, Alec Edgington, and Ross Duncan.
\newblock ``t{$|$ket$\rangle$}: a retargetable compiler for {NISQ} devices''.
\newblock \href{https://dx.doi.org/https://doi.org/10.1088/2058-9565/ab8e92}{Quantum Science and Technology {\bf 6}, 014003}~(2020).

\bibitem{anderson61}
P.~W. Anderson.
\newblock ``Localized magnetic states in metals''.
\newblock \href{https://dx.doi.org/10.1103/PhysRev.124.41}{Phys. Rev. {\bf 124}, 41--53}~(1961).

\bibitem{epperly22}
Ethan~N. Epperly, Lin Lin, and Yuji Nakatsukasa.
\newblock ``A theory of quantum subspace diagonalization''.
\newblock \href{https://dx.doi.org/10.1137/21M145954X}{SIAM Journal on Matrix Analysis and Applications {\bf 43}, 1263--1290}~(2022).

\bibitem{kirby23}
William Kirby, Mario Motta, and Antonio Mezzacapo.
\newblock ``Exact and efficient {L}anczos method on a quantum computer''.
\newblock \href{https://dx.doi.org/10.22331/q-2023-05-23-1018}{{Quantum} {\bf 7}, 1018}~(2023).

\bibitem{inquanto_docs}
Andrew Tranter, Cono Di~Paola, David Mu\~{n}oz Ramo, David~Zsolt Manrique, Duncan Gowland, Evgeny Plekhanov, Gabriel Greene-Diniz, Georgia Christopoulou, Georgia Prokopiou, Harry D~J Keen, Iakov Polyak, Irfan~T Khan, Jerzy Pilipczuk, Josh J~M Kirsopp, Kentaro Yamamoto, Maria Tudorovskaya, Michal Krompiec, Michelle Sze, and Nathan Fitzpatrick.
\newblock ``\href{https://inquanto.quantinuum.com}{In{Q}uanto {U}ser {G}uide}''~(2024).

\bibitem{qiskit}
{Qiskit contributors}.
\newblock ``\href{https://github.com/Qiskit/qiskit}{Qiskit: An {O}pen-source {F}ramework for {Q}uantum {C}omputing}''~(2023).

\bibitem{baker21}
Thomas~E. Baker.
\newblock ``Lanczos recursion on a quantum computer for the green's function and ground state''.
\newblock \href{https://dx.doi.org/10.1103/PhysRevA.103.032404}{Phys. Rev. A {\bf 103}, 032404}~(2021).

\end{thebibliography}


\end{document}